\documentclass[amsmath,amssymb,aps,prd,onecolumn,floatfix,natbib,longbibliography]{revtex4-2}
\pdfoutput=1
\usepackage{lipsum}
\usepackage{autobreak}
\usepackage{graphicx}
\usepackage{appendix}
\usepackage{amsfonts}
\usepackage{comment}
\usepackage[bookmarks=false,colorlinks]{hyperref}
\hypersetup{urlcolor=blue, citecolor=blue}
\usepackage[utf8x]{inputenc}
\usepackage[normalem]{ulem}

\newcommand{\ben}{\begin{enumerate}}
\newcommand{\een}{\end{enumerate}}
\newcommand{\beq}{\begin{equation}}
\newcommand{\eeq}{\end{equation}}
\newcommand{\bal}{\begin{align}}
\newcommand{\eal}{\end{align}}

\newcommand{\bea}{\begin{eqnarray}}
\newcommand{\eea}{\end{eqnarray}}
\newcommand{\nn}{\nonumber\\}

\def\z#1{\zeta_{#1}}

\newcommand{\Ca}{C_A}
\newcommand{\Cf}{C_F}
\newcommand{\nf}{n_F}
\newcommand{\nfv}{n_{fv}}

\newcommand{\df}{{\rm d}}
\def\Dm1{{{\delta(1-z)}}}

\def\nf{{n_{\! f}}}

\def\A#1{{A_{#1}}}
\newcommand{\Lqrfrt}{L_{qr}L_{fr}^{2}}
\newcommand{\Lqrfr}{L_{qr}L_{fr}}
\newcommand{\Lqrtfr}{L_{qr}^{2}L_{fr}}

\def\g0#1DY{{g_{_{0#1}}}}

\newcommand{\ncfodyzh}{$\sigma^{\text{DY,ZH}}_{\text{N$^3$LO}}$}
\newcommand{\nclldyzh}{$\sigma^{\text{DY,ZH}}_{\text{N$^3$LO+N$^3$LL}}$}
\newcommand{\ncfodywh}{$\sigma^{\text{DY,WH}}_{\text{N$^3$LO}}$}
\newcommand{\nclldywh}{$\sigma^{\text{DY,WH}}_{\text{N$^3$LO+N$^3$LL}}$}

\newcommand{\dRAoNR }{\frac{d_R^{abcd}d_A^{abcd}}{N_R}}
\newcommand{\dRFoNR }{\frac{d_R^{abcd}d_F^{abcd}}{N_R}}

\def\gNB#1{{\bar{g}_{#1}}}
\newcommand{\Lqr}{L_{qr}}
\newcommand{\Lfr}{L_{fr}}

\newcommand{\iW}{\omega^{-1}}
\def\LogmW1{{{\ln (1-\omega)}}}
\def\w{{\omega}}
\def\WbimW{{\frac{\omega}{(1-\omega)}}}
\def\LogomWtIMW{{\frac{\ln(1-\omega)}{(1-\omega)}}}
\def\LogomWtIMWt{{\frac{\ln (1-\omega)}{(1-\omega)^2}}}
\def\LogtmWtIMWt{{\frac{\ln (1-\omega)^2}{(1-\omega)^2}}}
\def\LogttmWtIMWt{{\frac{\ln (1-\omega)^3}{(1-\omega)^2}}}

\def\LogtmWtIMW{{\frac{\ln (1-\omega)^2}{(1-\omega)}}}
\def\WttmWtimWt{{\frac{\omega(2-\omega)}{(1-\omega)^2}}}
\def\WtbimWt{{\frac{\omega^2}{(1-\omega)^2}}}

\def\btzAIII{{\frac{A_3}{\beta_0^2}}}
\def\btzAII{{\frac{A_2}{\beta_0}}}
\def\btzAI{{A_1}}

\def\btzDII{{\frac{D_2}{\beta_0}}}
\def\btzDI{{D_1}}

\def\btztAIV{{\frac{A_4}{\beta_0^2}}}
\def\btztAIII{{\frac{A_3}{\beta_0}}}
\def\btztAII{{A_2}}
\def\btztAI{{\beta_0 A_1}}

\def\btztDIII{{\frac{D_3}{\beta_0}}}
\def\btztDII{{D_2}}
\def\btztDI{{\beta_0 D_1}}

\def\bobt{{\frac{\beta_1 \beta_2}{\beta_0^5}}}

\def\AAo{{\frac{A_1}{\beta_{0}}}}
\def\AAt{{\frac{A_2}{\beta_{0}^2}}}

\def\DDo{{\frac{D_1}{\beta_{0}}}}

\newcommand{\btoo}{{\bigg(\frac{\beta_{2}}{\beta_0^{3}}\bigg)}}
\newcommand{\bthr}{{\bigg(\frac{\beta_{3}}{\beta_0^{4}}\bigg)}}
\newcommand{\overbar}[1]{\,\overline{\!{#1}}}
\newcommand{\Nbar}{\overbar{N}}
\newcommand{\gbar}{\overbar{g}}
\newcommand{\as}{a_S}
\newcommand{\muf}{\mu_F}
\newcommand{\mur}{\mu_R}

\newcommand{\eq}[1]{Eq.\ (\ref{#1})}
\newcommand{\fig}[1]{Fig.\ \ref{#1}}
\newcommand{\tab}[1]{Table\ \ref{#1}}
\newcommand{\sect}[1]{Sec.\ \ref{#1}}

\begin{document}

\preprint{
        TTK-22-34
        \\
        \vspace{-2cm}
        \hspace{13cm}P3H-22-106}

\title{{Threshold enhanced cross sections for colorless productions}}
\author{Goutam Das}
\email{goutam@physik.rwth-aachen.de}
\affiliation{Institut f{\"u}r Theoretische Teilchenphysik und Kosmologie,\\ RWTH Aachen University,D-52056 Aachen, Germany}
\author{Chinmoy Dey}
\email{d.chinmoy@iitg.ac.in}
\affiliation{Department of Physics, Indian Institute of Technology Guwahati, Guwahati-781039, Assam, India}
\author{M. C. Kumar}
\email{mckumar@iitg.ac.in}
\affiliation{Department of Physics, Indian Institute of Technology Guwahati, Guwahati-781039, Assam, India}
\author{Kajal Samanta}
\email{kajal003@ucla.edu}
\affiliation{Department of Physics and Center for Field Theory and Particle Physics, \\
Fudan University, Shanghai 200433, China}
\affiliation{Department of Physics and Astronomy, University of California, Los Angeles, CA-90095,  USA}


\begin{abstract}
	We study the threshold effect for neutral and 
charged Drell-Yan productions, associated production of 
Higgs boson with a massive vector boson and Higgs 
production in bottom quark annihilation at the Large Hardon Collider to the 
third order in QCD. Using the third order soft-virtual 
results for these processes and exploiting the universality 
of the threshold logarithms, we extract the 
process-dependent coefficients for these processes 
and resum large threshold logarithms to 
next-to-next-to-next-to leading logarithmic (N$^3$LL) 
accuracy. By matching our results to the recently available 
N$^3$LO results, we provide the most 
precise theoretical predictions for these processes.
We present numerical results for invariant mass 
distribution and total production cross sections.  
We find the conventional scale uncertainties of about 
$0.4\%$ at N$^3$LO level in the fixed order results get 
reduced to as small as less than $0.1\%$ at N$^3$LO+N$^3$LL 
level in the high invariant mass region.
\end{abstract}


\maketitle
\section{Introduction} \label{sec:introduction}
Color singlet production processes are 
important to understand properties of elementary 
particles, the proton structure as well as 
to extract fundamental quantities like coupling 
and masses at the colliders \cite{Heinrich:2020ybq} 
like the Large Hadron Collider (LHC).
From theoretical point of view, they are relatively 
easier to understand, and experimentally they 
often provide clear signatures. Processes 
like the Drell-Yan (DY) productions \cite{PhysRevLett.25.316} are  
standard candles at the hadron colliders and are 
important for luminosity monitoring. On the other 
hand processes like Higgs-strahlung where 
a Higgs boson ($H$) is produced in association 
with a massive vector boson ($V=Z,W$) are important 
to understand Higgs boson properties and also 
to search new physics beyond the Standard Model (SM).
In this article we focus our study on a few such processes.  
In particular, we studied the neutral Drell-Yan (\textit{nDY})
production where a $Z$ boson or virtual 
$\gamma$ decays to a lepton-antilepton pair, 
charged DY (\textit{cDY}) 
where a $W$ boson decays to a lepton and neutrino,
associated production of 
Higgs boson with a vector boson like $Z$ or $W$, and
Higgs boson production through
bottom quark annihilation. All these processes
are produced through quark annihilation at the 
born level. 
The next-to-leading order (NLO) QCD corrections to the DY production
process were computed about 40 years ago \cite{Altarelli:1979ub,Abad:1978gx,Kubar-Andre:1978eri,Harada:1979bj,Humpert:1979qk,Kubar:1980zv}, 
and the next-to-next-to-leading order (NNLO) QCD results
are now known \cite{Hamberg:1990np} for quite some time.
The higher order QCD corrections have also been studied in the 
BSM context \cite{Mathews:2004xp,Ahmed:2016qhu,Banerjee:2017ewt}.

Recently the fixed order (FO) predictions 
have been improved to 
next-to-next-to-next-to leading order (N$^3$LO) 
\cite{Anastasiou:2015vya,Dreyer:2016oyx,Duhr:2020seh,Duhr:2020sdp,Duhr:2019kwi,Duhr:2021vwj,Chen:2019fhs,Dreyer:2018qbw} 
and public codes are available 
\cite{Dreyer:2016oyx,Bonvini:2016frm,Harlander:2016hcx,Dulat:2018rbf,Baglio:2022wzu} 
which provide
the flexibility of studying N$^3$LO cross section 
with different parameters, parton distribution functions
(PDFs) and scales.
There is also recent progress on the EW corrections
which could be competitive as the 
QCD correction in the higher orders.
At NLO, the EW corrections for massive gauge bosons 
have been performed in
\cite{Dittmaier:2001ay,
Baur:2001ze,
Baur:2004ig,
Arbuzov:2005dd,
CarloniCalame:2006zq,
Zykunov:2005tc,
CarloniCalame:2007cd,
Arbuzov:2007db,
Dittmaier:2009cr},
whereas 
mixed EW-QCD corrections are also 
studied recently at NNLO
\cite{
Bonciani:2020tvf,
Bonciani:2021zzf,
Bonciani:2021iis,
Armadillo:2022bgm}
and are known to give contribution as 
large as $-1.5\%$ of NLO QCD corrections in the
high invariant mass region.
The mixed QCD-QED corrections 
for $b\bar{b}H$ can be found in \cite{Ajjath:2019ixh}

While improving the FO calculations, it is essential 
to push precision frontiers, for often they are not 
sufficient to correctly describe the data.
In particular, FO predictions are plagued 
by the large logarithmic contributions which 
appear at each order of perturbation theory.
A particular class of these enhanced logarithms 
appear in the threshold region when 
partonic threshold variable $z\to 1$. 
Truncated FO predictions
at a relatively lower order are insufficient to 
capture the dominant contributions from the 
large threshold logarithms. On the other hand,
the universality of these logarithms allows us 
to resum them to all orders and capture 
significant contribution which is essential 
to compare to the experimental predictions.

The singular part of the fixed order contribution 
\textit{viz.}\ the soft-virtual (SV) corrections to 
these processes are known for a long time through 
N$^3$LO and beyond  
\cite{Anastasiou:2014vaa,Anastasiou:2014lda,Ahmed:2014cha,Ajjath:2019neu,Kumar:2014uwa,Ahmed:2014cla,Ahmed:2019udm,Li:2014bfa,Catani:2014uta,Das:2019uvh,Das:2019btv,Das:2020adl}.
Some of these results have been used to perform threshold 
resummation up to N$^3$LL. Threshold resummation 
has been well studied from the early days of QCD for 
a wide range of processes 
\cite{Sterman:1986aj,Catani:1989ne}.
This is possible due to refactorization 
\cite{
Catani:1996yz,
Contopanagos:1996nh,
Magnea:2000ss,
Catani:2003zt,
Manohar:2003vb,
Eynck:2003fn,
Moch:2005ba,
Moch:2005ky,
Laenen:2005uz,
Ravindran:2005vv,
Ravindran:2006cg,
Idilbi:2006dg,
Becher:2006mr}
of partonic cross sections in terms of 
soft and hard functions in the threshold region.
For color-singlet productions,
they are extensively studied to N$^3$LL accuracies 
\cite{
deFlorian:2012za,
Catani:2014uta,
Bonvini:2014joa,
Harlander:2014wda,
Bonvini:2016frm,
Ajjath:2019neu,
Ajjath:2020rci,
Ajjath:2021bbm,
Ajjath:2021lvg,
Ajjath:2022kpv} 
and even to N$^4$LL 
\cite{Das:2019btv,Das:2019uvh,Das:2020adl} for 
some processes in the SM.
In general the inclusion of threshold resummation results
into better perturbative convergence with an improved 
scale uncertainty. In \cite{Ajjath:2020rci}, it was 
pointed out that the threshold resummation is important 
in the DY invariant mass distribution in the moderate 
and high invariant regions and results $2\%$ correction 
to the fixed order.
The scale uncertainty reduces to below $1\%$ in the 
high invariant mass region $Q>1500$ GeV where a 
matching at third order was limited only to the SV part.
A general framework has been provided for 
arbitrary color singlet production in 
\cite{Forslund:2020lnu,Ahmed:2020nci}.
Beyond the SM (BSM), these are also studied for 
a wide range of models and stringent bounds have 
been obtained with models parameters with 
precise N$^3$LL results
\cite{Ahmed:2016otz,Schmidt:2015cea,deFlorian:2007sr,Das:2019bxi,Das:2020gie,Das:2020pzo,Beenakker:2009ha,Beenakker:2016lwe,Fiaschi:2022odp,Kramer:2019wvd,Bhattacharya:2021hae}.
However, one also needs to 
match the threshold logarithmic contributions 
with the FO results to capture the subleading
terms which become important at higher orders as 
well. The recent results on the FO frontiers 
allow us to study this with the availability of 
a fixed order code \texttt{n3loxs} \cite{Baglio:2022wzu} 
shipped with a range of color-singlet processes at 
N$^3$LO.

The purpose of the present article is thus to perform
a complete study with properly matching N$^3$LL 
resummed results with the newly available N$^3$LO
results. 
We extract all the process-dependent and universal 
coefficients needed for N$^3$LL resummation following
the formalism developed in 
\cite{Moch:2005ba,
Moch:2005ky,
Laenen:2005uz,
Ravindran:2005vv,
Ravindran:2006cg}.
It is worth mentioning that the Higgs boson
production in gluon fusion at the LHC has been studied
extensively in the literature. 
The mass of the Higgs boson being around $125$ GeV, 
the underlying parton fluxes at high energy hadron 
colliders are large. In addition to the fact that 
the lowest order dominant contribution comes from the gluon
fusion channel, the QCD corrections cannot be 
neglected even beyond NNLO in QCD.
This led to the computation of higher order QCD 
corrections to N$^3$LO QCD in the fixed order
and to N$^3$LL accuracy in the context of resummation, 
in order to achieve
the robust precision studies for this process. 
The details of these precision
studies can be found in \cite{Harlander:2002wh,Ravindran:2003um,Anastasiou:2014vaa,Anastasiou:2014lda,Bonvini:2014joa,Bonvini:2014tea,Bonvini:2015ira,Bonvini:2016frm,Ahmed:2016otz,Bonvini:2018xvt,Mistlberger:2018etf}.  
Hence, we will not repeat them here but
rather focus on the other color-singlet production 
processes at hadron colliders.
The article is organized as follows: in 
\sect{sec:theory}, we briefly lay out the theoretical 
framework providing the essential formulas to perform 
threshold resummation and the matching. In 
\sect{sec:numerics} we present the phenomenological 
results for different color-singlet processes in the 
context of LHC, and finally we conclude in 
\sect{sec:conclusion}. 

\section{Theoretical Framework} \label{sec:theory}
The hadronic cross section for colorless
production
at the hadron collider is given by,
\begin{align}\label{eq:had-xsect}
	\sigma(Q^2)
	&=\sum_{a,b={q,\overline q,g}} 
	\int_0^1 \df x_1
	\int_0^1 \df x_2~ f_a(x_1,\muf^2) ~
	f_b(x_2,\muf^2) 
	 \int_0^1 \df z \,\,
	\hat{\sigma}_{ab}(z,Q^2,\muf^2)
	\delta(\tau-z x_1 x_2)\,,
\end{align}
where $\sigma(Q^2) \equiv Q^2\df \sigma/\df Q^2$ for 
the DY-type processes and 
$\sigma(Q^2)\equiv\sigma(M_H^2)$ 
for $b\bar{b}H$ process.
Note that for the total production cross section 
for $VH$, we integrate this over the invariant mass $Q$
of the final state $VH$.
The hadronic and partonic threshold variables $\tau$ 
and $z$ are defined as
\begin{align}
\tau=\frac{Q^2}{S}, \qquad z= \frac{Q^2}{\hat{s}} \,,
\end{align}
where $S$ and $\hat{s}$ are the hadronic and partonic 
center of mass energies respectively.
$\tau$ and $z$ are thus related by $\tau = x_1 x_2 z$.
The partonic coefficient $\hat{\sigma}_{ab}$
can be further decomposed as follows,
\begin{align}\label{eq:partonic-decompose}
	\hat{\sigma}_{ab}(z,Q^2,\mu_F^2)
	=
	\sigma^{(0)}(Q^2) \Big( 
		\Delta_{ab}^{\rm sv}\left(z,\muf^2\right) 
		+ \Delta_{ab}^{\rm reg}\left(z,\muf^2\right)
			\Big) \,.
\end{align}
The term 
$\Delta_{ab}^{\rm sv}$ is known as the soft-virtual (SV)
partonic coefficient and captures all the singular terms
in the $z \to 1$ limit. The 
$\Delta_{ab}^{\rm reg}$ term contains regular contributions 
in the variable $z$.
The overall 
normalization factor 
$\sigma^{(0)}$ depends on the process 
under study. In particular, the 
processes under consideration take the following forms,
\begin{align}
	\sigma_{DY}^{(0)}(Q^2) 
	&= 
	\frac{\pi }{n_c } 
	\bigg[ {\cal F}_{DY}^{(0)}(Q^2) \bigg]\,, 
	\qquad \text{with } DY \in \{nDY, cDY, ZH, WH \} \,,
	\nn
	\sigma_{b\bar{b}H}^{(0)} 
	&= 
	\frac{\pi m_b^{2}(\mu_R^2) \tau}{6 M_H^2 v^2} \,,
\end{align}
where,
\begin{align}
	{\cal F}_{nDY}^{(0)}(Q^2) =&
	{4 \alpha^2 \over 3 S} \Bigg[Q_q^2 
	- {2 Q^2 (Q^2-M_Z^2) \over  \left((Q^2-M_Z^2)^2
	+ M_Z^2 \Gamma_Z^2\right) c_w^2 s_w^2} Q_q g_e^V g_q^V 
\nn
	&+ {Q^4 \over  \left((Q^2-M_Z^2)^2+M_Z^2 \Gamma_Z^2\right) c_w^4 s_w^4}\Big((g_e^V)^2
	+ (g_e^A)^2\Big)\Big((g_q^V)^2+(g_q^A)^2\Big) \Bigg]\,,
\nn
	{\cal F}_{cDY}^{(0)}(Q^2) =&
	{4 \alpha^2 \over 3 S} \Bigg[
	{Q^4 |V_{qq^{'}}|^2 \over  \left((Q^2-M_W^2)^2+M_W^2 \Gamma_W^2\right) s_w^4}\Big((g_e^{'V})^2
	+ (g_e^{'A})^2\Big)\Big((g_q^{'V})^2+(g_q^{'A})^2\Big) \Bigg]\,,
\nn
	\mathcal{F}^{(0)}_{ZH}(Q^2) =& 
	\frac{\alpha^2}{S}\Bigg[ 
		\frac{M_Z^2Q^2\lambda^{1/2}(Q^2,M_H^2,M_Z^2)
		\bigg( 1 + \frac{\lambda(Q^2,M_H^2,M_Z^2)}{12 M_Z^2/Q^2}\bigg)}{((Q^2-M_Z^2)^2+ M_Z^2\Gamma_Z^2)c_w^4s_w^4}
        \bigg((g_q^V)^2+(g_q^A)^2 \bigg)
	\Bigg]\,,
\nn
	\mathcal{F}^{(0)}_{WH}(Q^2) =& 
	\frac{\alpha^2}{S}\Bigg[ 
		\frac{M_W^2Q^2 |V_{qq^{'}}|^2\lambda^{1/2}(Q^2,M_H^2,M_W^2)
		\bigg( 1 + \frac{\lambda(Q^2,M_H^2,M_W^2)}{12 M_W^2/Q^2}\bigg)}{((Q^2-M_W^2)^2+M_W^2\Gamma_W^2)s_w^4}
        \bigg((g_q^{'V})^2+(g_q^{'A})^2 \bigg)
	\Bigg]\,.
\end{align}
Here, 
$V_{qq^{'}}$ are the CKM matrix coefficients with 
$Q_{q} + Q_{q'}=\pm 1$ 
and 
\begin{align}
g_a^A =& -\frac{1}{2} T_a^3 \,, \qquad 
g_a^V = \frac{1}{2} T_a^3  - s_w^2 Q_a \,,
\nn
g_a^{'A} =& -\frac{1}{2\sqrt{2}}\,, \qquad 
g_a^{'V} = \frac{1}{2\sqrt{2}}\,,
\end{align} 
where $Q_a$ is the electric charge and $T_a^3$ is the 
weak isospin of the fermions. Here, $M_V$ and $M_H$ are
the masses of the weak gauge boson and Higgs boson respectively,
$m_b$ is the mass of the bottom quark and $v$ being the vacuum expectation value. 
The function $\lambda$ which appears in the $VH$ case
is defined as
\begin{align}
	\lambda(z,y,x) = 
	\bigg(1-\frac{x}{z}-
	\frac{y}{z} \bigg)^2 -4\frac{xy}{z^2}\,.
\end{align}
The singular part of the partonic coefficient
has a universal structure which gets 
contributions from the underlying hard form 
factor 
\cite{Moch:2005tm,
Moch:2005id,
Baikov:2009bg,
Gehrmann:2010ue,
Gehrmann:2014vha}, 
mass factorization kernels 
\cite{Moch:2004pa,Vogt:2004mw}, 
and soft radiations 
\cite{Ravindran:2005vv,
Ravindran:2006cg,
Sudakov:1954sw,
Mueller:1979ih,
Collins:1980ih,
Sen:1981sd}. All of these are 
infrared divergent which, when regularized 
and combined give finite contributions.
The finite singular part of these 
has the universal structure in terms of 
$\delta(1-z)$ and plus-distributions 
${\cal D}_i = [\ln(1-z)^i/(1-z)]_+$.
These large distributions can be resummed to all 
orders in the threshold limit ($z \to 1$).
Threshold resummation is conveniently performed 
in the Mellin ($N$) space where the convolution 
structures become a simple product. 

The partonic coefficient in the Mellin space is 
organized as follows:
\begin{align}\label{eq:resum-partonic}
	\hat{\sigma}_N^{\rm N^{n}LL}
	= \int_0^1 \df z ~ z^{N-1} \Delta^{\rm sv}(z) 
	\equiv g_{0} \exp \left( G_N \right)
\end{align}
The factor $g_{0}$ is independent of the Mellin variable,
whereas the threshold enhanced large logarithms 
are resummed through the exponent $G_{N}$.
The resummed accuracy is determined through the 
successive terms from the exponent $G_N$ which up to 
N$^3$LL takes the form,
\begin{align}\label{eq:gn}
	G_N = 
	\ln (\Nbar) ~\gbar_1(\Nbar)
	+\gbar_2(\Nbar)
	+\as ~\gbar_3(\Nbar)
	+ \as^2~\gbar_4(\Nbar)\,,
\end{align}
where $\Nbar = N \exp (\gamma_E)$.
These coefficients are universal 
and only depend on the partonic flavors being 
either quark or gluon. Their explicit form can be 
found in e.g. \cite{Catani:2003zt,Moch:2005ba} 
and are also given in the Appendix \ref{appendixa}.

In order to achieve complete resummed accuracy 
one also needs to know the $N$--independent 
coefficient $g_0$ up to sufficient accuracies.
In particular, up to N$^3$LL, it takes the form,
\begin{align}\label{eq:g0}
	g_0
	=
	1
	+ \as ~g_{_{01}}
	+ \as^2 ~g_{_{02}}
	+ \as^3 ~g_{_{03}} \,.
\end{align}
All these coefficients $g_{0i}$ are given in Appendix \ref{appendixa}.
It is also possible to resum part (or all) 
of the $g_0$ by including them in the exponent 
\cite{Bonvini:2014joa,Bonvini:2016frm,Eynck:2003fn,Das:2019btv,Ajjath:2020rci},
which however have subleading effect as these 
contributions are not dominated in the threshold 
region.  

To obtain the results in $z$ space, one needs to do the Mellin inversion as
\begin{align}
	\sigma^{\rm N^{n}LL}
	=&
	\sigma^{(0)} 
	\sum_{a,b \in \{q,\bar{q}\}}
	\int_{c-i\infty}^{c+i\infty}
	\frac{\df N}{2\pi i}
	\tau^{-N}
	f_{a,N}(\muf)
	f_{b,N}(\muf)
		~ \hat{\sigma}_N^{\rm N^{n}LL} \,. 
\end{align}
While doing this complex integral, one encounters the Landau pole at 
$N =\exp\big(1/(2 a_S \beta_0) -\gamma_E\big)$, 
and hence the choice of the contour becomes crucial. We choose the value of $c$ in 
the \textit{minimal prescription} \cite{Catani:1996yz} so that 
the Landau pole will be on the right of the integration contour and
all other singularities will be on the left side of the contour.
The Mellin inversion can then be performed \cite{Vogt:2004ns} along the 
contour $N=c + x~ \exp(i\phi)$, where $x$ is real variable, and $c$ and 
$\phi$ determine the contour. We choose $c=1.9$ and $\phi=3\pi/4$ to obtain a stable result.
The final matched result can be written as,
\begin{align}
	\sigma^{\rm N^{n}LO+N^{n}LL}
	=&
	\sigma^{\rm N^{n}LO}
	+
	\sigma^{(0)}
	\sum_{a,b \in \{q,\bar{q}\}}
	\int_{c-i\infty}^{c+i\infty}
	\frac{\df N}{2\pi i}
	\tau^{-N} 
	f_{a,N}(\muf)
	f_{b,N}(\muf)
	\bigg( 
		\hat{\sigma}_N^{\rm N^{n}LL} 
		- 	
		\hat{\sigma}_N^{\rm N^{n}LL} \bigg|_{\rm tr}
	\bigg) \,.
\end{align}
The $f_{a,N}$ are the Mellin transformed 
PDF similar to the partonic coefficient in 
\eq{eq:resum-partonic} and can be evolved e.g.
using the publicly available code QCD-PEGASUS \cite{Vogt:2004ns}. 
However, for practical purposes, it can be 
also approximated 
by directly using $z$--space PDF following 
\cite{Catani:1989ne,Catani:2003zt,Kulesza:2002rh}.
The last term in the bracket denotes that the resummed 
partonic coefficient in \eq{eq:resum-partonic} 
has been truncated 
to the fixed order to avoid double counting the 
singular terms already present in the fixed order
through	$\sigma^{\rm N^{n}LO}$.

\section{Numerical Results}\label{sec:numerics}
In this section, we present numerical results for various 
color singlet production processes discussed in the 
previous section in the context of the LHC. Unless 
specified otherwise, in our numerical analysis  
we use MMHT2014 \cite{Harland-Lang:2014zoa} parton 
distribution functions (PDFs) throughout taken from the 
{\tt LHAPDF} \cite{Buckley:2014ana}. 
The LO and NLO cross sections are obtained by convolving 
the respective coefficient functions with MMHT2014lo68cl 
and MMHT2014nlo68cl PDFs, while the NNLO and N$^3$LO 
cross sections are obtained with MMHT2014nnlo68cl PDF sets. 
In all these cases, the central set (iset=0) is the 
standard choice. Our default choice of the $\as$ is the 
same as the one used in the {\tt n3loxs} code, and it 
varies order by order in the perturbation theory. The 
fine structure constant is $\alpha \simeq 1/132.184142$. 
The masses of the weak gauge bosons being $m_Z = 91.1876$ 
GeV and $m_W = 80.379$ GeV with the corresponding total 
decay widths of $\Gamma_Z = 2.4952$ GeV and 
$\Gamma_W = 2.085$ GeV. The Weinberg angle is 
$\text{sin}^2\theta_\text{w} = (1 - m_W^2/m_Z^2)$ and is 
calculated internally. This corresponds to the weak 
coupling $G_F \simeq 1.166379^{-5} \text{ GeV}^{-2}$. 
The mass of the Higgs boson is taken to be $m_H = 125.1$ GeV 
and the vacuum expectation value is $v = 246.221$ GeV. 
Finally, the pole masses of the bottom and top quarks are 
taken to be $m_b = 4.78$ GeV and $m_t = 172.76$ GeV, 
while their running masses are $m_b(m_b) = 4.18$ GeV
and $m_t(m_t) = 162.7$ GeV. The default choice of center 
mass energy of the incoming protons is $13$ TeV, unless it
is mentioned otherwise.

The unphysical renormalization and factorization scales 
are chosen to be $\mu_R = \mu_F = Q$, where $Q$ is 
the invariant mass of the dilepton or the invariant mass 
of the vector and Higgs bosons in the final state. For 
the case of Higgs production in bottom annihilation, 
however, the default scale choice is $\mu_R = m_H$ 
and $\mu_F = m_H/4$ following the suggestion from 
Ref.\cite{Duhr:2019kwi}. For all the processes we have 
considered here, the scale uncertainties are estimated by 
varying the unphysical scales in the range so that 
$|\,\text{ln}(\mu_R/\mu_F)\,| < \text{ln}\,4$. 
The symmetric scale uncertainty is calculated from 
the maximum absolute deviation of the cross section from that
obtained with the central/default scale choice.
To estimate the impact of the higher order corrections 
from FO and resummation, we define the following ratios 
of the cross sections which are useful in the experimental 
analysis:
\begin{eqnarray}
\text{K}_{\text{N}^i\text{LO}} = \frac{\sigma_{\text{N}^i\text{LO}}}{\sigma_{\text{LO}}} 
\quad \text{ and } \quad 
\text{R}_{\text{ij}} = \frac{\sigma_{\text{N}^i\text{LO} + \text{N}^i\text{LL}}}{\sigma_{\text{N}^j\text{LO}}} 
\quad \text{ with } 
\quad i, \, j= 0, 1, 2 \text{ and } 3 \, \cdot
	\label{eq:ratio}
\end{eqnarray}
%
\subsection{Neutral DY production}
\onecolumngrid
\begin{figure}[ht!]
	\centerline{
		\includegraphics[width=7.0cm, height=7.0cm]{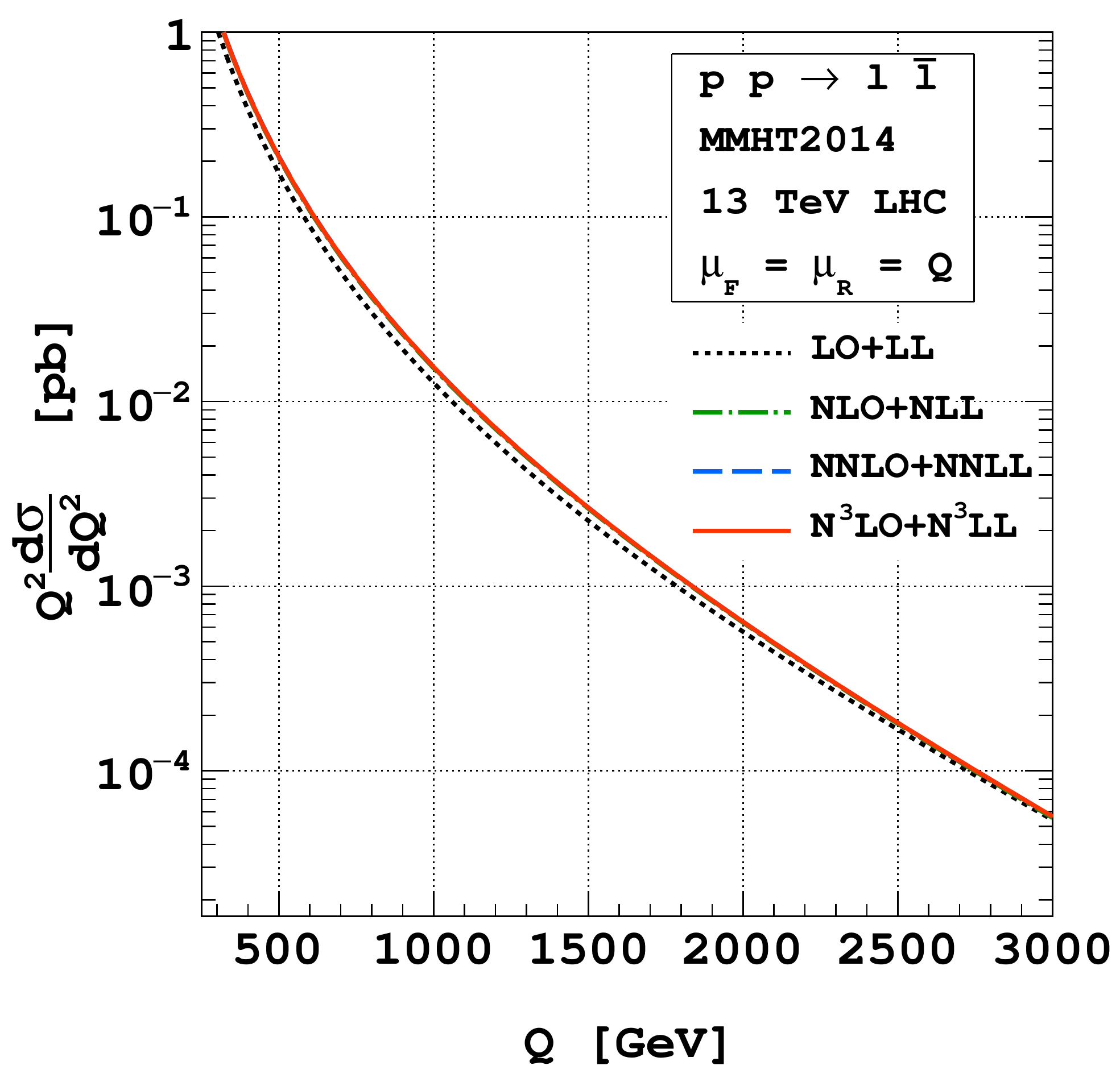}
		\includegraphics[width=7.0cm, height=7.0cm]{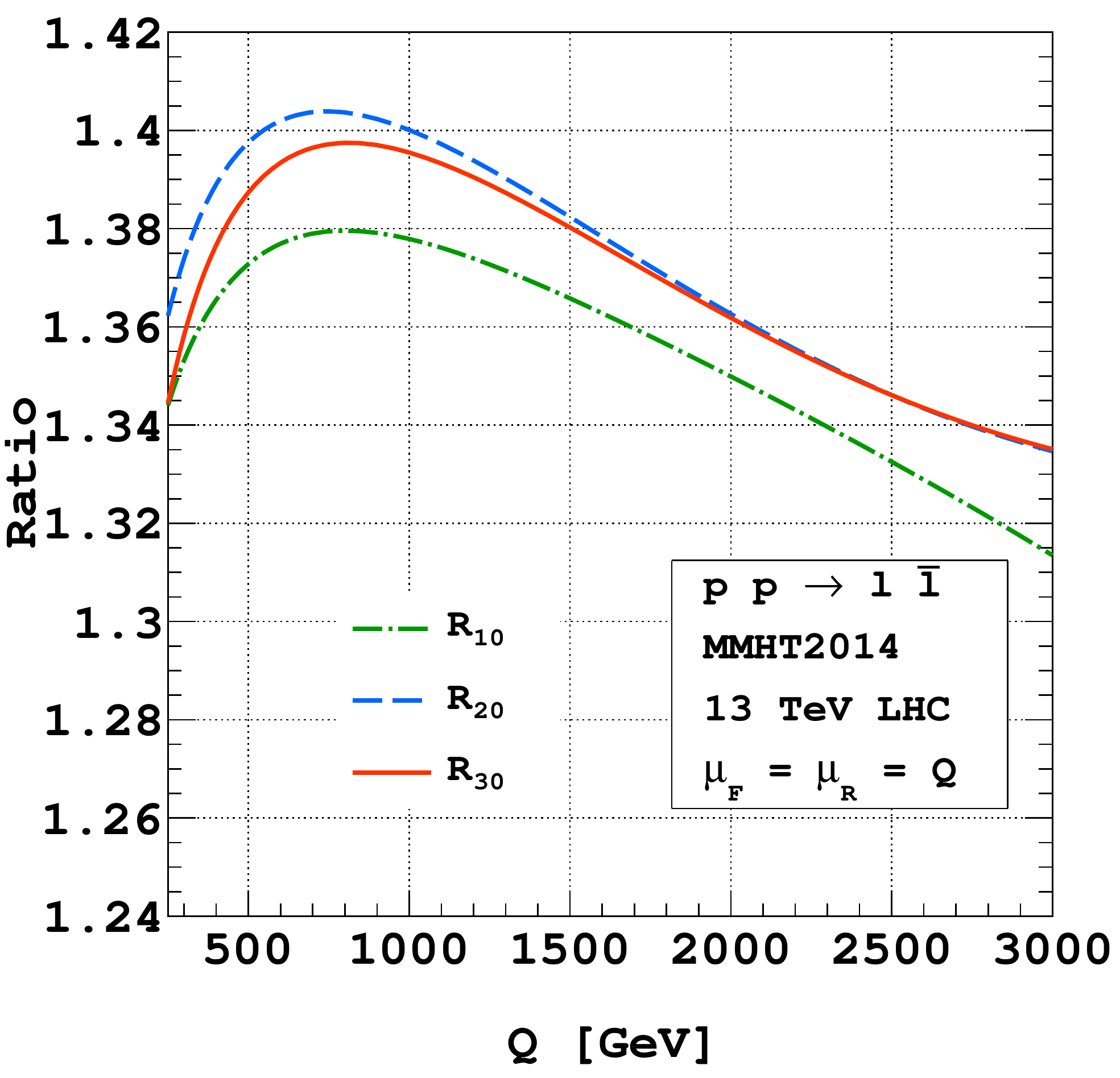}
	}
	\vspace{-2mm}
	\caption{\small{Invariant mass distribution for the enhancement in the resummed cross section of $DY$ for $13$ TeV LHC (left panel) and the resummed cross section over fixed order LO are shown here (right panel) through $R_{ij}$ is defined in \eq{eq:ratio}.}}
	\label{fig:matched_kfac_DY}
\end{figure}
For the neutral DY case, the fixed order results and the 
associated uncertainties have been discussed in 
Ref.\cite{Baglio:2022wzu} 
in greater detail. Hence, we will not repeat them here, 
instead we focus on the resummed results. In the left panel 
of \fig{fig:matched_kfac_DY}, we present the invariant 
mass distribution $Q^2 d\sigma / dQ^2$ up to 
N$^3$LO+N$^3$LL by varying $Q$ from  $250$ GeV to $3000$ GeV.
The corresponding $R_{i0}$-factors defined above are given 
in the right panel. It can be seen that $R_{20}$ is 
larger than $R_{30}$ here up to about $Q=2000$ GeV, 
and then they slowly converge to each other, while 
$R_{10}$ being smaller than these two for the entire 
$Q$ region considered. We also present the $R_{ii}$-factors 
which estimate the contribution of higher order threshold 
logarithms over the respective FO results. The effect of 
threshold resummation is prominent at NLO. However, 
its contribution at N$^3$LO level is very small. 
This is expected as the FO results for DY case are 
converging already at NNLO level onwards, unlike the 
case of Higgs production in gluon fusion channel 
\cite{Harlander:2002wh,Ravindran:2003um,Anastasiou:2015vya,Mistlberger:2018etf}.
Here, the resummed results at N$^3$LO+N$^3$LL demonstrate
excellent convergence of the perturbation theory.

At this level of precision, it is also important to consider effects from power corrections.
As a first step, one can consider the next-to-soft-virtual (NSV) corrections
by taking into account the logarithms of the kind $\text{ln}^{i}(1-z)$.
In the context of the $nDY$ process, a detailed phenomenology
at LHC has been studied in Refs. \cite{Ajjath:2021lvg, vanBeekveld:2021hhv}. In Ref. \cite{Ajjath:2021lvg},
these NSV logarithms have been resummed to next-to-next-to-leading
logarithmic accuracy and a comparison of the same has been carried out
with the SV resummed result. It is found that the differences between the SV and NSV resummations are
about 4\% at LL, 2.84\% at NLL and 0.85\% at NNLL accuracy for $Q=1000$ GeV.
At higher invariant mass, the differences do not change much, {\it e.g.} at $Q=2000$ GeV,
the differences are about 4\%, 2.8\% and 0.89\% respectively.
A general observation is that the difference between SV and NSV resummations is less
sensitive to the value of $Q$ and that this difference decreases with higher
logarithmic accuracy. If this trend were to continue, at N$^3$LL the difference between
SV and NSV resummed results could be about 0.1\%.
This translates to a rough estimate of the order of
0.1\% theory uncertainty due to the missing power corrections. It is also worth noting that the
NSV logarithms from other partonic channels will also contribute to the cross section.
However, a detailed study of such power corrections is beyond the scope of the present work.

\subsection{Charged DY production}
\begin{figure}[ht!]
	\centerline{
		\includegraphics[width=7.0cm, height=7.0cm]{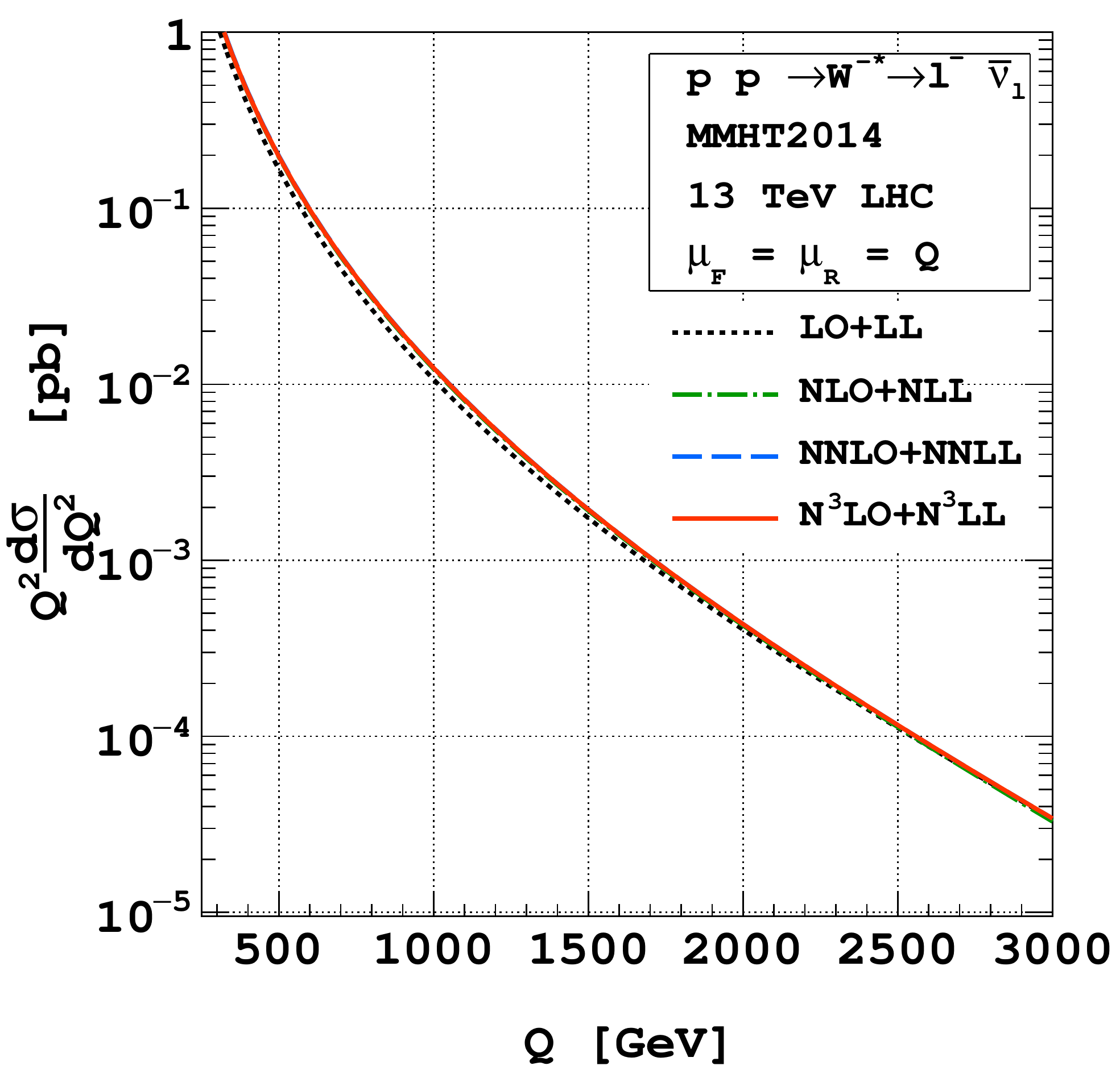}
		\includegraphics[width=7.0cm, height=7.0cm]{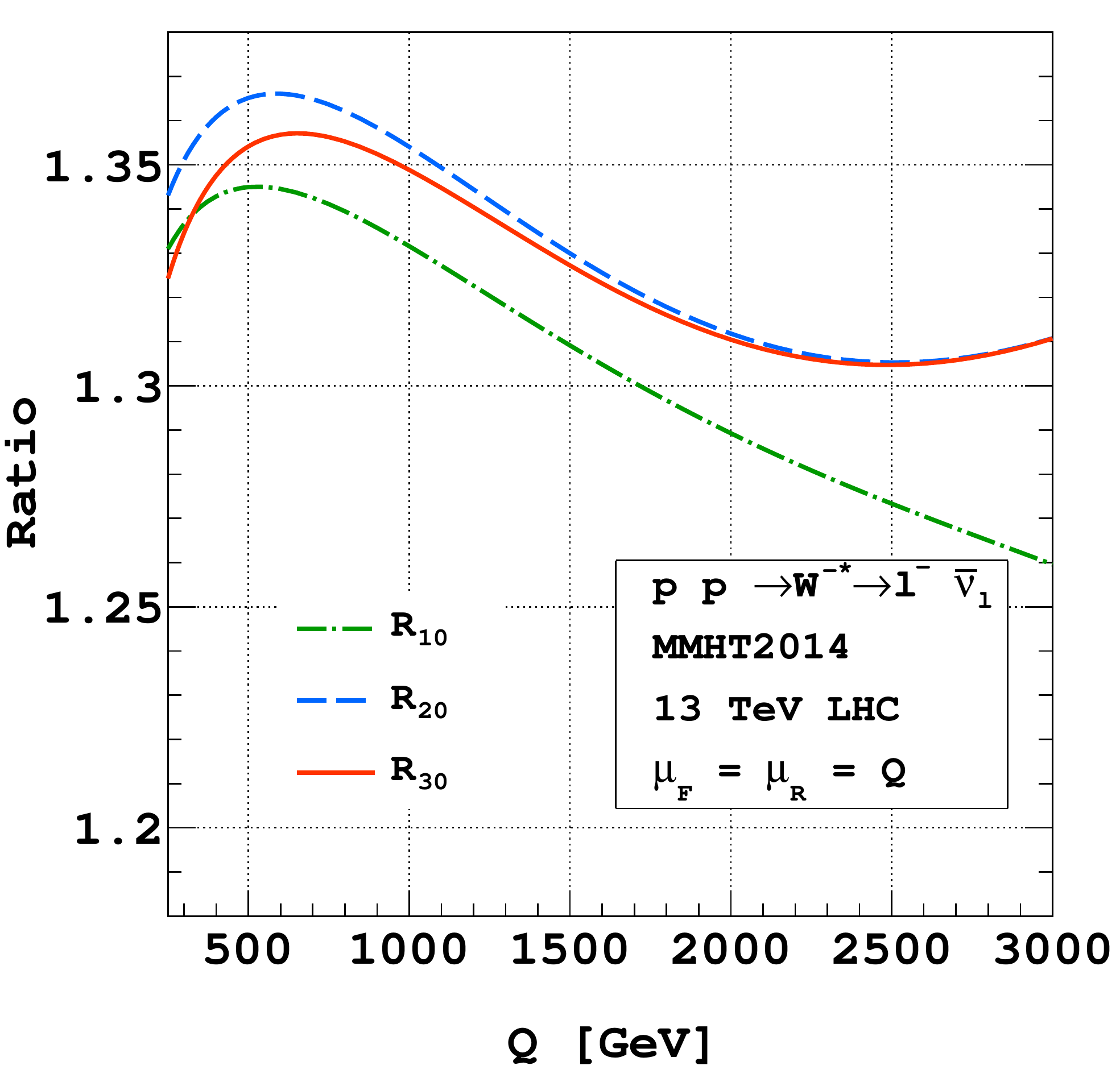}
	}
	\vspace{-2mm}
	\caption{\small{Invariant mass distribution for the enhancement in the resummed cross section of charge current DY ($W^{-*}$) for $13$ TeV LHC (left panel) and the resummed cross section 
			over fixed order LO are shown here 
			(right panel) through $R_{ij}$ is defined in 
		\eq{eq:ratio}.}}
	\label{fig:matched_kfac_WmDY}
\end{figure}
Similar to the neutral DY case, we present in 
\fig{fig:matched_kfac_WmDY}, the invariant mass 
distribution (left panel) for charge DY($W^{-*}$) case 
up to N$^3$LO+N$^3$LL accuracy and the corresponding 
$R_{i0}$ factors (right panel). For the case of charged DY, 
the corresponding parton fluxes are different from those 
of neutral DY case. This will clearly result in a slightly 
different behavior of the higher order corrections. 
This together with the NLL enhancement of the cross sections 
can explain the behavior of the $R$-factors noticed.
Quantitatively, the impact of QCD corrections in the high 
invariant mass region ($Q \ge 2500$ GeV) are smaller than
the corresponding ones for neutral DY case.
\onecolumngrid
\begin{figure}[ht!]
	\centerline{
		\includegraphics[width=7.0cm, height=7.0cm]{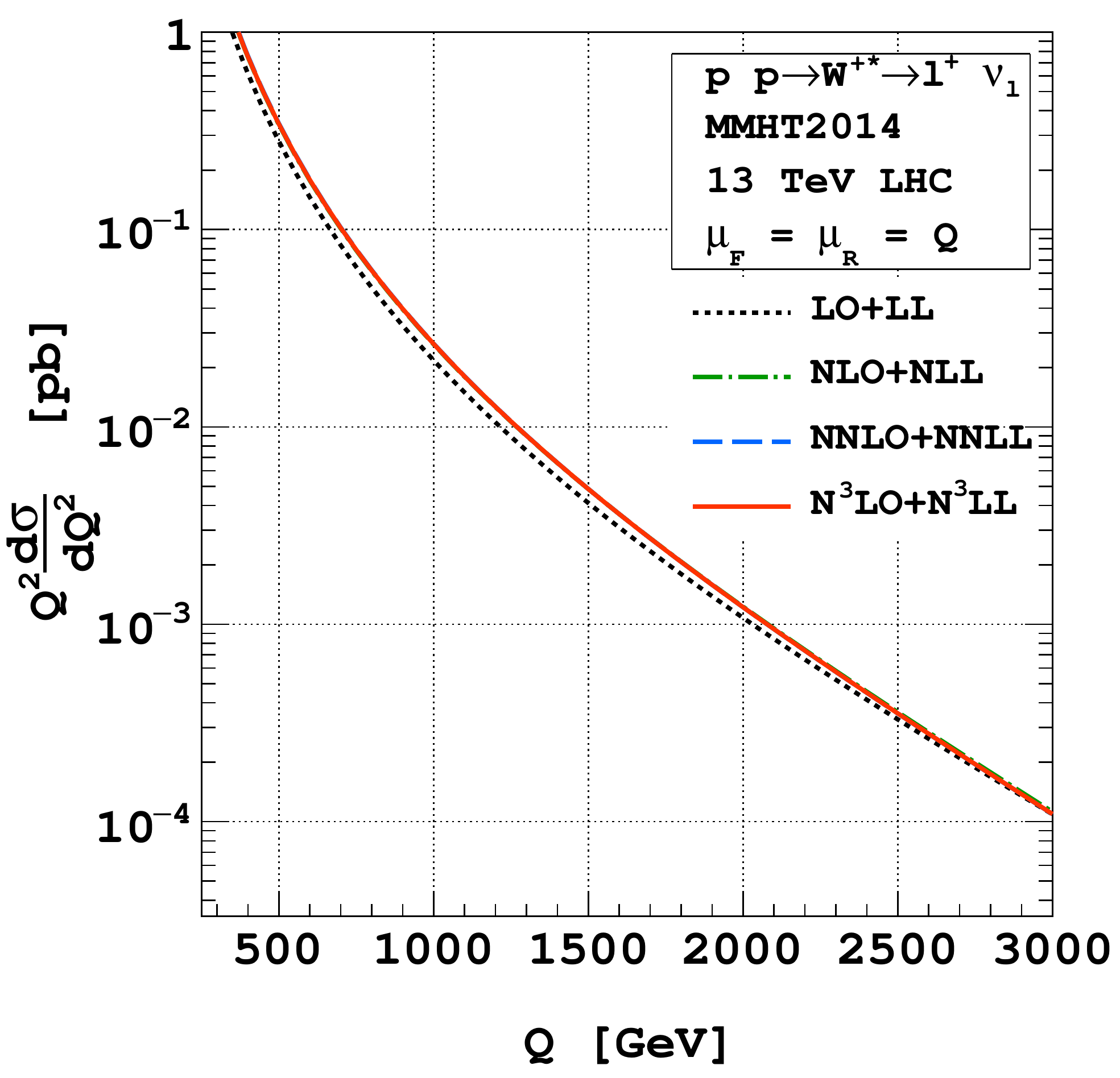}
		\includegraphics[width=7.0cm, height=7.0cm]{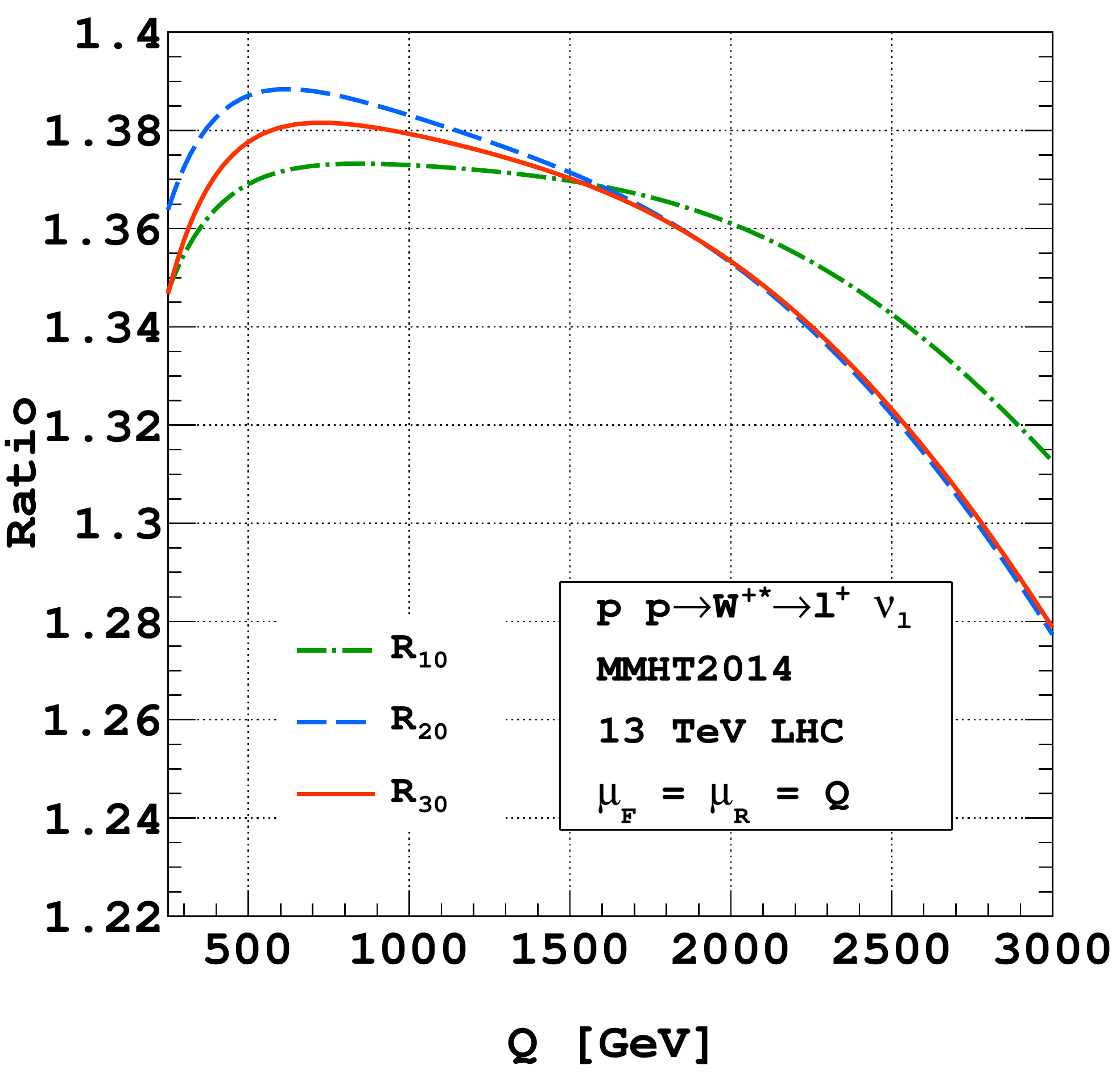}
	}
	\vspace{-2mm}
	\caption{\small{Invariant mass distribution for the enhancement in the resummed cross section of charge current DY ($W^{+*}$) for $13$ TeV LHC (left panel) and the resummed cross section 
			over fixed order LO are shown 
			here (right panel) through $R_{ij}$ is 
			defined in \eq{eq:ratio}.}}
	\label{fig:matched_kfac_WpDY}
\end{figure}
\onecolumngrid
\begin{figure}[ht!]
	\centerline{
		\includegraphics[width=5.5cm, height=5.5cm]{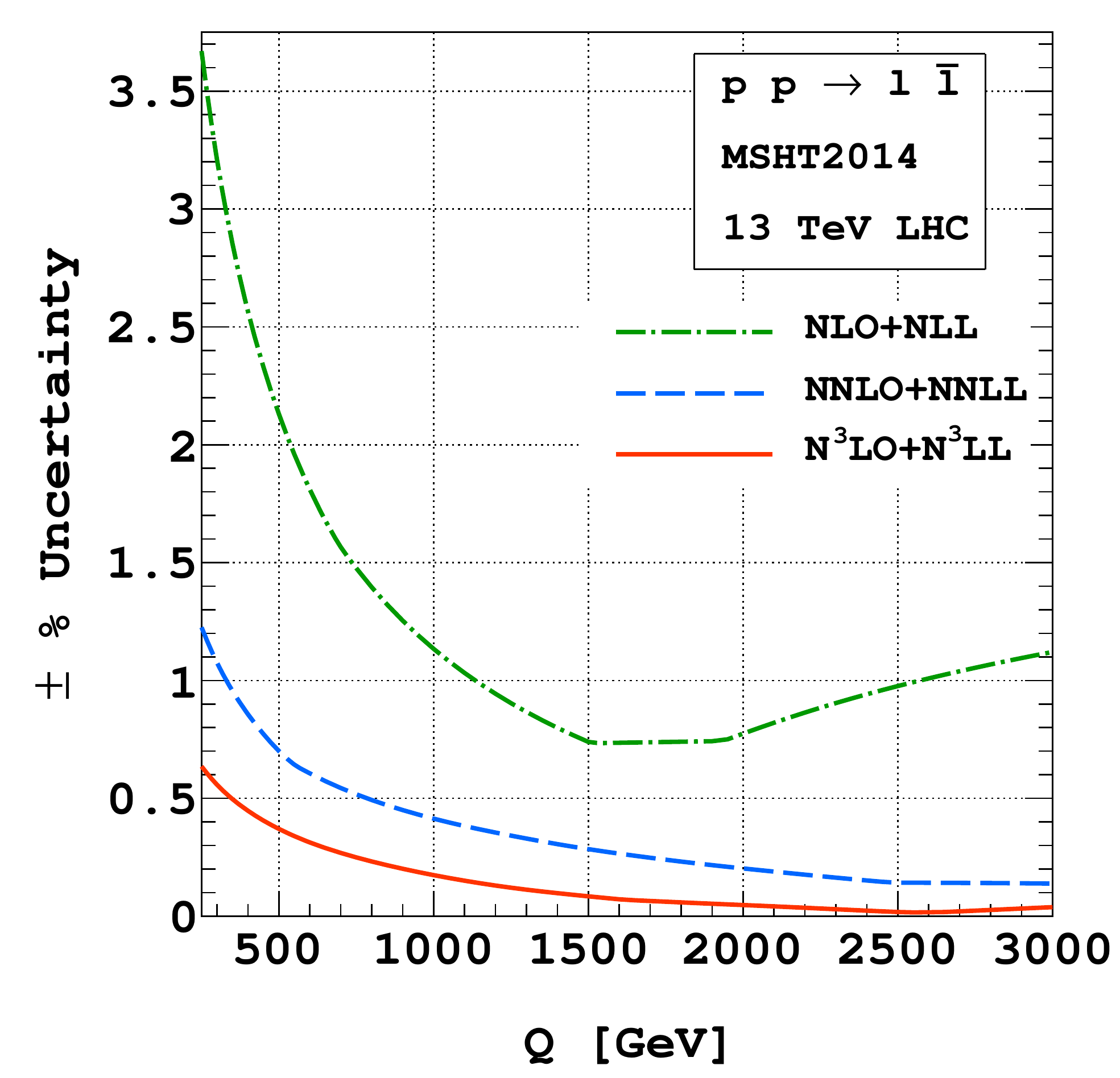}
		\includegraphics[width=5.5cm, height=5.5cm]{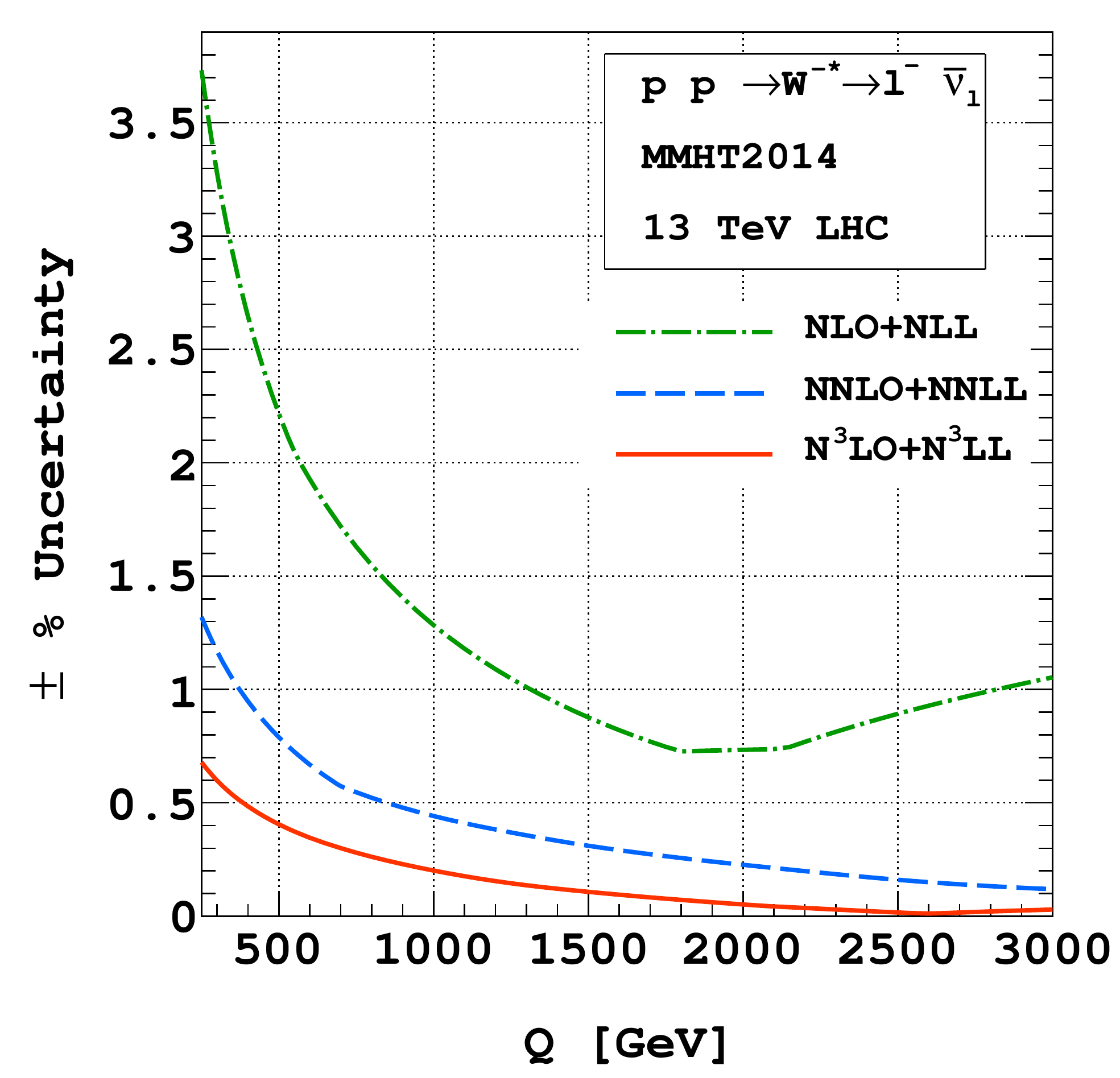}
		\includegraphics[width=5.5cm, height=5.5cm]{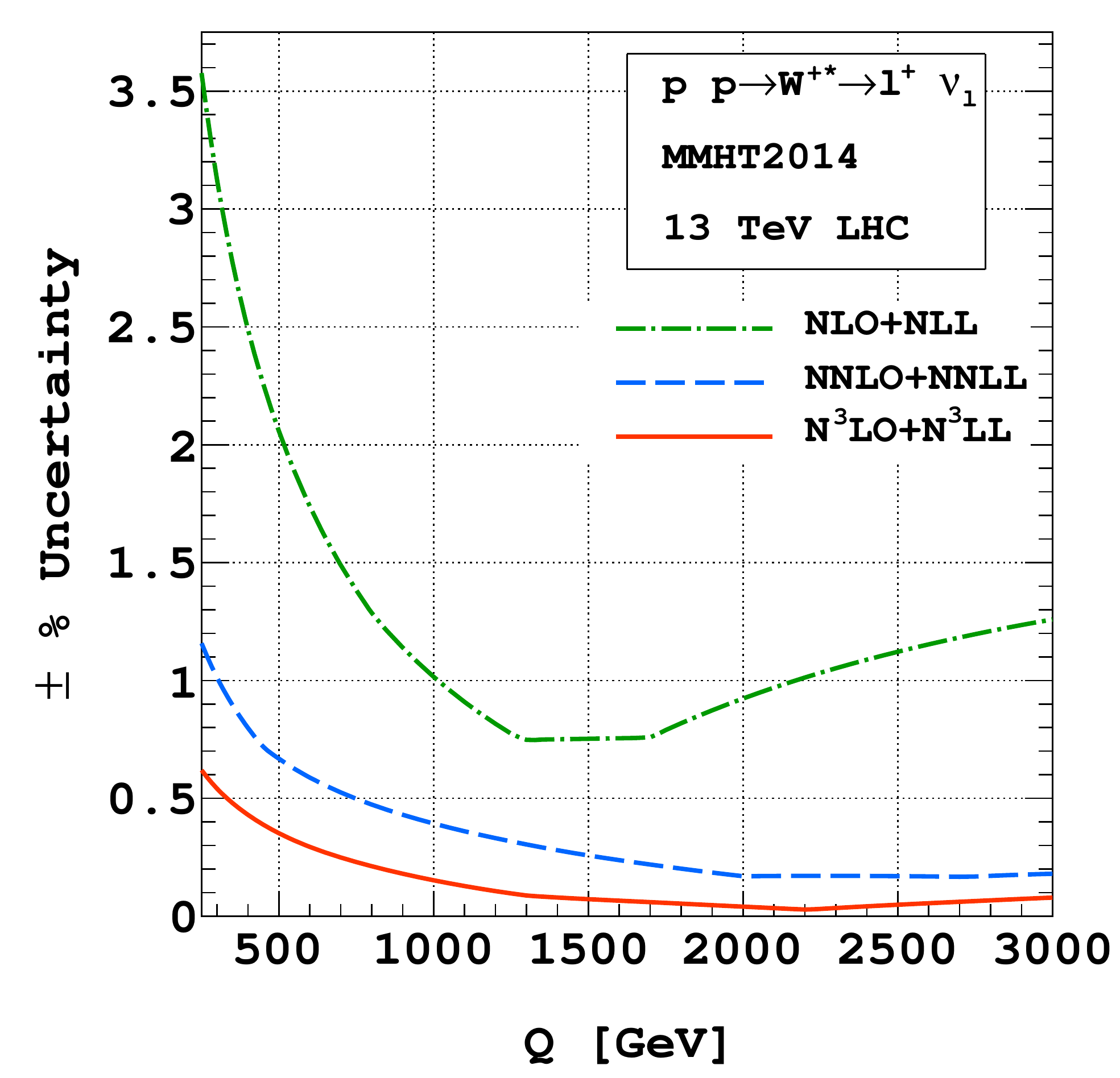}
	}
	\vspace{-2mm}
	\caption{\small{The $7$-point scale uncertainty for neutral DY [$nDY$] (left panel), charged DY [$cDY$] $W^{-*}$ (middle panel) and $W^{+*}$(right panel)}}
	\label{fig:match_uncertainty_7points_DY}
\end{figure}

\begin{figure}[ht!]
	\centerline{
		\includegraphics[width=5.5cm, height=5.5cm]{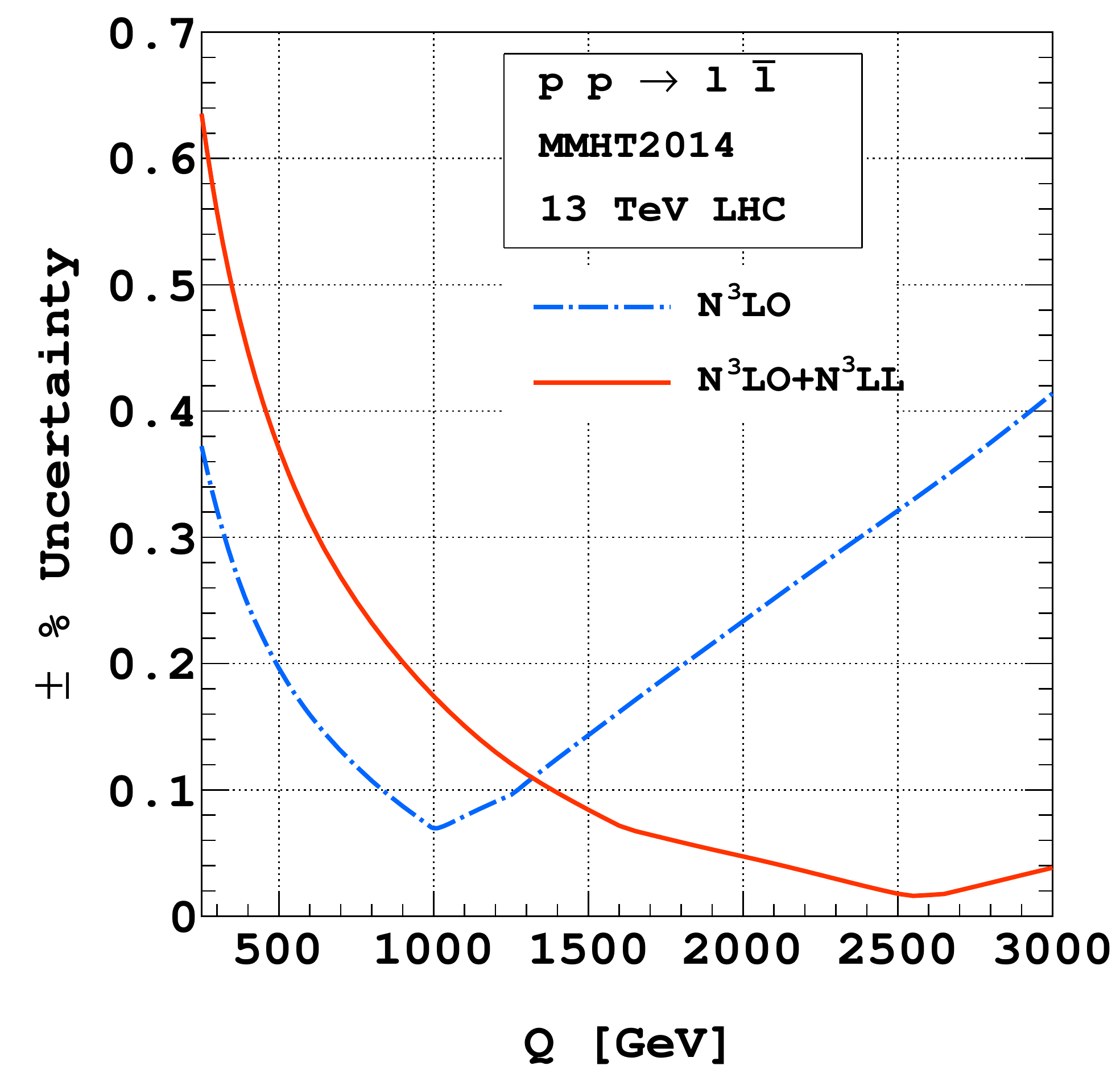}
		\includegraphics[width=5.5cm, height=5.5cm]{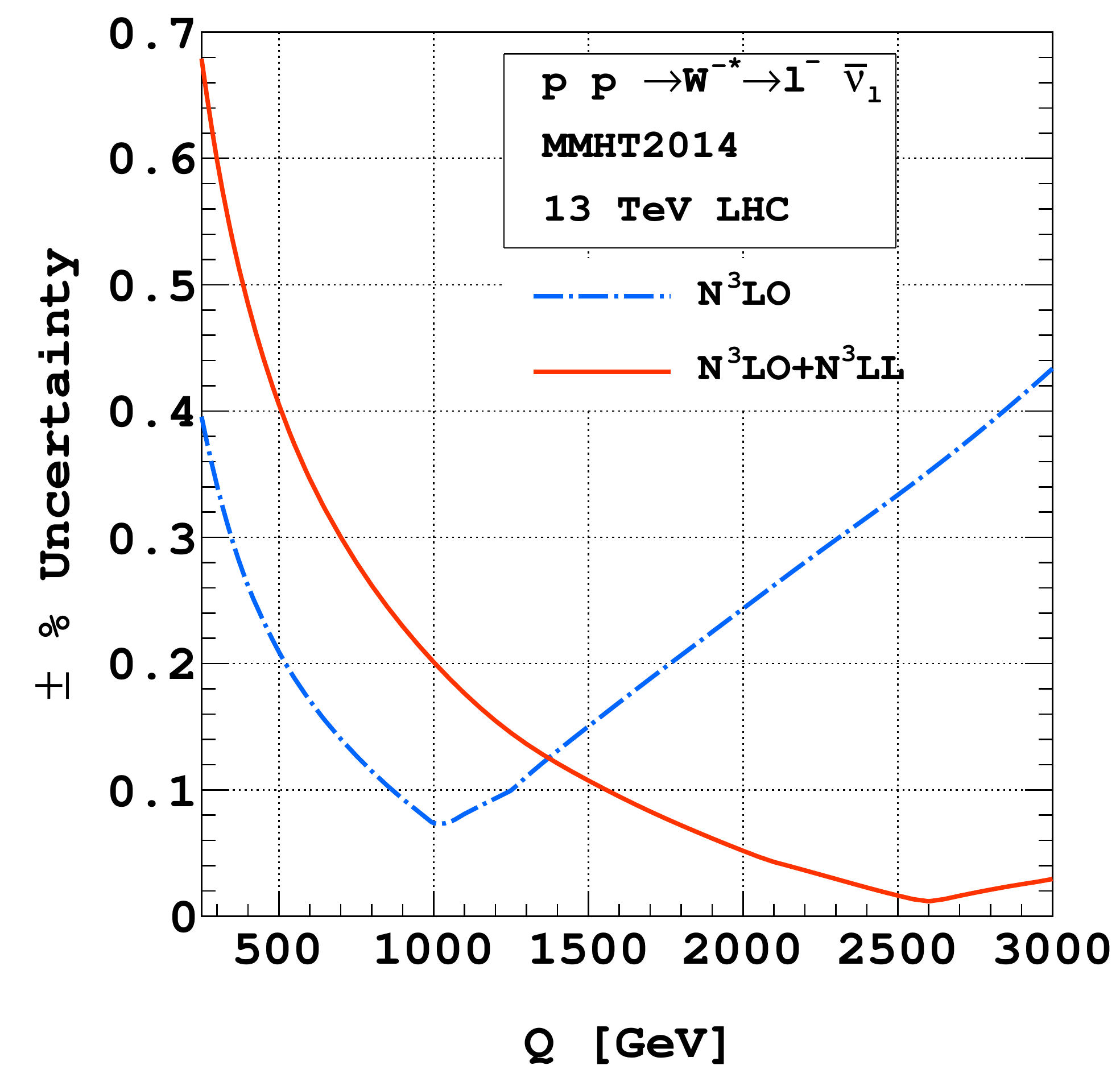}
		\includegraphics[width=5.5cm, height=5.5cm]{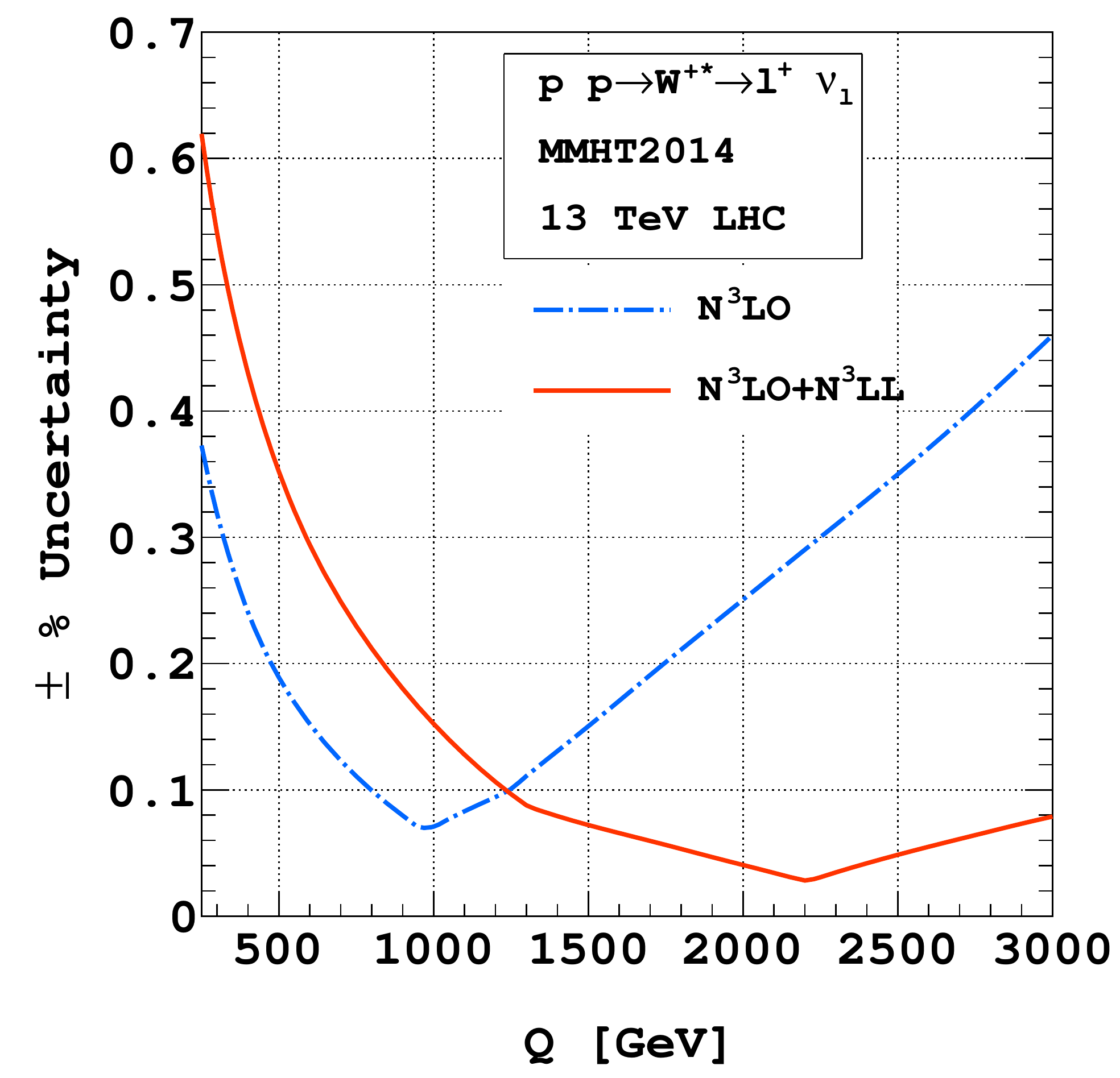}
	}
	\vspace{-2mm}
	\caption{\small{Comparison of $7$-point scale uncertainty between resum and fixed order results for neutral DY [$nDY$] (left panel), charged DY [$cDY$] $W^{-*}$ (middle panel) and $W^{+*}$(right panel).}}
	\label{fig:comparison_uncertainty_7points_DY}	
\end{figure}
\begin{figure}[ht!]
        \centerline{
                \includegraphics[width=5.5cm, height=5.5cm]{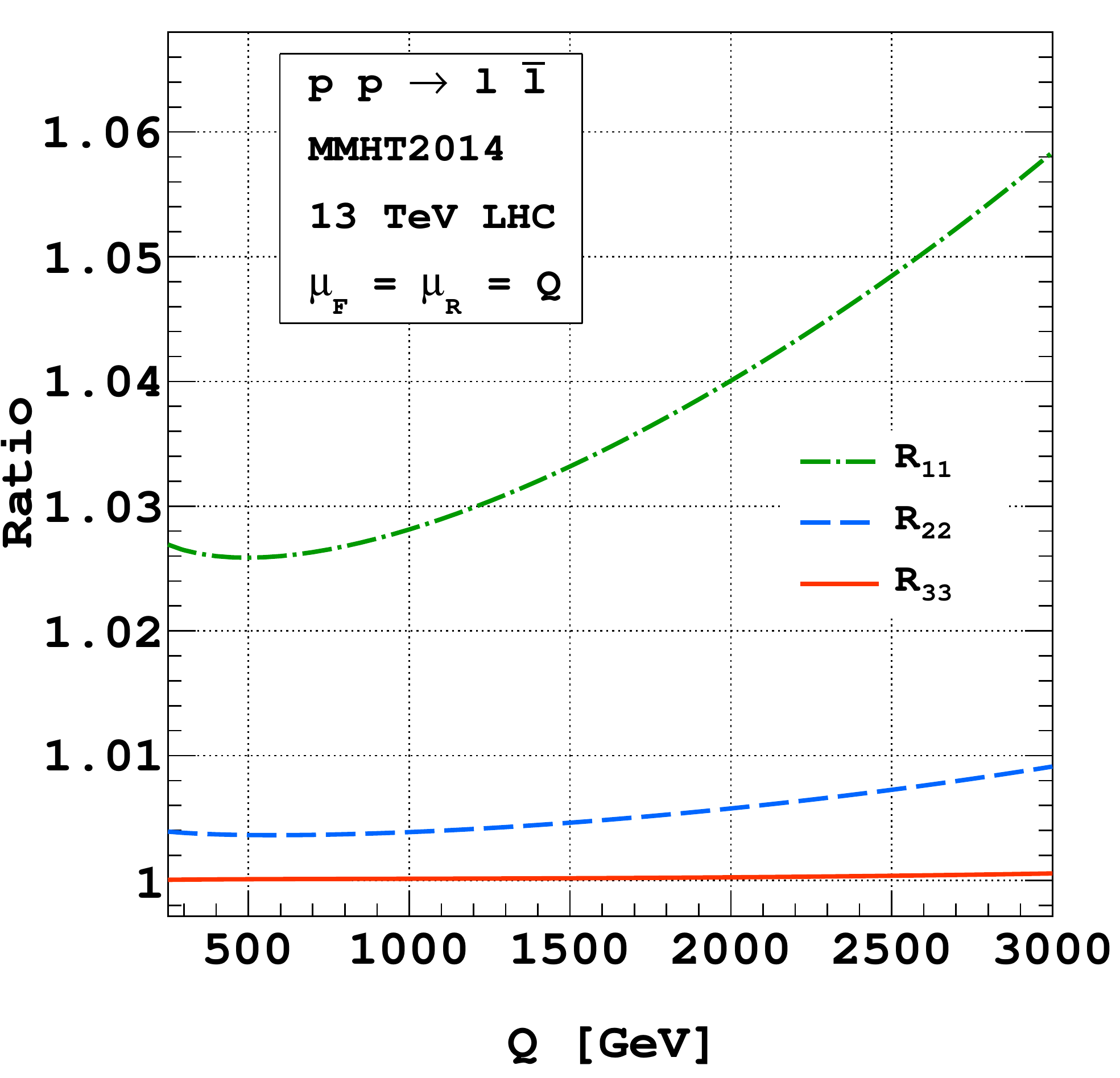}
                \includegraphics[width=5.5cm, height=5.5cm]{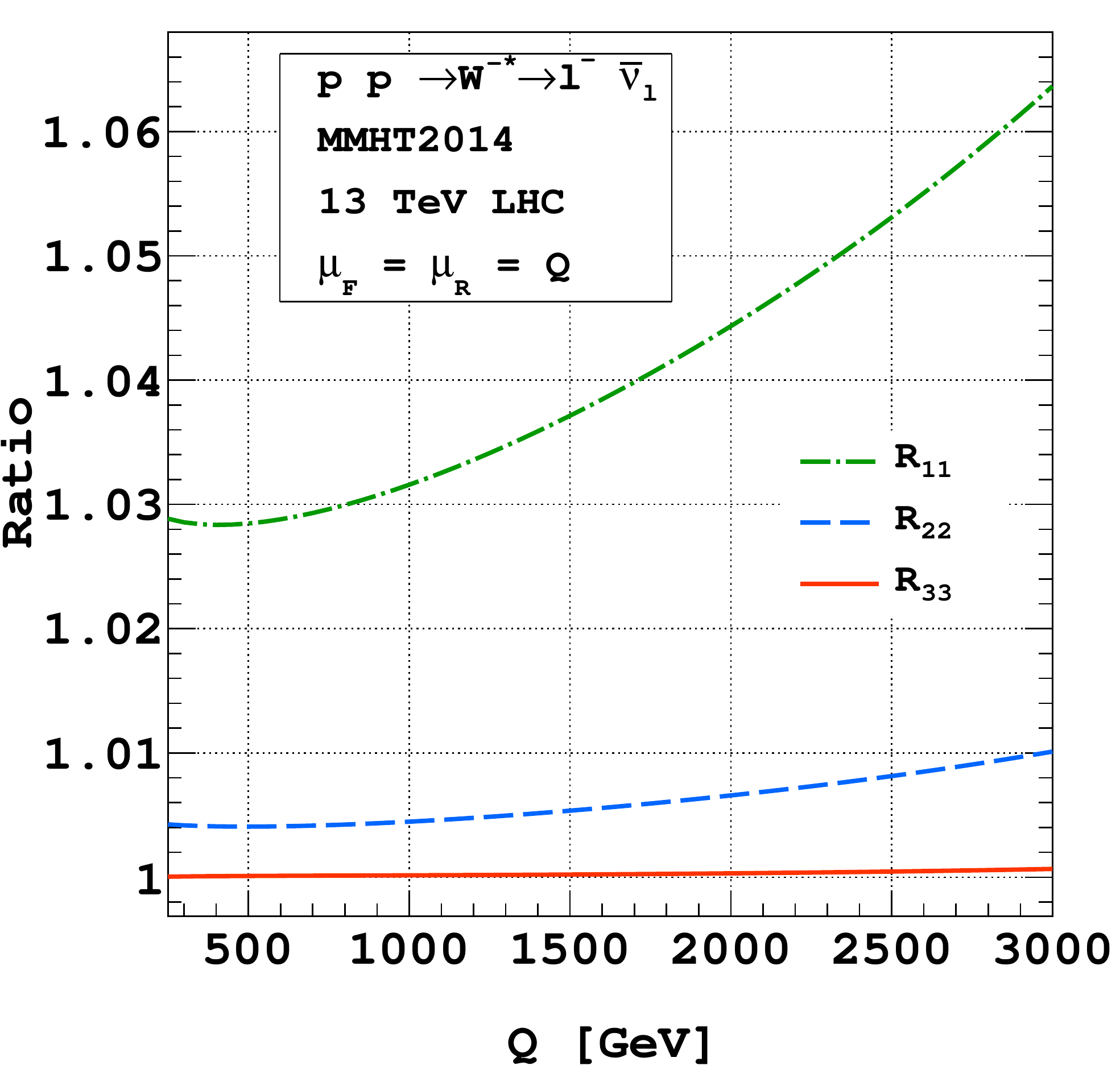}
                \includegraphics[width=5.5cm, height=5.5cm]{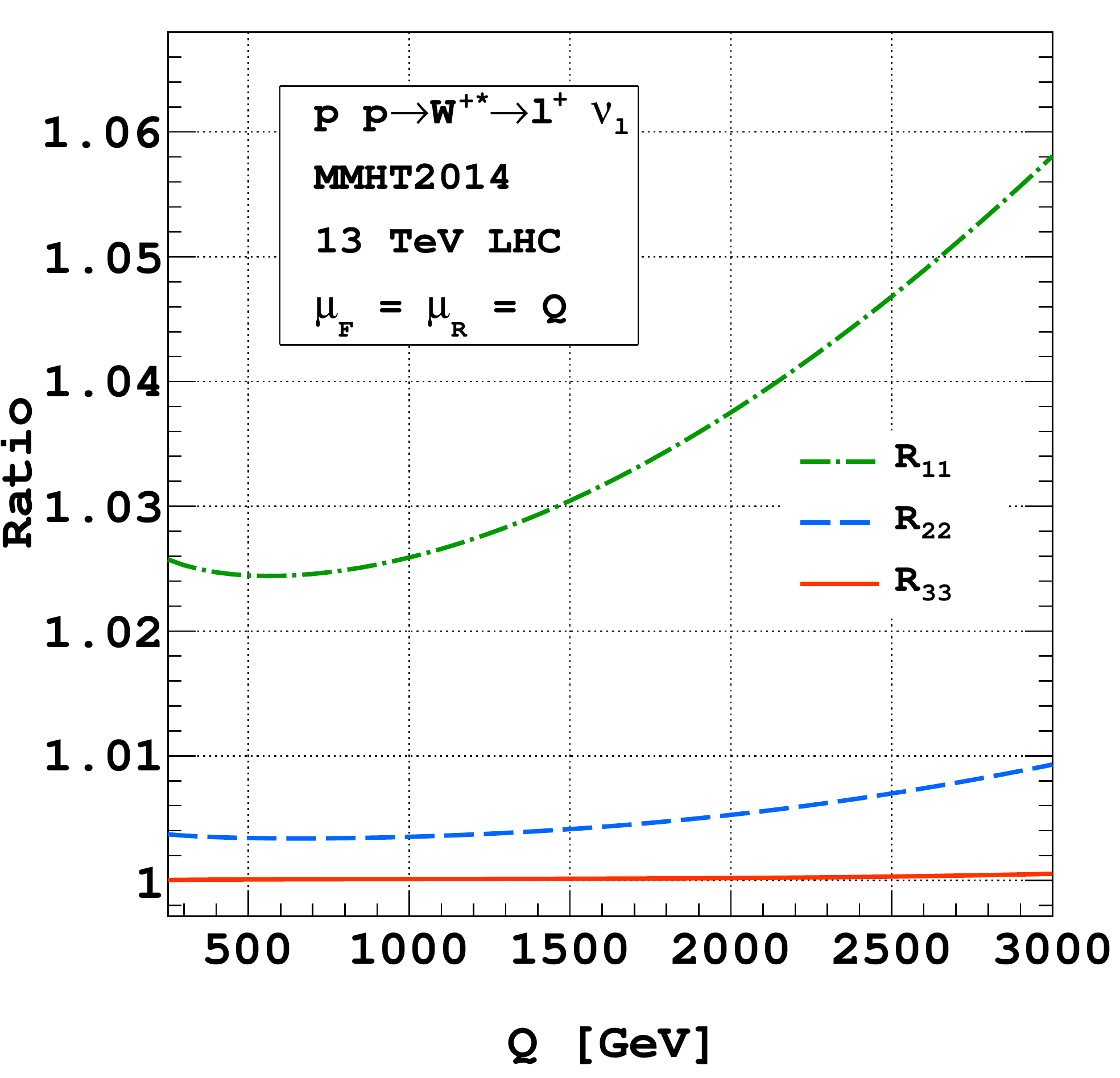}
        }
        \vspace{-2mm}
        \caption{\small{Resummed enhancement over respective 
		fixed order are shown here for neutral 
		DY (left panel), charged DY $W^{-*}$ (middle panel) 
		and $W^{+*}$ (right panel) through $R_{ij}$ defined in 
		\eq{eq:ratio}.}}
        \label{fig:match_fo_ratio_DY}
\end{figure}

Similar results have been obtained 
for $W^{+*}$ and the corresponding $R_{i0}$ factors 
are shown in \fig{fig:matched_kfac_WpDY}.
Again, owing to the different underlying parton fluxes 
that are different starting from the Born level, 
the behavior of these $R_{i0}$-factors is expected to be 
different from those of both neutral DY and $W^{-*}$. 

Besides, the resummed results at N$^3$LO+N$^3$LL play a 
significant role in reducing the conventional $7$-point 
scale uncertainties.  The scale uncertainties in the 
resummed predictions up to N$^3$LO+N$^3$LL accuracy are 
given in \fig{fig:match_uncertainty_7points_DY} for 
neutral DY (left panel), for $W^{-*}$ (middle panel) 
and for $W^{+*}$ (right panel). While the FO scale 
uncertainties for neutral DY case at $Q=3000$ GeV are 
found to get reduced from about $1.5\%$ at NNLO 
level to about $0.4\%$ at N$^3$LO level, the same in the 
resummed results are found to get significantly reduced 
from about $0.2\%$ at NNLO+NNLL level to almost about 
$0.1\%$ at N$^3$LO+N$^3$LL level. Thus, DY process is one 
of the processes for which the theoretical predictions 
available to-date are the most precise, any uncertainties 
that can still be present in these cross sections
are only due to the PDFs.  We will discuss the uncertainty 
due to PDFs later. It should be noted here that the scale 
uncertainties in the resummation context will not show 
improvement over the FO results in the low $Q$-region, 
say below $1000$ GeV, where the threshold logarithms are 
not the sole dominant contributions to the cross sections.  
However, the scale uncertainties in the resummed results 
will get reduced in the high $Q$ region.  To elaborate on 
this, we compare and contrast the $7$-point scale 
uncertainties in FO and resummed results at third order 
in QCD (see \fig{fig:comparison_uncertainty_7points_DY}).
The scale uncertainties in the low $Q$ region (where 
regular terms and other parton channel contributions are 
non-negligible) 
are smaller for FO case, while they are smaller for 
resummed case in the high $Q$ region 
(where the threshold logarithms are important).
%

To see the effects of resummation over FO results at a 
given order, it is quite useful to use the
factors $R_{ii}$ defined in \eq{eq:ratio}. We plot 
in \fig{fig:match_fo_ratio_DY} these factors 
$R_{11}, \, R_{22}$ and $R_{33}$ for all 
three different DY processes
as a function of the invariant mass of the dileptons. 
In all these plots, we can see the $R_{11}$ contribution 
is dominant particularly in the high $Q$ region, while 
the $R_{33}$ is almost unity except for a small 
contribution in the high invariant mass distribution. 
\subsection{$VH$ production}
In this section, we present the numerical results for the 
Higgs production in association with a massive vector 
boson $V=Z, \, W^-, \, W^+$.  We present results for both 
the invariant mass distribution and the total production 
cross sections at hadron colliders for different center 
of mass energies. However, for invariant mass distribution 
our results are confined to only DY process i.e. the 
production of an off-shell gauge boson followed by its 
decay to on-shell $V$ and $H$.
\onecolumngrid
\begin{figure}[ht!]
	\centerline{
		\includegraphics[width=7.0cm, height=7.0cm]{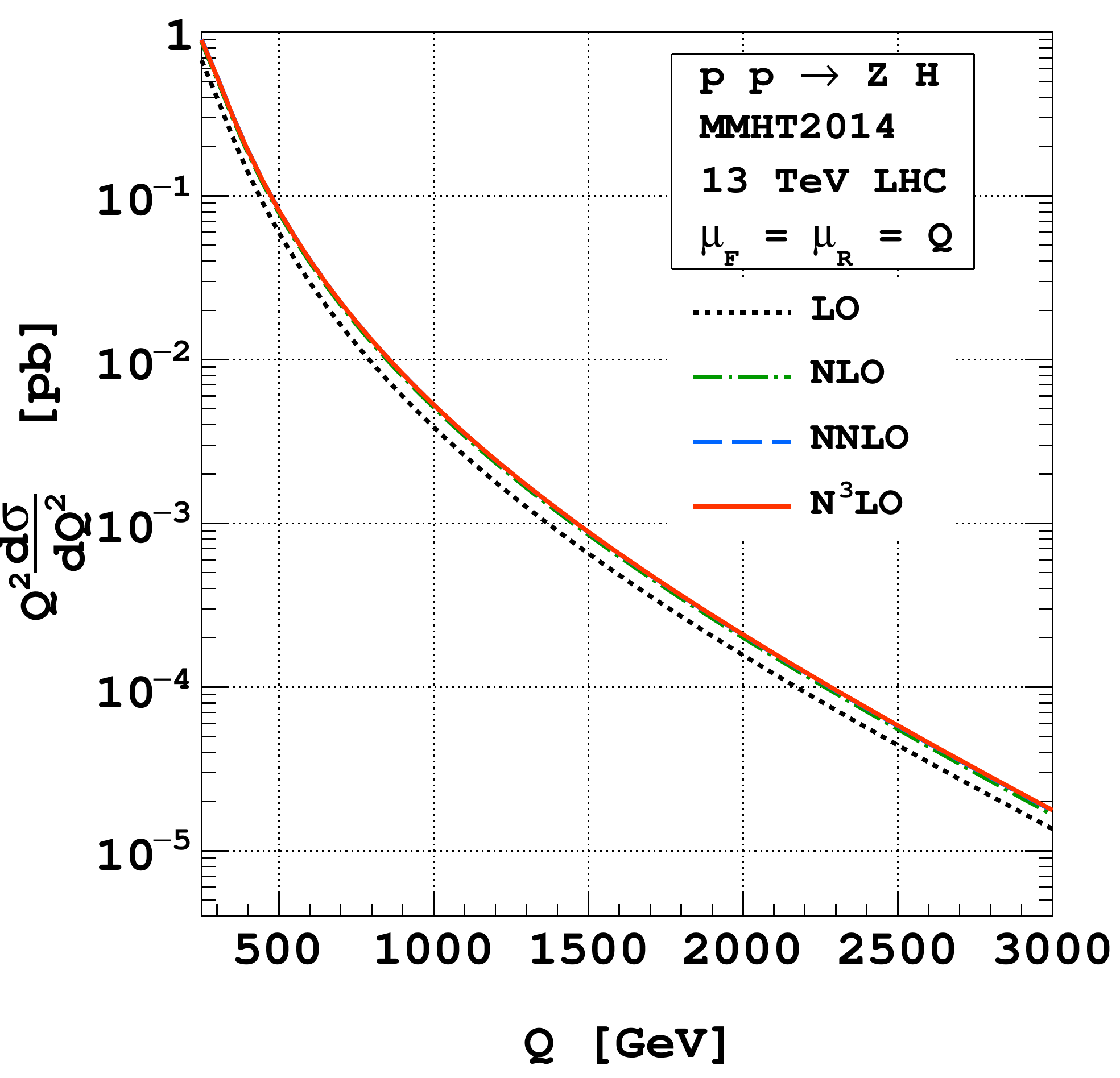}
		\includegraphics[width=7.0cm, height=7.0cm]{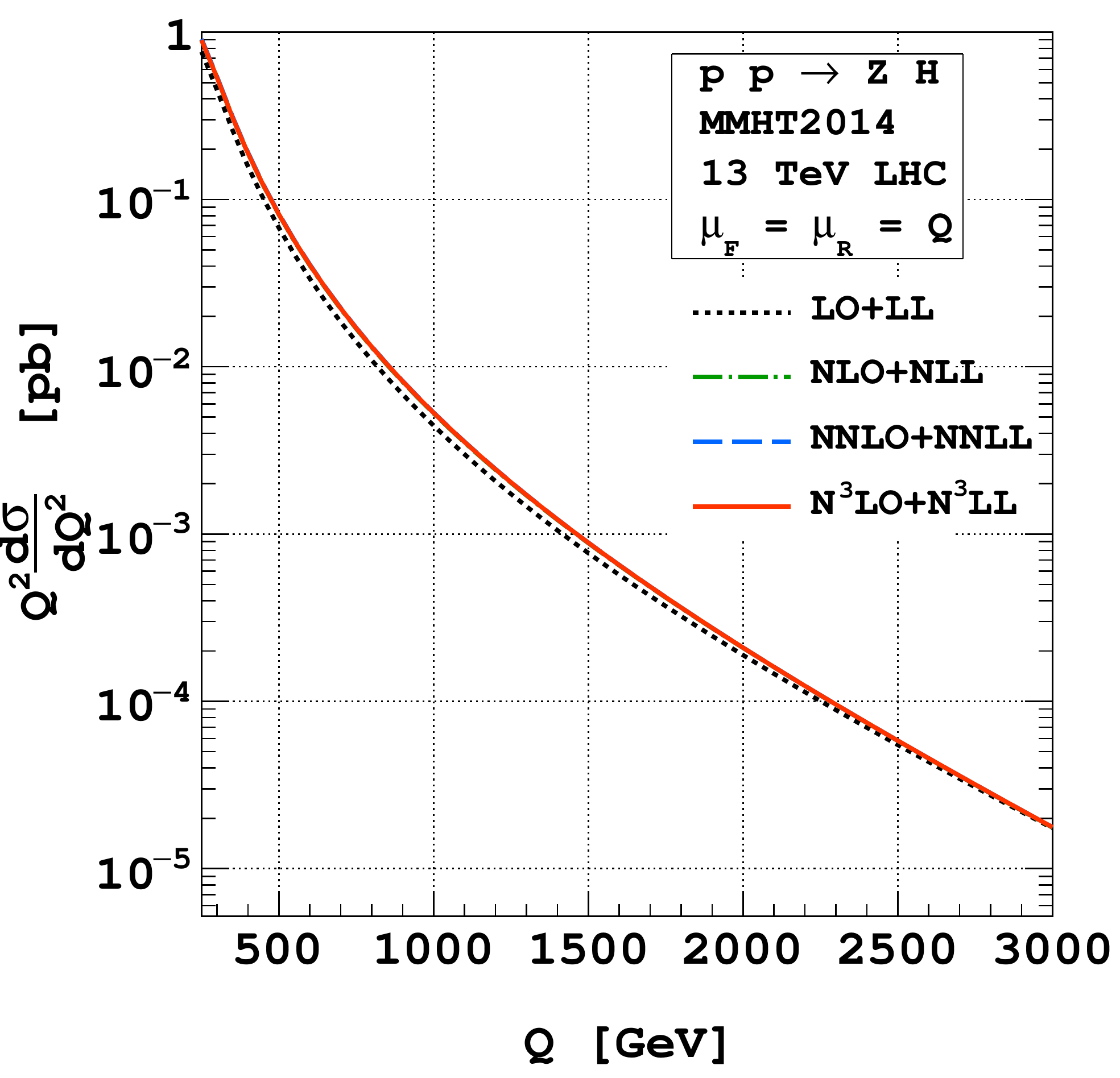}
	}
	\vspace{-2mm}
	\caption{\small{Invariant mass distribution of 
	$ZH$ for $13$ TeV LHC fixed order (left panel) 
	and the resummed (right panel).}}
	\label{fig:fo_resum_ZH}
\end{figure}

\begin{figure}[ht!]
	\centerline{
		\includegraphics[width=7.0cm, height=7.0cm]{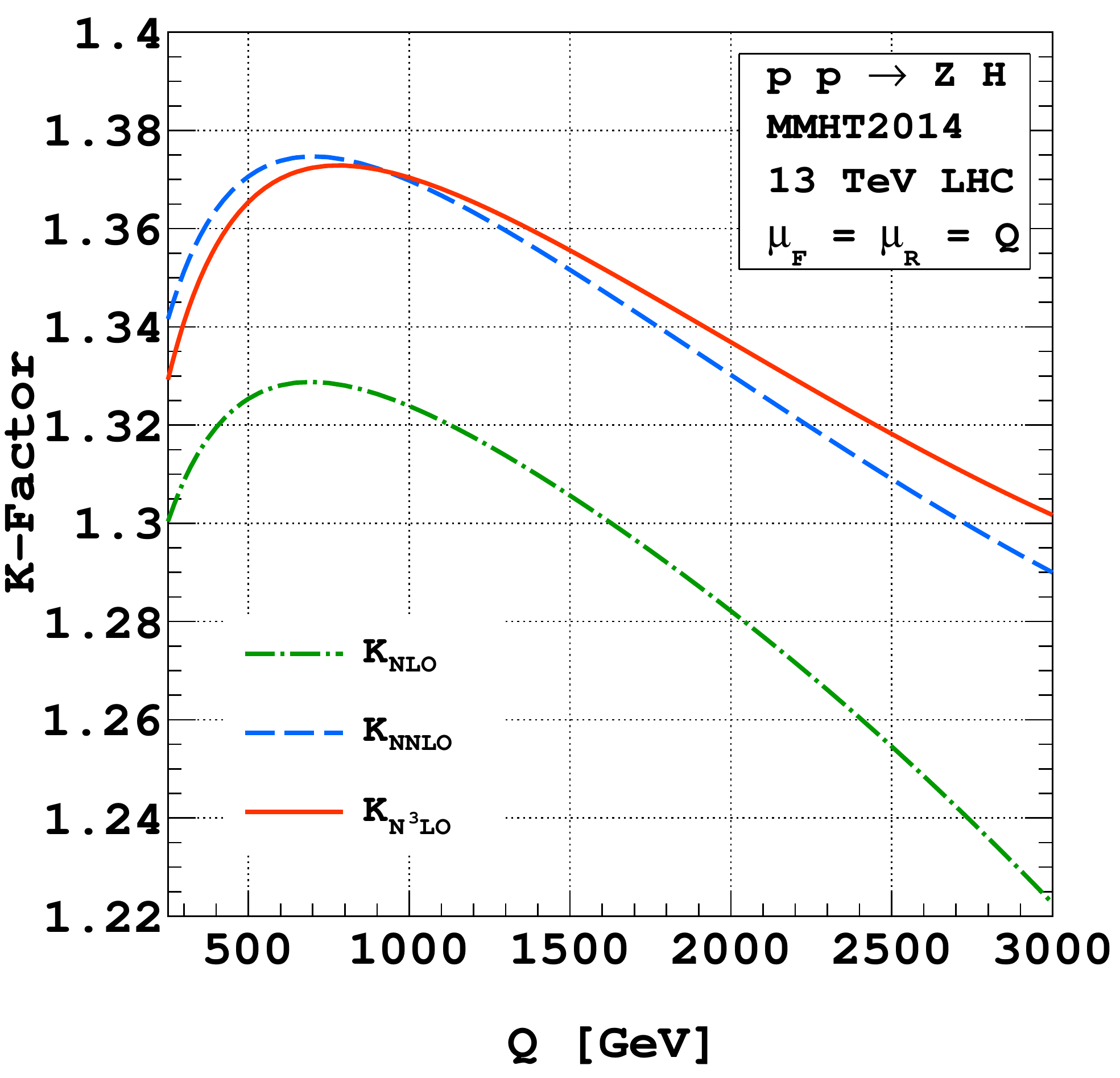}
		\includegraphics[width=7.0cm, height=7.0cm]{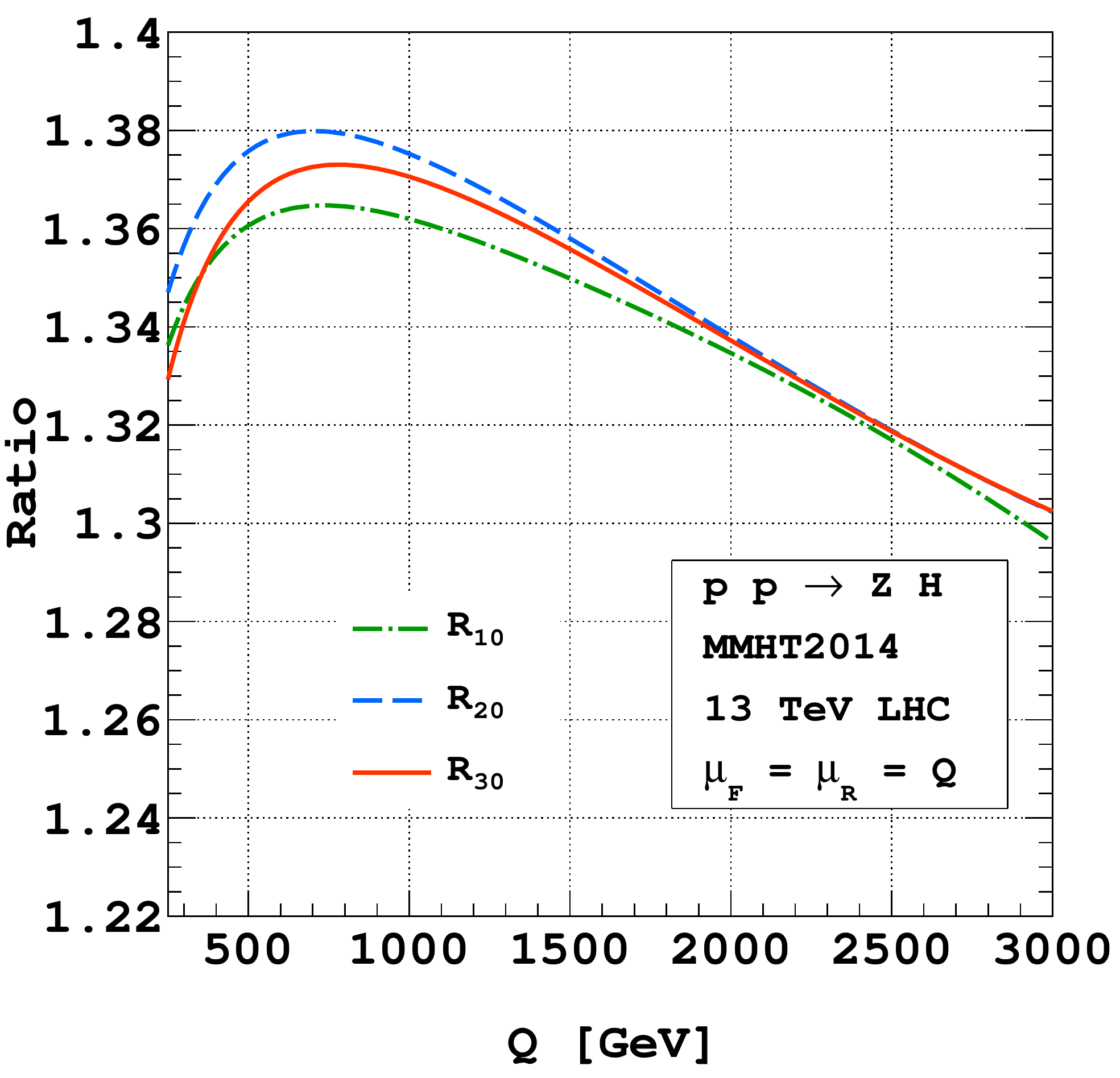}
	}
	\vspace{-2mm}
	\caption{\small{K-factor for fixed order of $ZH$ for $13$ TeV LHC (left panel) and the enhancement in the resummed cross section 
			over fixed order LO are shown here 
			(right panel) through $R_{ij}$ 
			is defined in \eq{eq:ratio}.}}
	\label{fig:matched_kfac_ZH}
\end{figure}

\begin{figure}[ht!]
	\centerline{
		\includegraphics[width=5.5cm, height=5.5cm]{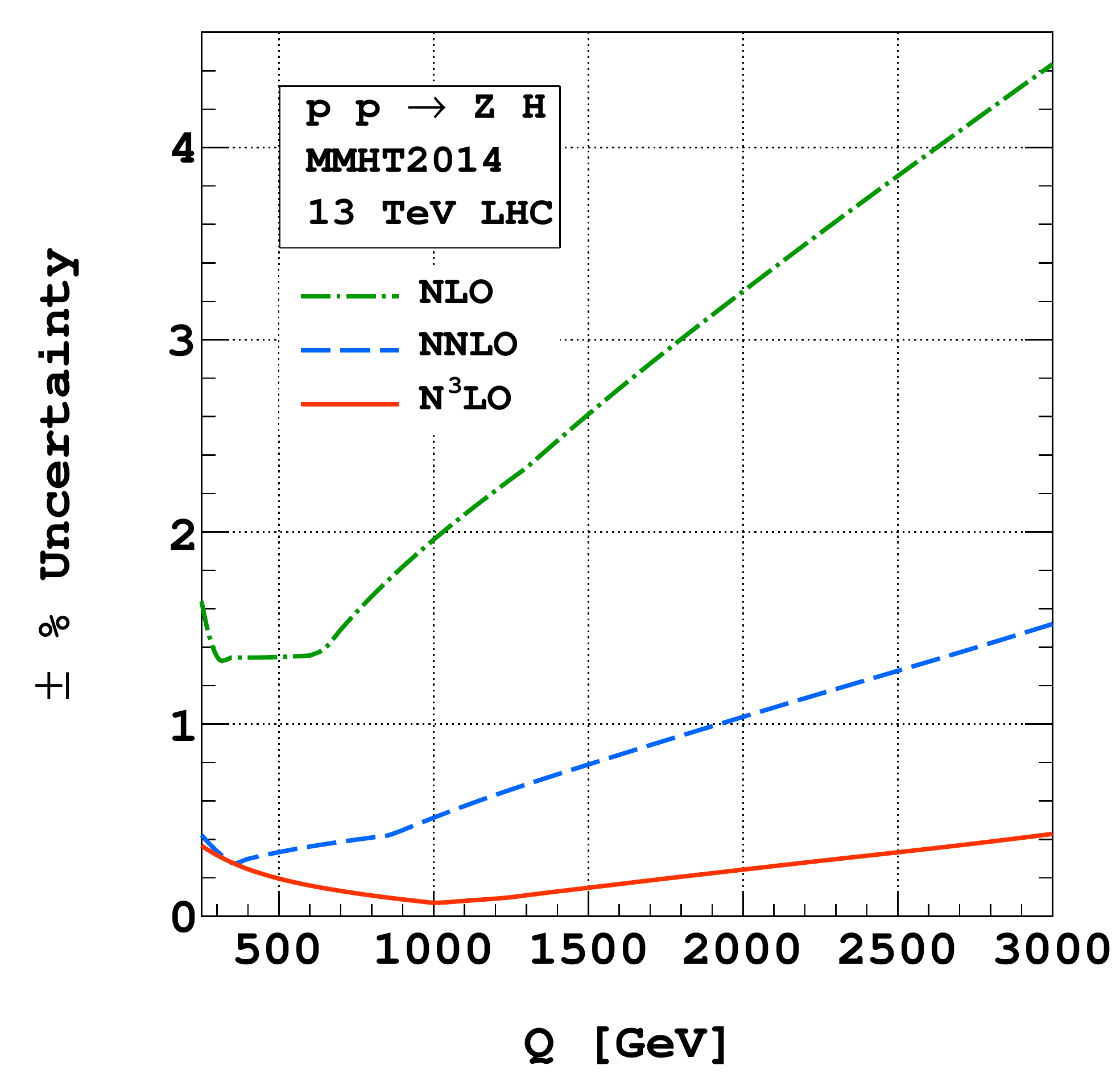}
		\includegraphics[width=5.5cm, height=5.5cm]{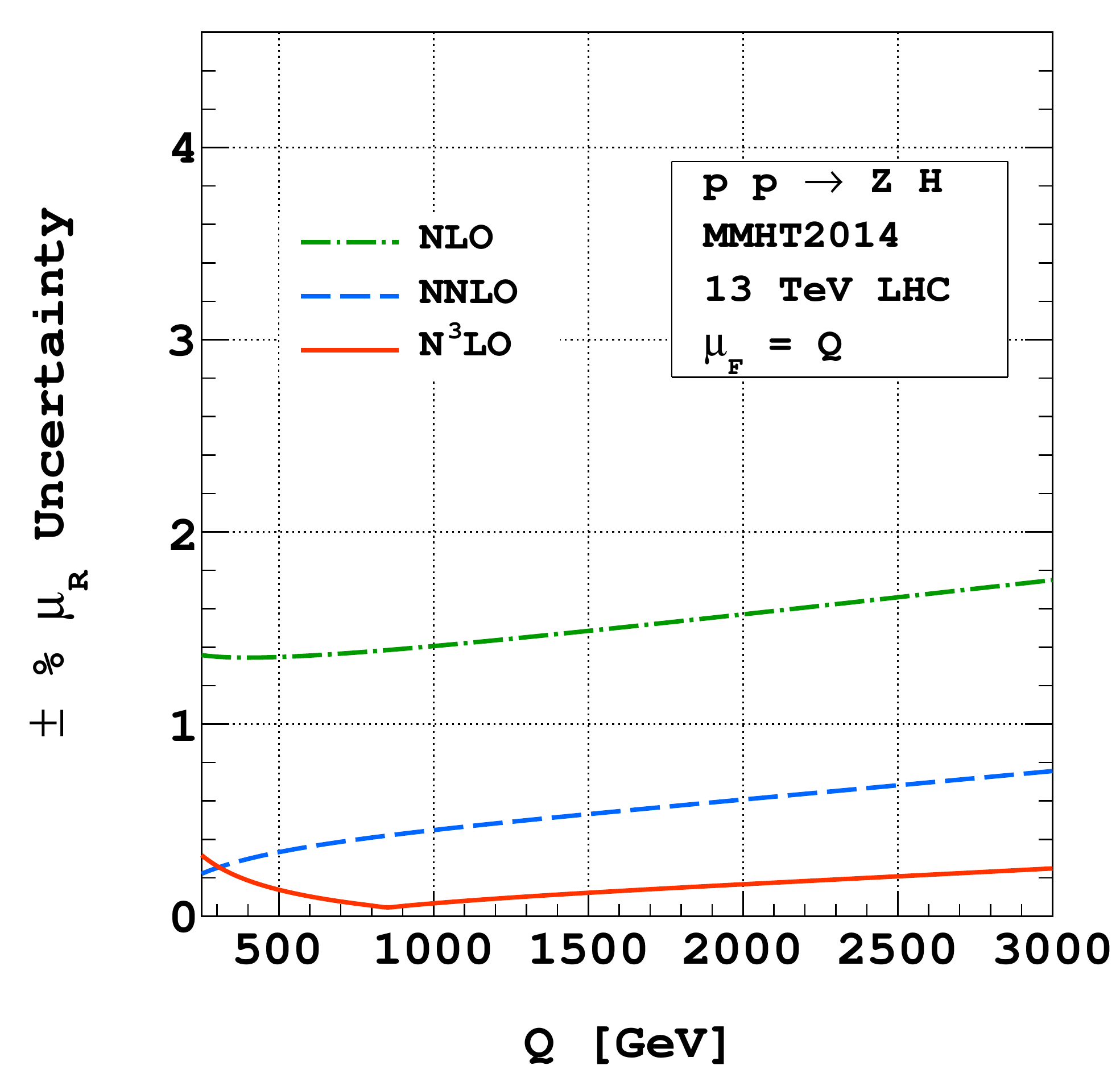}
		\includegraphics[width=5.5cm, height=5.5cm]{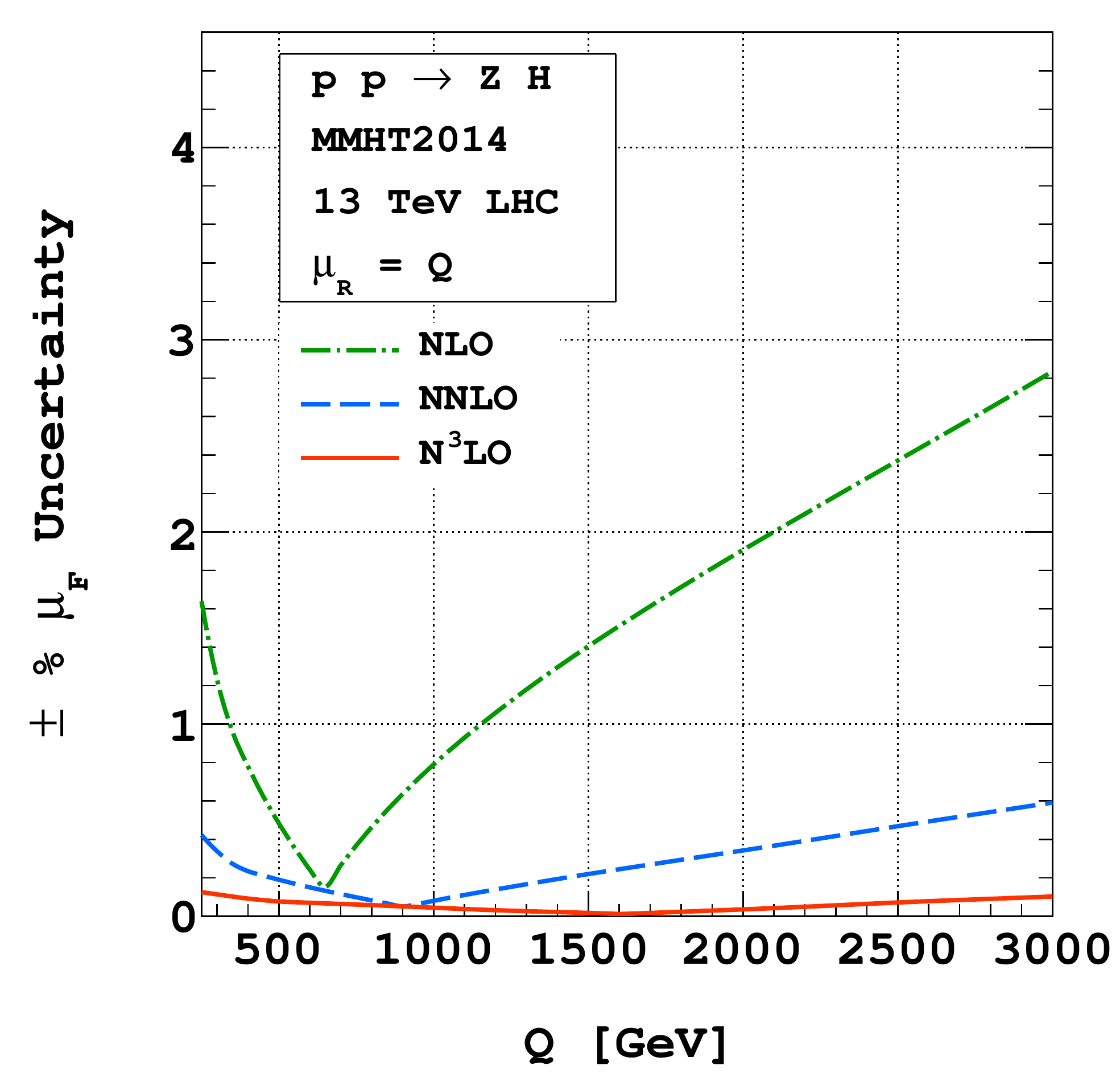}
	}
	\vspace{-2mm}
	\caption{\small{The $7$-point scale uncertainty (left panel),  $\mu_R$} scale uncertainty (middle panel) and $\mu_F$ scale uncertainty (right panel) up to N$^3$LO for $ZH$ production.}
	\label{fig:fo_uncertainty_ZH}
\end{figure}

\begin{figure}[ht!]
	\centerline{
		\includegraphics[width=5.5cm, height=5.5cm]{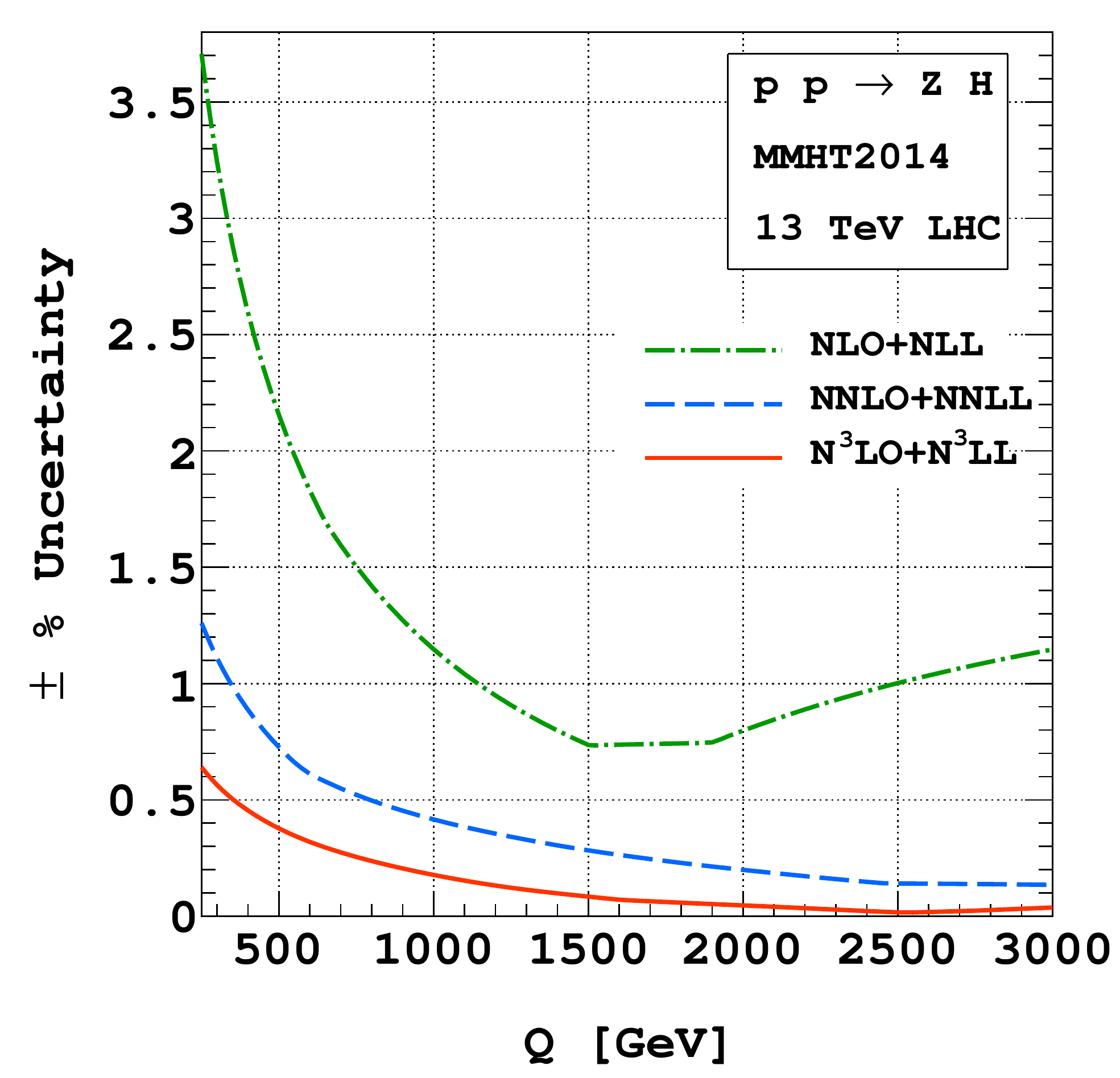}
		\includegraphics[width=5.5cm, height=5.5cm]{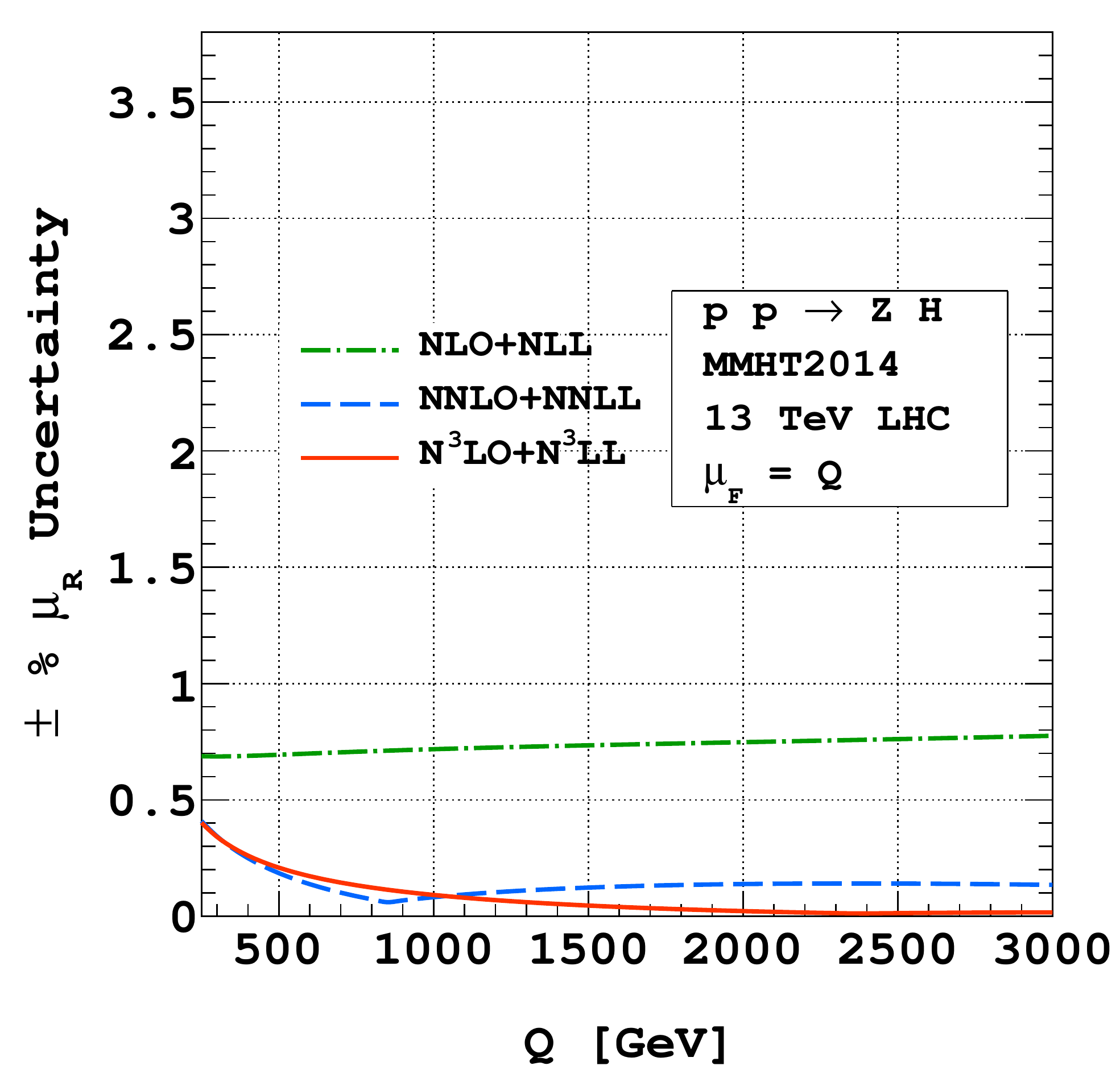}
		\includegraphics[width=5.5cm, height=5.5cm]{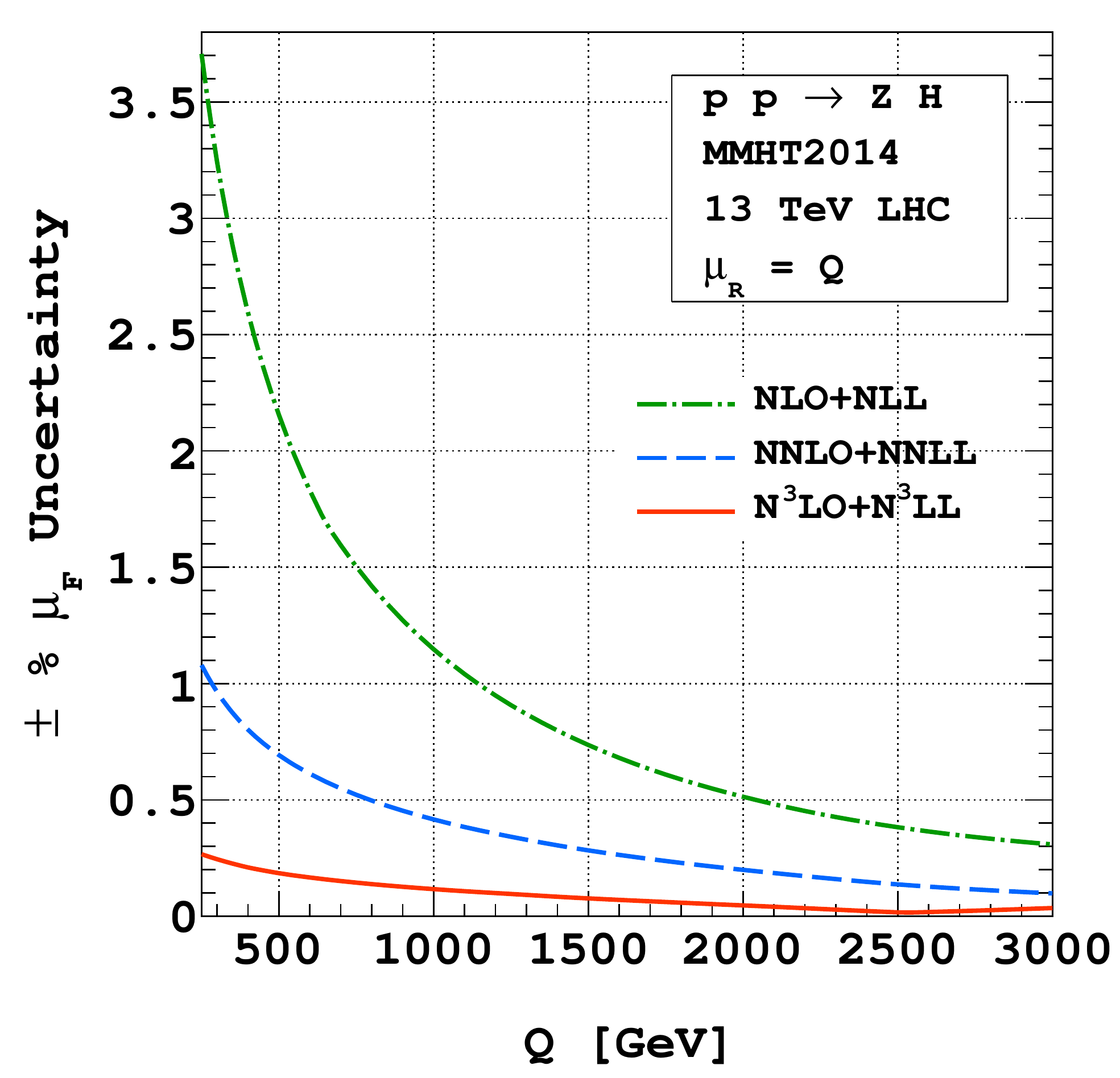}
	}
	\vspace{-2mm}
	\caption{\small{The 7-point scale uncertainty (left panel), $\mu_R$ scale uncertainty (middle panel) $\mu_F$ scale uncertainty (right panel) up to N$^3$LO+N$^3$LL for $ZH$ production.}}
	\label{fig:match_uncertainty_ZH}
\end{figure}
%
It is worth noting that the total production cross sections 
for these processes up to NNLO are available through the 
public code {\tt vh@nnlo} \cite{Brein:2012ne}. The updated 
version {\tt vh@nnlo 2.1} \cite{Harlander:2018yio} of the 
code can handle both the SM Higgs and other BSM 
scenarios where Higgs can be produced in association with 
a gauge boson. In this version, the code is also interfaced 
with {\tt MCFM} \cite{Campbell:2016jau} to produce 
invariant mass distributions up to NNLO level.

In the present context, we provide the invariant mass 
distribution up to third order (N$^3$LO+N$^3$LL). To 
achieve this we use the invariant mass distributions of 
DY processes available at N$^3$LO level through 
{\tt n3loxs} \cite{Baglio:2022wzu} code, and we 
numerically incorporate the decay of the off-shell 
vector boson to $V$ and $H$ instead of to leptons, 
using the Eq.(2) and Eq.(3) of Ref.\cite{Harlander:2018yio}.
For consistency, we reproduce the results obtained 
from {\tt vh@nnlo 2.1} up to NNLO for $VH$ invariant 
mass distribution. We then extend our analysis to the fixed 
order N$^3$LO level. For the resummation case, we use 
our in-house numerical code, similar to the one used 
for dilepton production case discussed in previous 
sections. 

In \fig{fig:fo_resum_ZH}, we plot the invariant mass 
distribution of $VH$ at FO (left panel) and at resum 
level (right panel) up to third order (N$^3$LO+N$^3$LL) 
for $ZH$ production process by varying $Q$ from $250$ GeV 
to $3000$ GeV.  Because the branching of off-shell $V^*$ 
to $VH$ is different from that to dileptons, the 
production cross sections will certainly be different 
from that of the neutral DY production of dileptons. 
However, the corresponding $K$-factors and $R$-factors 
are defined in \eq{eq:ratio}, expected to be 
almost same as those of neutral DY case due to the 
cancellation of the branching part in these ratios.
\onecolumngrid
\begin{table}[h!]
	\begin{center}
{\scriptsize		
\resizebox{15.0cm}{2.5cm}{
		\begin{tabular}{|c|c|c|c|c|c|}
\hline
			$\sqrt{S}$ (TeV) & $7.0$  & $8.0$  &  $13.0$  & 13.6 & 
			$100.0$  \\

\hline	
\hline
LO                       &  $0.2363 \pm 0.36\%$  &  $0.2908 \pm 1.00\%$  &  $0.5934 \pm 3.81\%$  & $0.6324 \pm 4.06\%$ & 
			$8.1105 \pm 13.57\%$\\
\hline
NLO                      &  $ 0.3164 \pm 1.55\%$  &  $ 0.3878 \pm 1.50\%$  &  $0.7754 \pm 1.36\%$  & $0.8245 \pm 1.36\%$ &
			$9.1445 \pm 4.40\%$\\
\hline
			NNLO                     &  $0.3280 \pm 0.41\%$  &  $0.4017 \pm 0.37\%$  &  $0.8005 \pm 0.35\%$  & $0.8508 \pm 0.36 \%$ &
			$9.1215 \pm 0.94\%$\\
\hline
			N$^3$LO           &  $0.3266 \pm 0.25\%$  &  $0.3996 \pm 0.27\%$  &  $0.7943 \pm 0.32\%$  & $0.8441 \pm 0.33 \%$ &
			$8.9790 \pm 0.49\%$\\
\hline
			LO+LL                    &  $0.2706 \pm 1.48\%$  &  $0.3319 \pm 1.45\%$  &  $0.6721 \pm 4.20\%$  & $0.7158 \pm 4.44\%$ & 
			$9.0406 \pm 13.86\%$\\
\hline
			NLO+NLL                  &  $0.3260 \pm 4.34\%$  &  $0.3992 \pm 4.33\%$  &  $0.7966 \pm 4.31\%$  & $0.8469 \pm 4.31 \%$ &
			$9.3550 \pm 5.39\%$\\
\hline
			NNLO+NNLL                &  $0.3295 \pm 1.40\%$  &  $0.4034 \pm 1.43\%$  &  $0.8036 \pm 1.53\%$  & $0.8542 \pm 1.53\%$ &
			$9.1500 \pm 1.77\%$\\
\hline
			N$^3$LO+N$^3$LL   &  $0.3266 \pm 0.46\%$  &  $0.3996 \pm 0.49\%$  &  $0.7943 \pm 0.57\%$  & $0.8441 \pm 0.58 \%$ &
			$8.9795 \pm 0.75\%$\\
\hline
\end{tabular}
 }
		\caption{\small{$ZH$ production cross section (in pb) for different $\sqrt{S}$} with 7-point scale uncertainty.} 
\label{tab:tableZHnew}
 }
 \end{center}
\end{table}
\begin{table}[h!]
	\begin{center}
{\scriptsize		
\resizebox{13.0cm}{2.00cm}{
		\begin{tabular}{|c|c|c|c|c|}
\hline
			$\sqrt{S}$ (TeV) &
			$\sigma_{\text {N}^3\text {LO}}^{DY,ZH}$  &  $\sigma_{\text {N}^3\text {LO+N}^3\text {LL}}^{DY,ZH}$ & $\sigma_{\text{N}^3\text{LO}}^{tot,ZH}$ 
			& $\sigma_{\text{N}^3\text{LO+N}^3\text{LL}}^{tot,ZH}$  \\

\hline
\hline
7.0                       
			&  $0.3266 \pm 0.25\%$  & $0.3266 \pm 0.46\%$ & $ 0.3534 \pm 1.25\% $   & $ 0.3534 \pm 1.17\% $ \\ 
\hline
8.0                     
			&  $ 0.3996 \pm 0.27\%$ & $0.3996 \pm 0.49\%$ & $ 0.4373 \pm 1.36\% $ & $ 0.4373 \pm 1.28\%$  \\
\hline
13.0                   
			&  $0.7943 \pm 0.32\%$  & $0.7943 \pm 0.57 \%$ & $ 0.9112 \pm 1.79\%$ & $ 0.9112 \pm 1.67\%$  \\
\hline
13.6     		
			&  $0.8441 \pm 0.33\%$  & $0.8441 \pm 0.58 \%$ & $0.9728 \pm 1.86\%$  &  $ 0.9728 \pm 1.75\%$  \\
\hline
100.0                   
			&  $8.9790 \pm 0.49\%$  & $8.9795 \pm 0.75\%$  & $ 13.2674 \pm 4.57\%$  &  $ 13.2678 \pm 4.39\%$   \\ 
\hline
\end{tabular}
 }
		\caption{\small{$ZH$ production cross section (in pb) of DY-type N$^3$LO, DY-type N$^3$LO+N$^3$LL, total N$^3$LO and total resummed N$^3$LO+N$^3$LL which are defined in Eq. (\ref{eq:ZHtotal_fo}) and Eq. (\ref{eq:ZHtotal_resum}) for different $\sqrt{S}$} with 7-point scale uncertainty.} 
\label{tab:tableZHcompare}
 }
 \end{center}
\end{table}
%
%
\onecolumngrid
\begin{figure}[ht!]
	\centerline{
		\includegraphics[width=7.0cm, height=7.0cm]{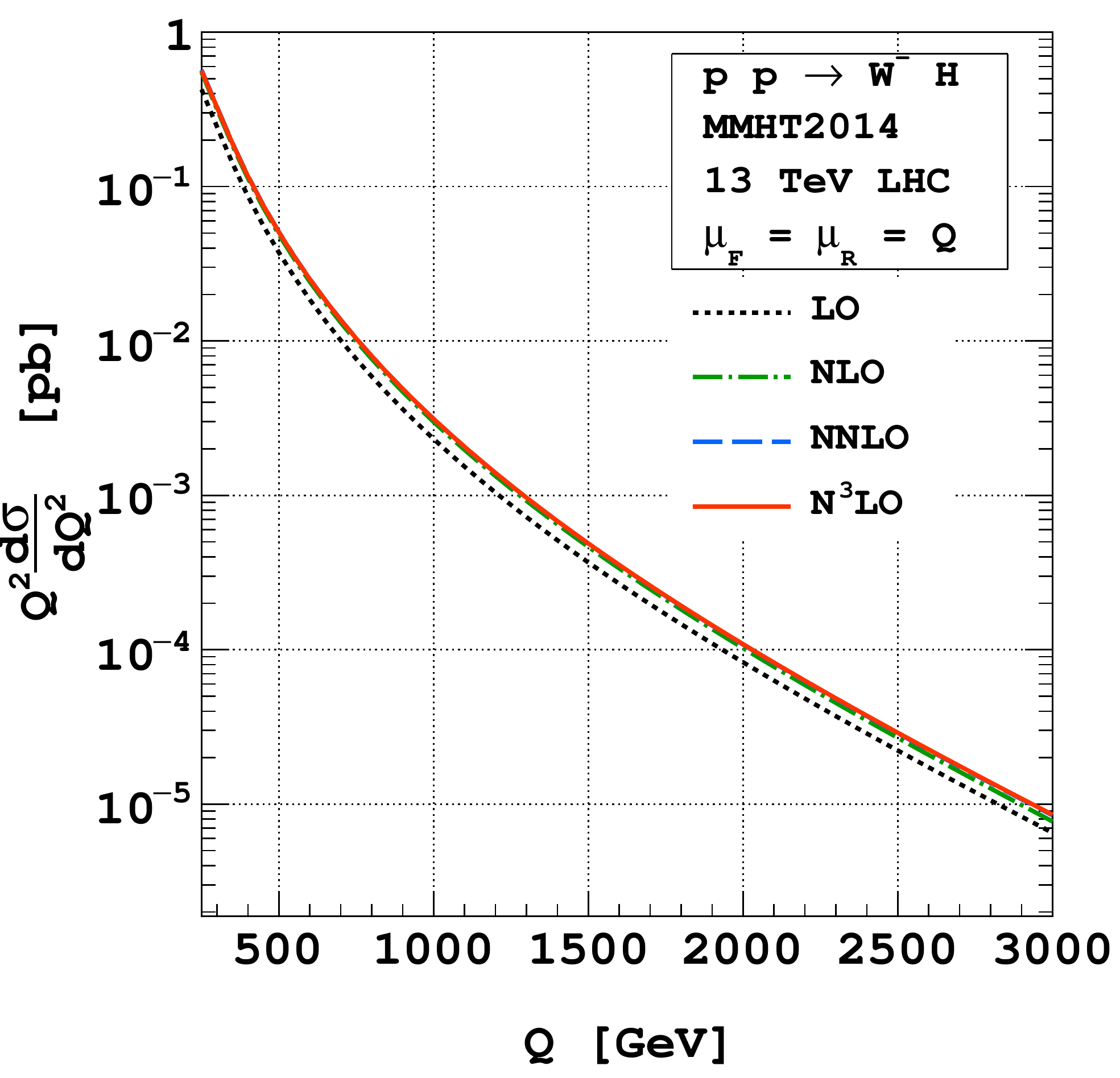}
		\includegraphics[width=7.0cm, height=7.0cm]{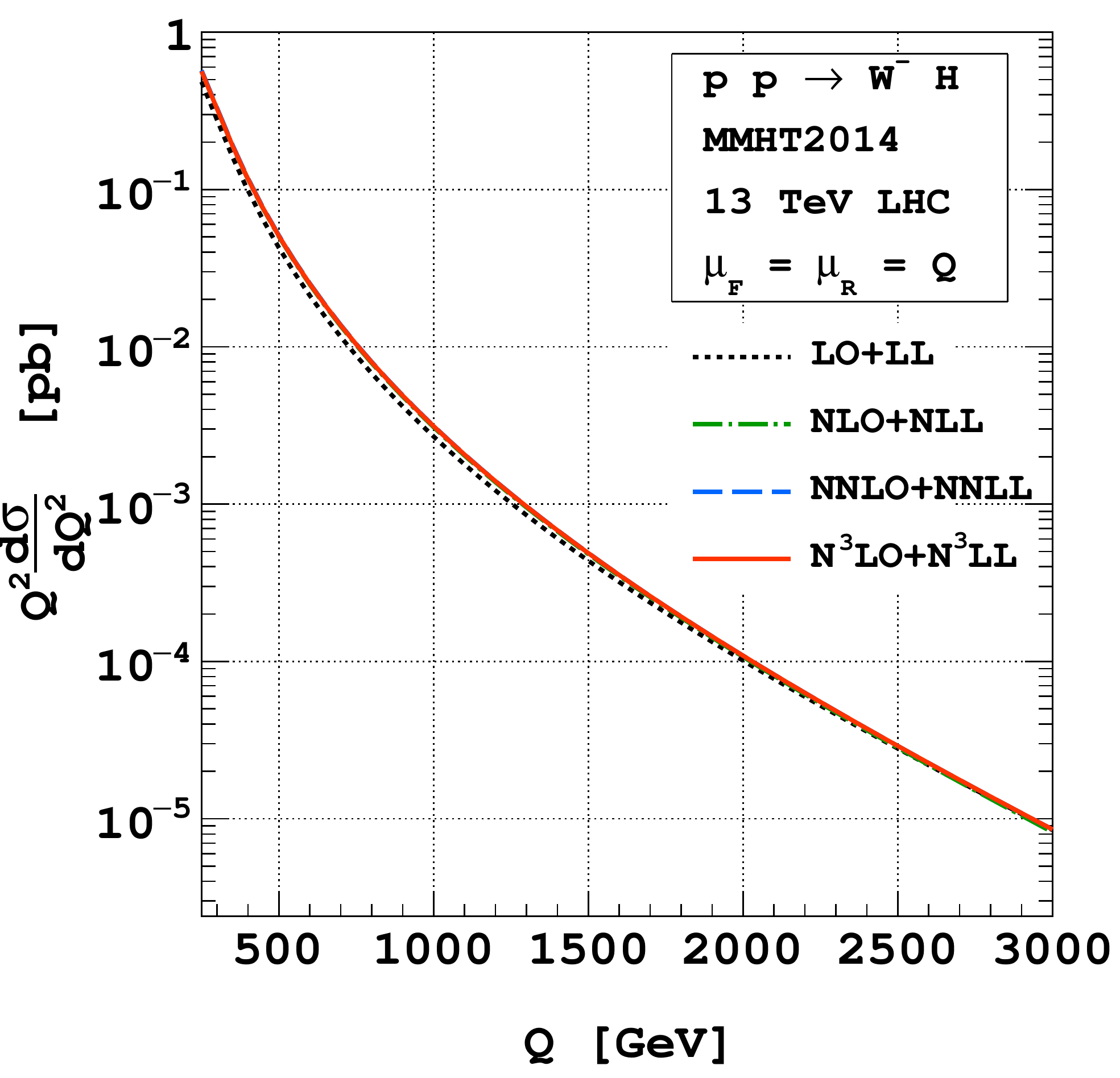}
	}
	\vspace{-2mm}
	\caption{\small{Invariant mass distribution of $W^{-}H$ for $13$ TeV LHC fixed order (left) and the resummed (right).}}
	\label{fig:fo_resum_WmH}
\end{figure}

\begin{figure}[ht]
	\centerline{
		\includegraphics[width=7.0cm, height=7.0cm]{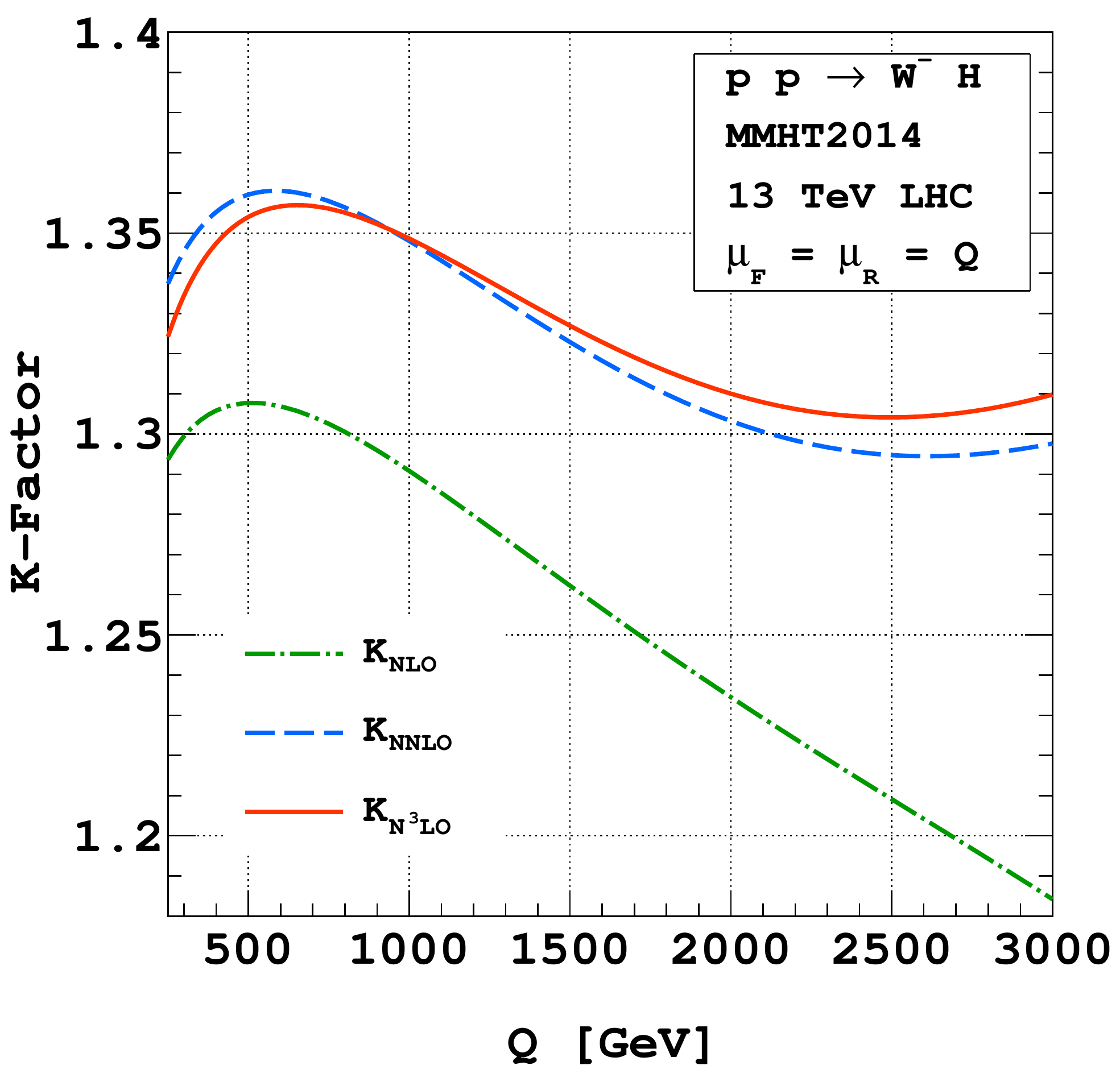}
		\includegraphics[width=7.0cm, height=7.0cm]{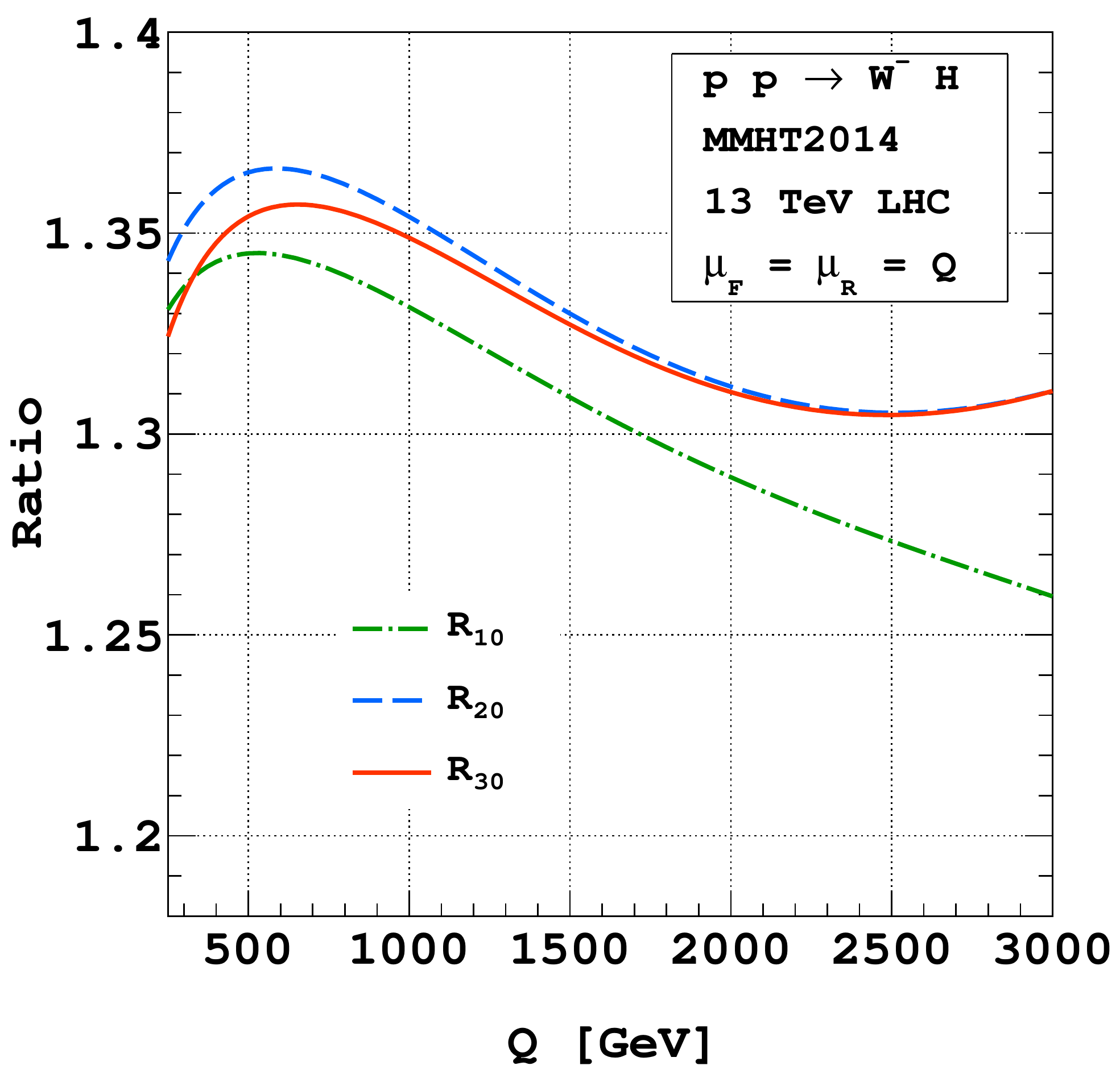}
	}
	\vspace{-2mm}
	\caption{\small{K-factor for fixed order of $W^{-}H$ for $13$ TeV LHC (left) and the enhancement in the resummed cross section 
			over fixed order LO are shown here (right) through $R_{ij}$ is defined in \eq{eq:ratio}.}}
	\label{fig:matched_kfac_WmH}
\end{figure}
In \fig{fig:matched_kfac_ZH}, we present these $K$-factors 
(left panel) and $R$-factors (right panel) up to third 
order (N$^3$LO+N$^3$LL). 
The corresponding scale uncertainties for the $ZH$ 
production case are shown for FO in 
\fig{fig:fo_uncertainty_ZH} and 
for the resum case in \fig{fig:match_uncertainty_ZH}. 
We notice that
the behavior of the scale uncertainties for FO is the 
same as that for the neutral DY case. 
For completeness, in \fig{fig:fo_uncertainty_ZH}, we show 
separately the uncertainties due to the $7$-point scale 
variations (left panel), those due to only $\mu_R$ for 
fixed $\mu_F=Q$ (middle panel) and those due to only 
$\mu_F$ for fixed $\mu_R=Q$ (right panel) for FO results. 
We find that the factorization 
scale uncertainties at higher orders, namely, at NNLO and 
N$^3$LO level are smaller than those due to the 
renormalization scale. In \fig{fig:match_uncertainty_ZH}, 
we show similar plots as those in 
\fig{fig:fo_uncertainty_ZH} but for resummed results. 
However, here we find that the factorization scale 
uncertainties are larger than the uncertainties due to 
the renormalization scale variations. This is expected 
because in the resummation case, the factorization scale 
dependence is included only from the threshold regions 
and that too in the $q\bar{q}$
channel. However, at the fixed order at N$^3$LO level, 
this is not the case as the full $\mu_F$ scale dependence 
is included in the coefficient functions at this order. 
We also notice that in general the scale uncertainties 
in FO results
in the high $Q$ region increase but very slowly. On the 
contrary, the scale uncertainties in the resummed 
predictions decrease with increasing $Q$. This is because 
in the high $Q$ region the bulk of the cross sections are 
dominated by threshold logarithms and are resummed to all 
orders in the perturbation theory.  

\begin{figure}[ht!]
	\centerline{
		\includegraphics[width=7.0cm, height=7.0cm]{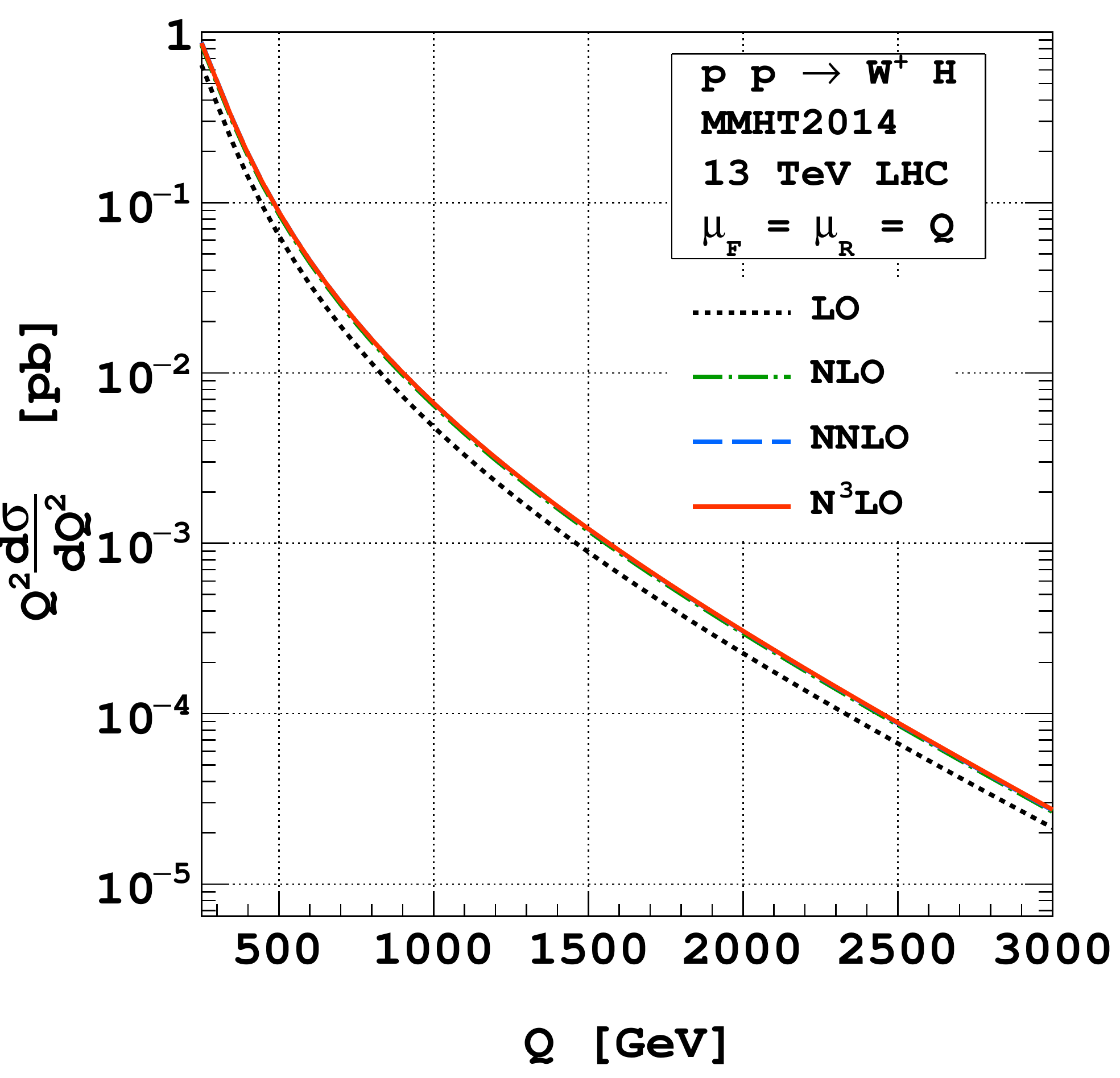}
		\includegraphics[width=7.0cm, height=7.0cm]{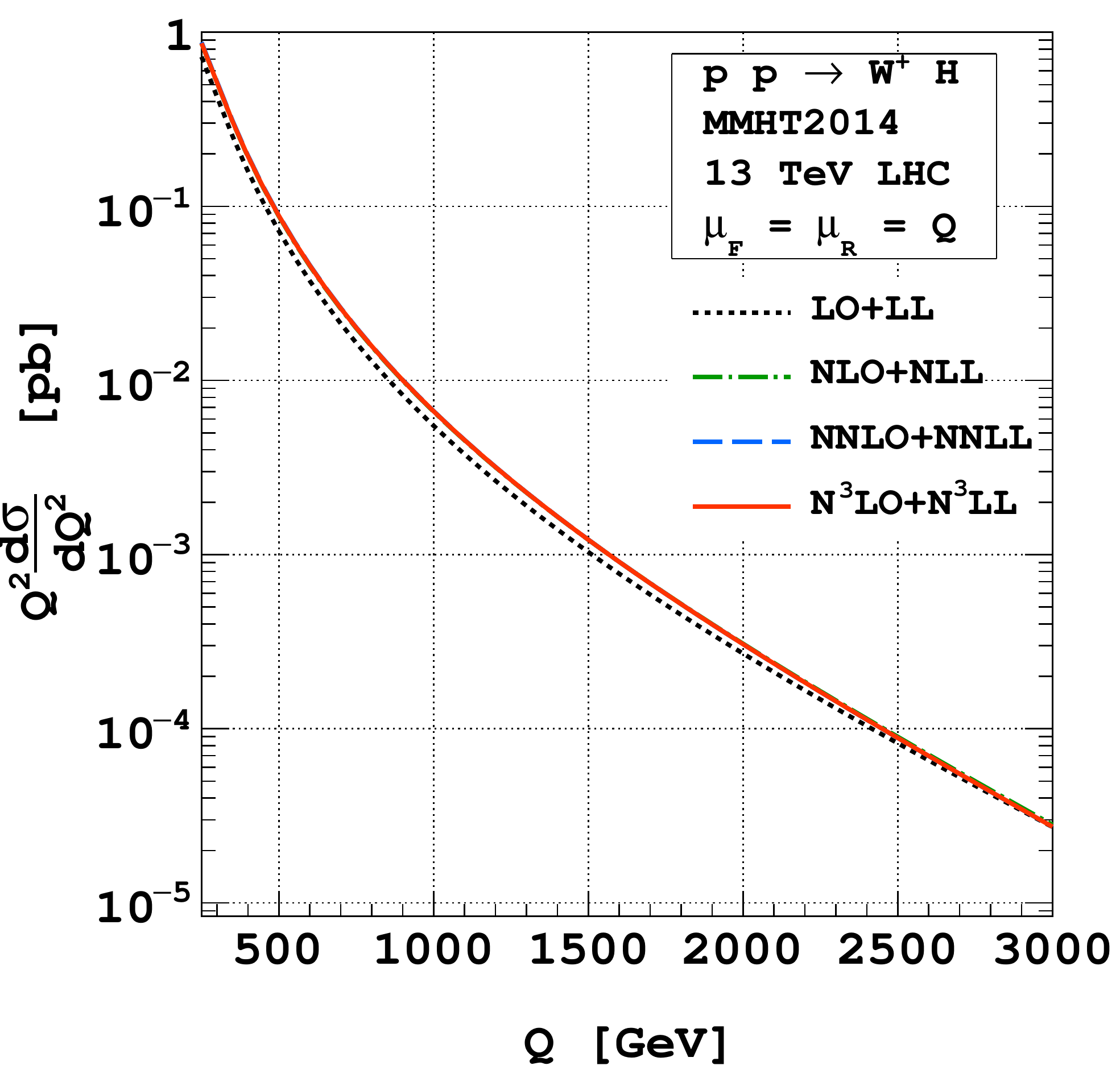}
	}
	\vspace{-2mm}
	\caption{\small{Invariant mass distribution 
	of W$^{+}$H for $13$ TeV LHC fixed order (left panel) 
	and the resummed (right panel).
	}}
	\label{fig:fo_resum_WpH}
\end{figure}

\begin{figure}[ht!]
	\centerline{
		\includegraphics[width=7.0cm, height=7.0cm]{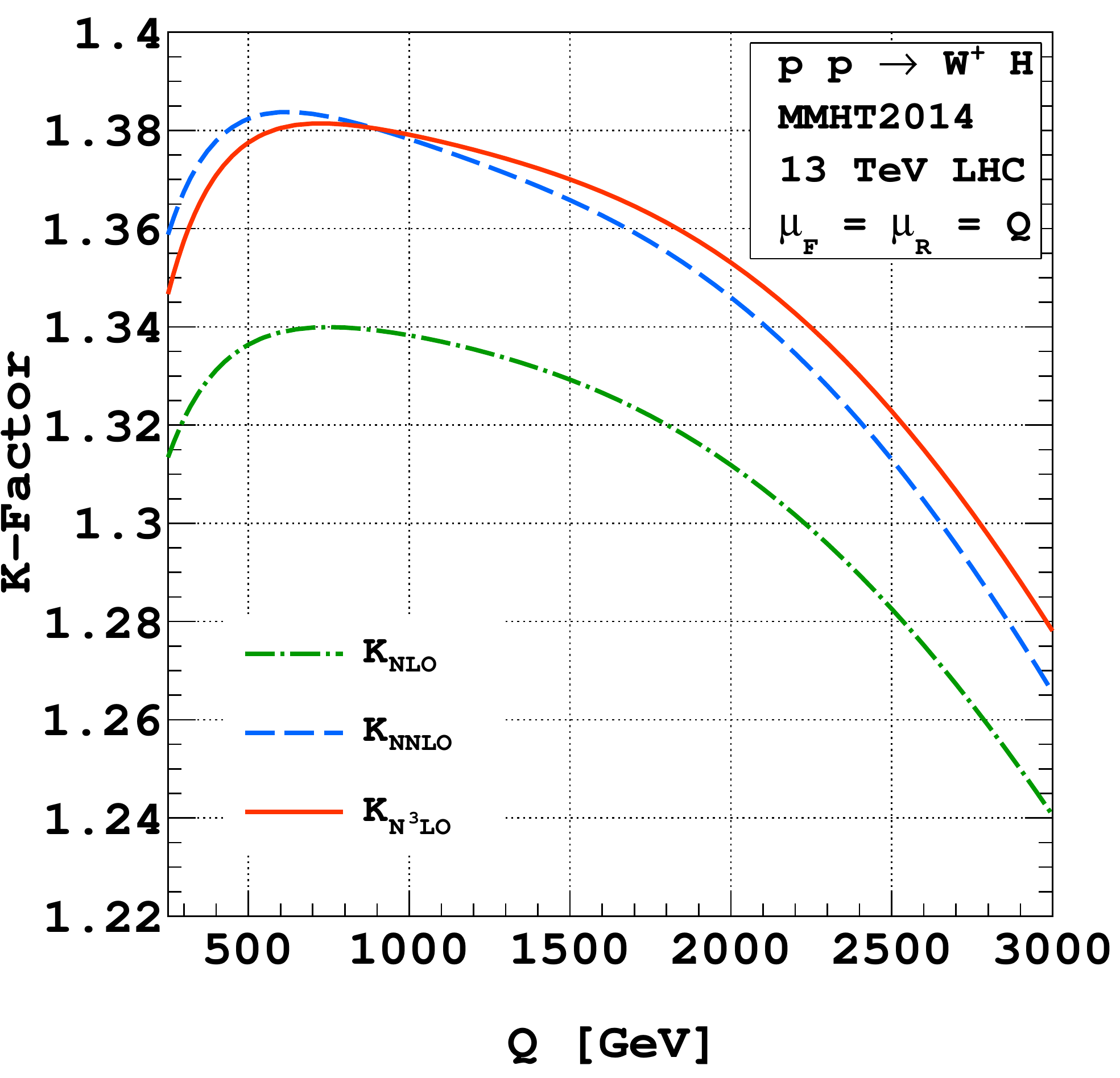}
		\includegraphics[width=7.0cm, height=7.0cm]{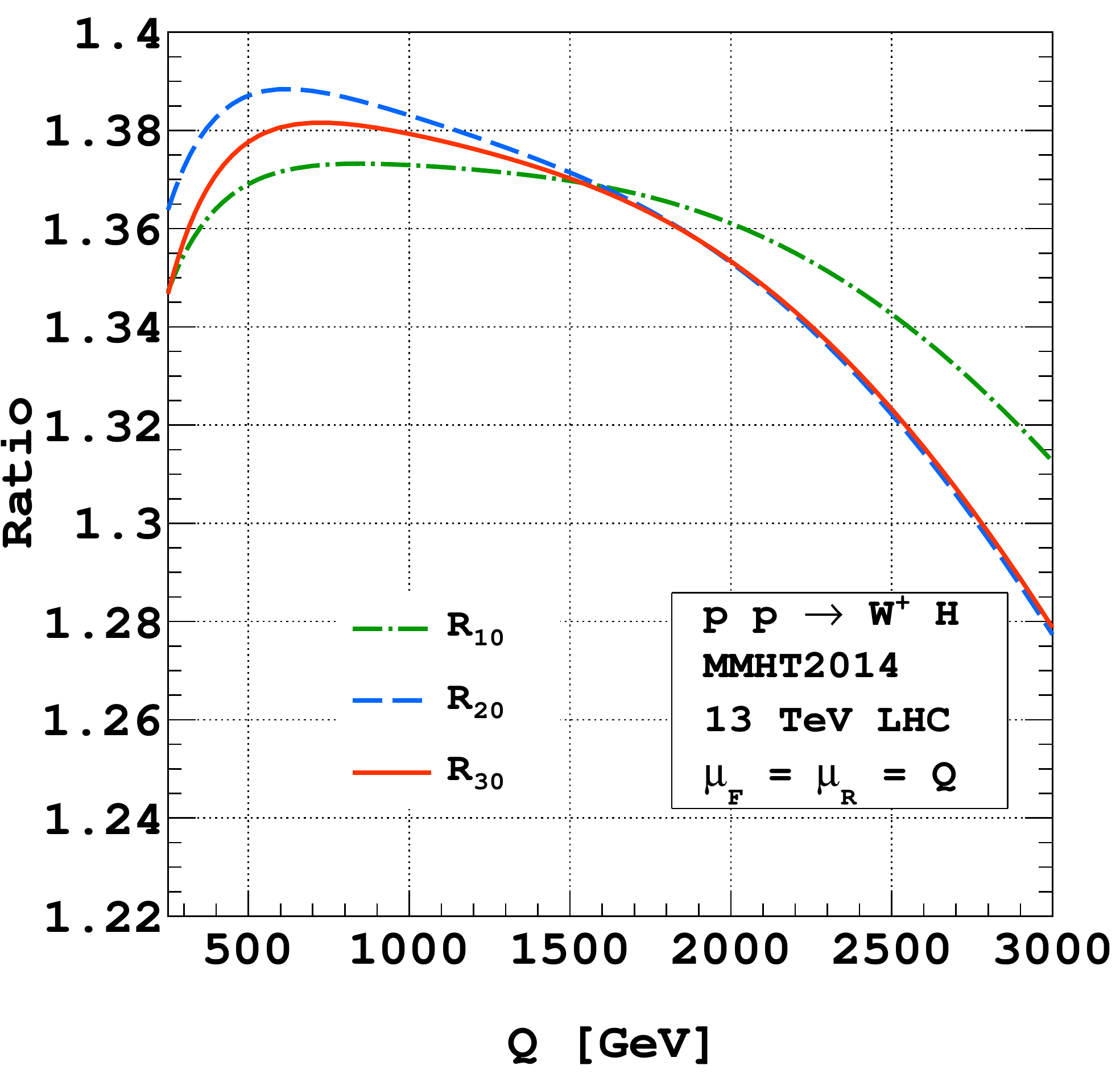}
	}
	\vspace{-2mm}
	\caption{\small{K-factor for fixed order of W$^{+}$H for $13$ TeV LHC (left panel) and the enhancement in the resummed cross section 
			over fixed order LO are shown here 
			(right panel) through $R_{ij}$ is 
			defined in \eq{eq:ratio}.
			}}
	\label{fig:matched_kfac_WpH}
\end{figure}

\begin{figure}[ht!]
	\centerline{
		\includegraphics[width=7.0cm, height=7.0cm]{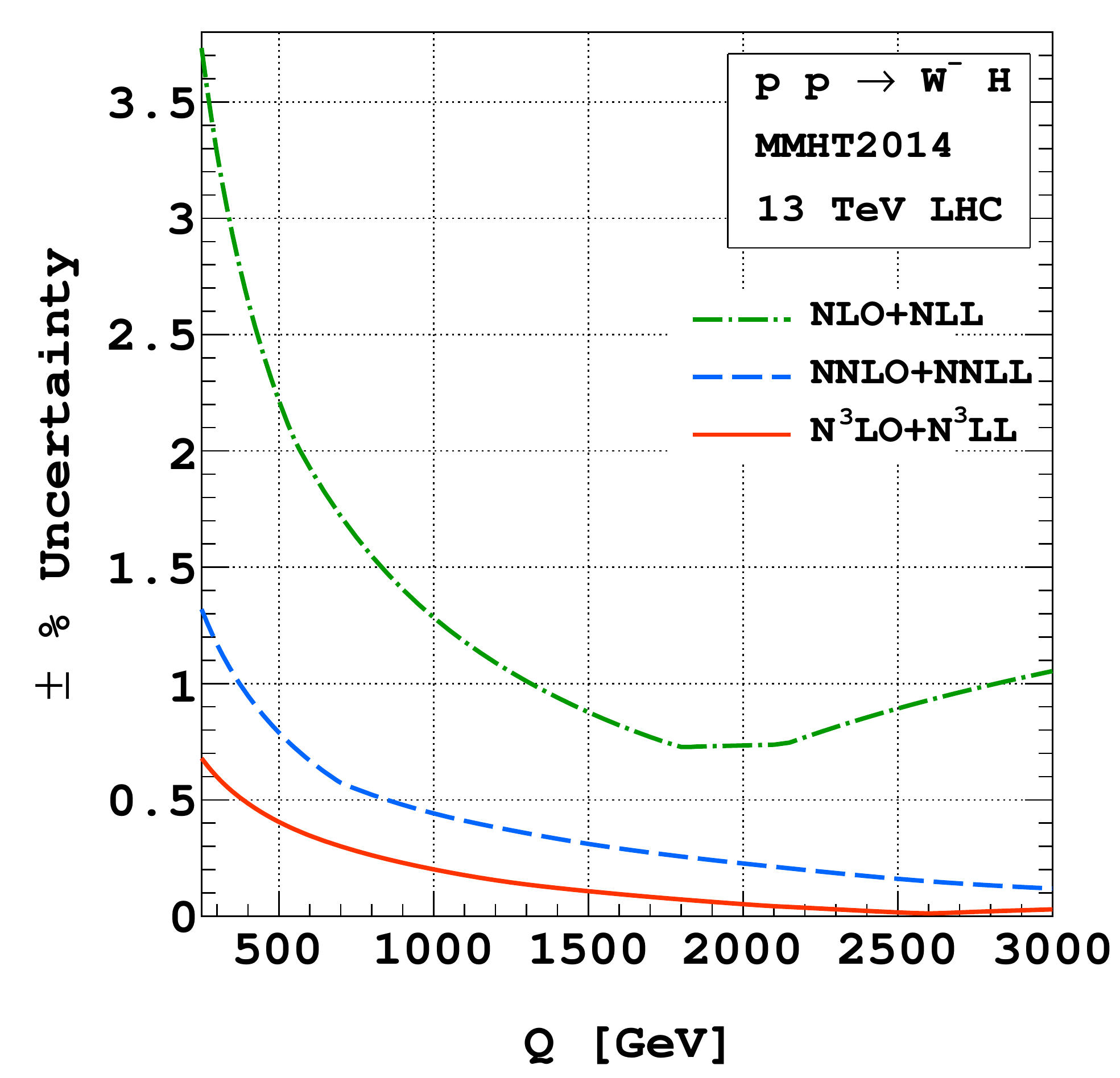}
		\includegraphics[width=7.0cm, height=7.0cm]{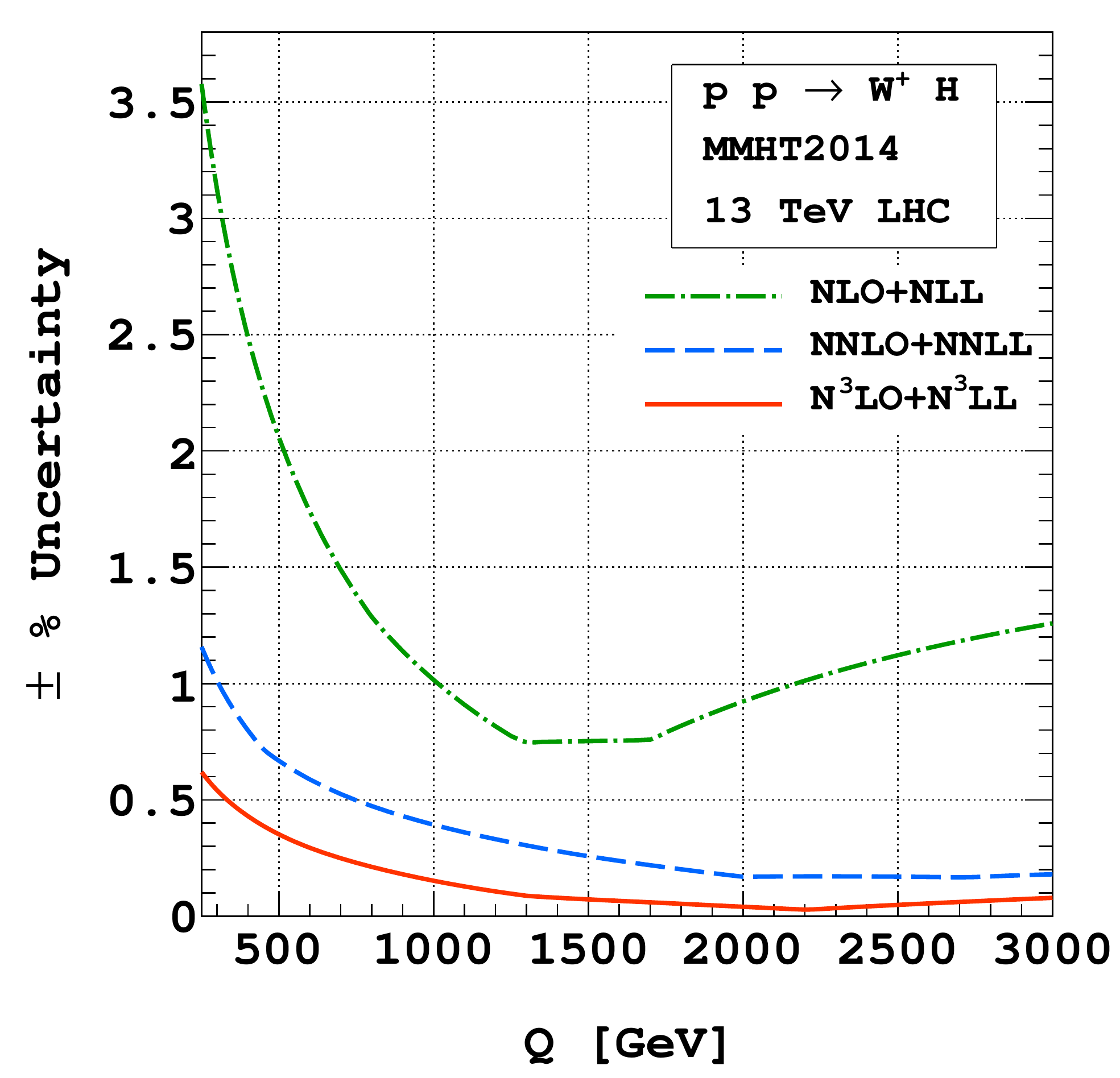}
	}
	\vspace{-2mm}
	\caption{\small{The resummed 7-point scale uncertainties for W$^-$H (left panel) and W$^+$H (right panel) up to N$^3$LO+N$^3$LL.}}
	\label{fig:match_uncertainty_7points}
\end{figure}

In \fig{fig:fo_resum_WmH} and \fig{fig:fo_resum_WpH}, we 
show the results for invariant mass distribution of $W^- H$ 
and $W^+ H$ production processes respectively.
The corresponding $K$-factors and $R_{i0}$-factors are 
shown in \fig{fig:matched_kfac_WmH} and 
\fig{fig:matched_kfac_WpH}. The results for the invariant 
mass distribution differ from those of the respective 
dilepton production processes through off-shell $W^-$ 
and $W^+$ gauge bosons. However, the ratios $K$ and 
$R$-factors are almost the same. Due to the underlying 
parton fluxes, these $K$-factors and 
$R$-factors for $WH$ production case, on the other hand,
will certainly be different from those of $ZH$ case. 
In \fig{fig:match_uncertainty_7points}, we show the
$7$-point scale uncertainties in the invariant mass 
distribution of $W^-H$ and $W^+H$ processes. 

\begin{figure}[ht!]
	\centerline{
		\includegraphics[width=5.5cm, height=5.5cm]{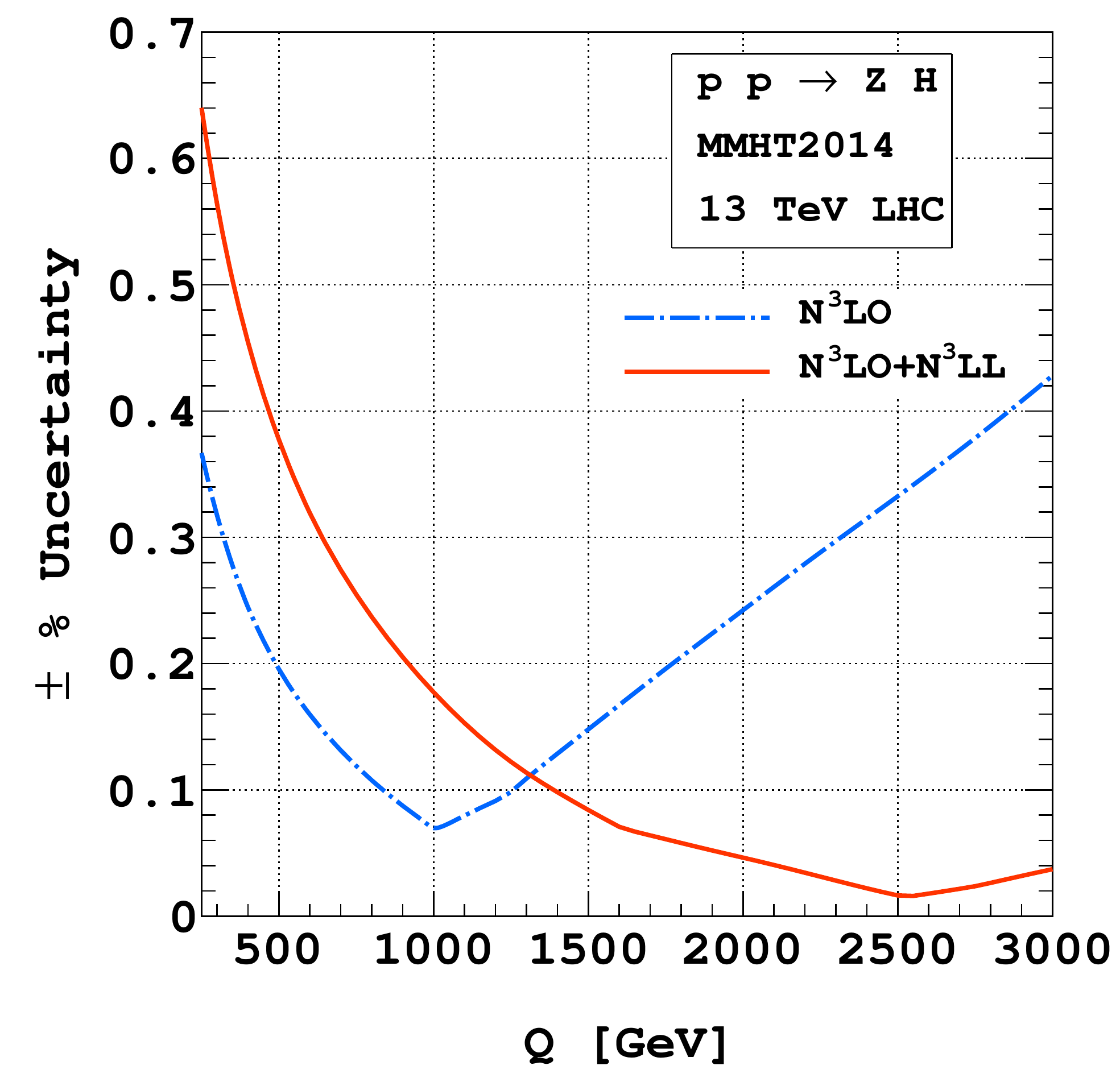}
		\includegraphics[width=5.5cm, height=5.5cm]{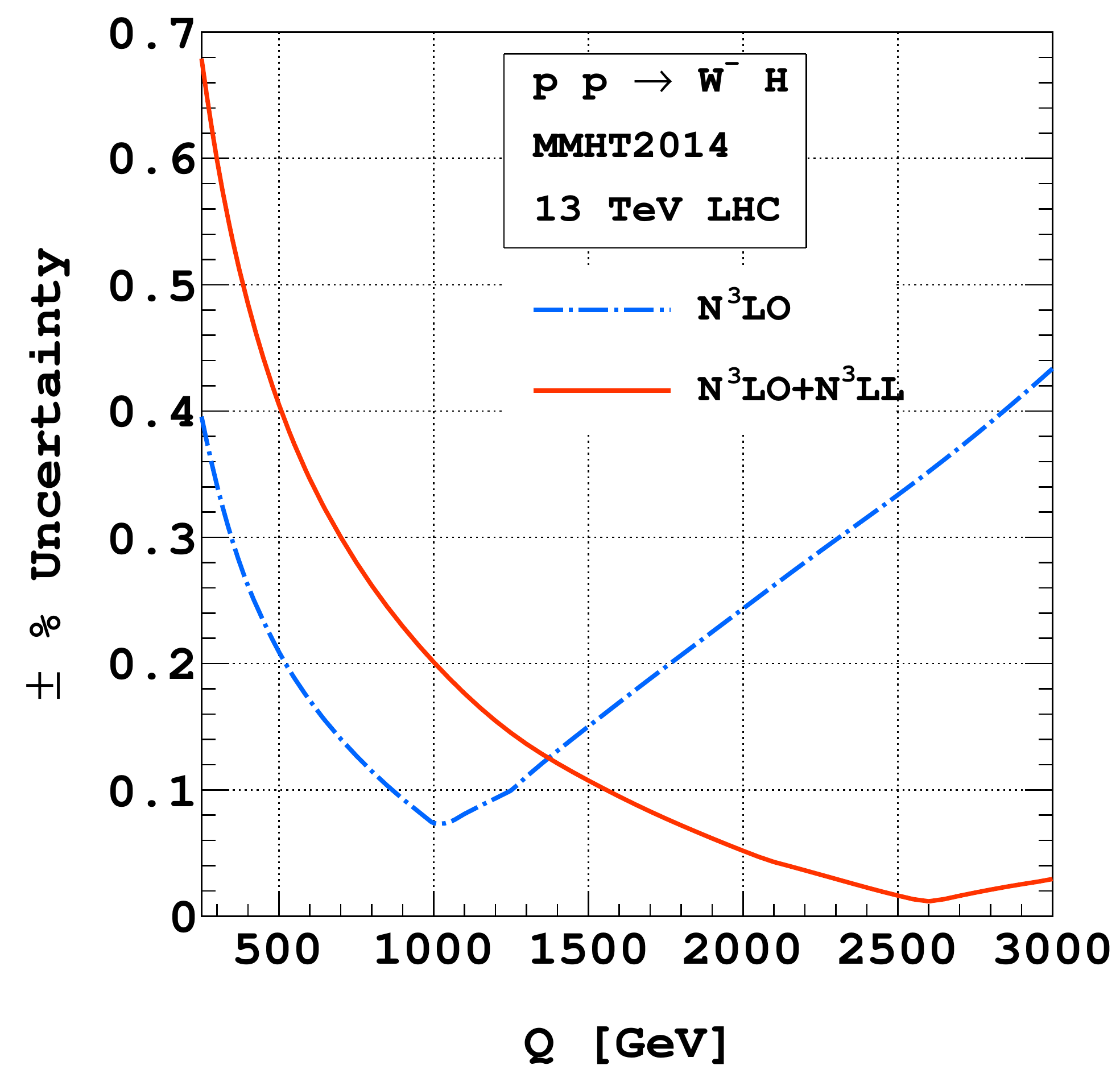}
		\includegraphics[width=5.5cm, height=5.5cm]{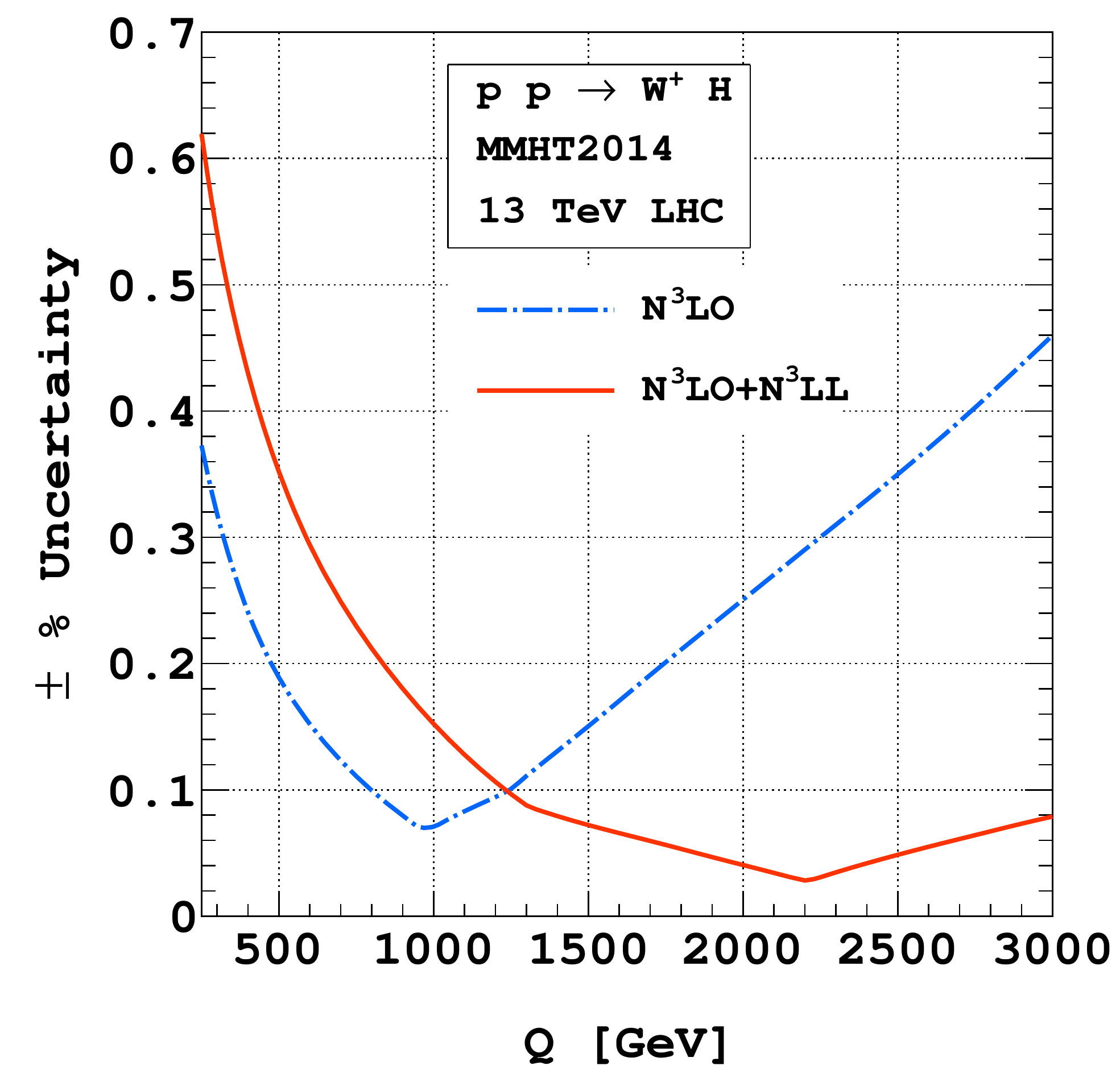}
	}
	\vspace{-2mm}
	\caption{\small{The 7-point scale uncertainty comparison for different Higgs associated production ZH (left panel), W$^-$H (middle panel) and W$^+$H (right panel) between resum and fixed order results.}}
	\label{fig:comparison_uncertainty_7points}	
\end{figure}

For these new results i.e. the invariant mass distribution 
for $VH$ process at N$^3$LO and N$^3$LO+N$^3$LL, 
we compare the $7$-point scale uncertainties in FO and 
in resum results \fig{fig:comparison_uncertainty_7points}. 
The behavior of the scale uncertainties 
is almost identical to that of the neutral DY case 
(see \fig{fig:comparison_uncertainty_7points_DY}), namely, 
the scale uncertainties are smaller in the low $Q$
region for FO case, while the same is true for resum case 
in the high $Q$ region. 
There is a very small and negligible 
difference between $ZH$ case and neutral DY case, which 
is mostly due to the presence of photon contribution 
in the latter case.

\begin{figure}[ht!]
	\centerline{	
		\includegraphics[width=5.5cm, height=5.5cm]{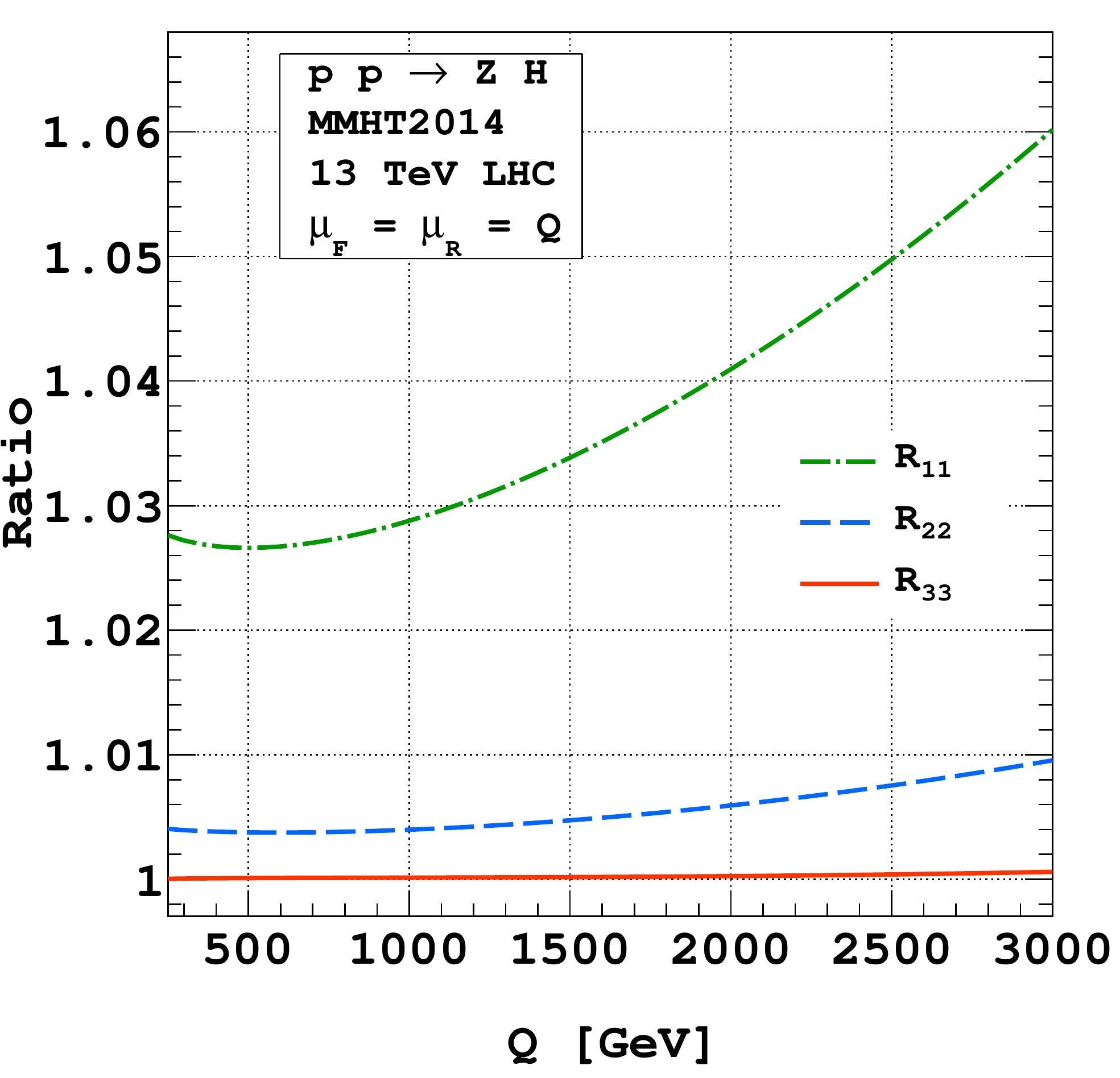}
		\includegraphics[width=5.5cm, height=5.5cm]{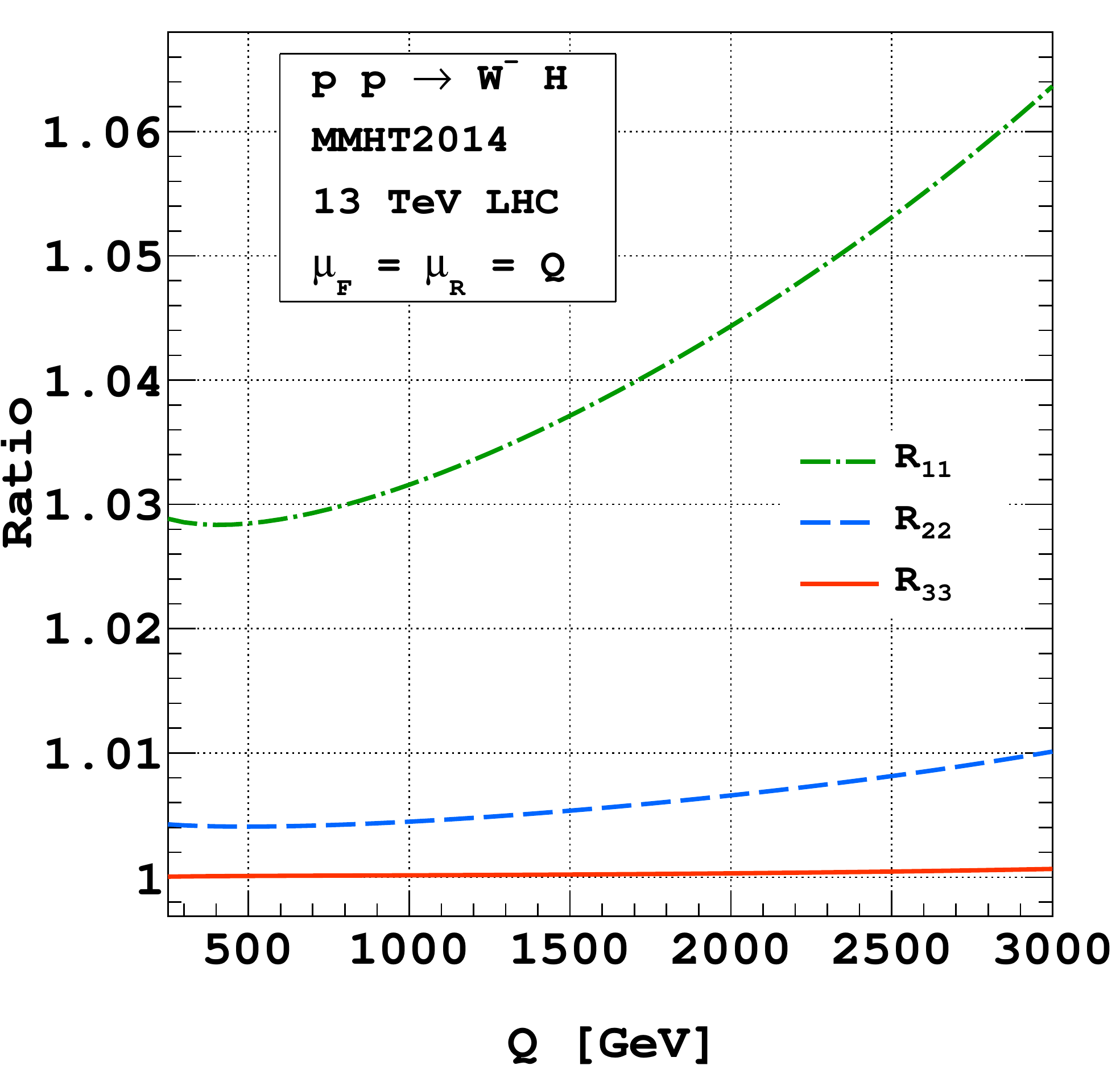}
		\includegraphics[width=5.5cm, height=5.5cm]{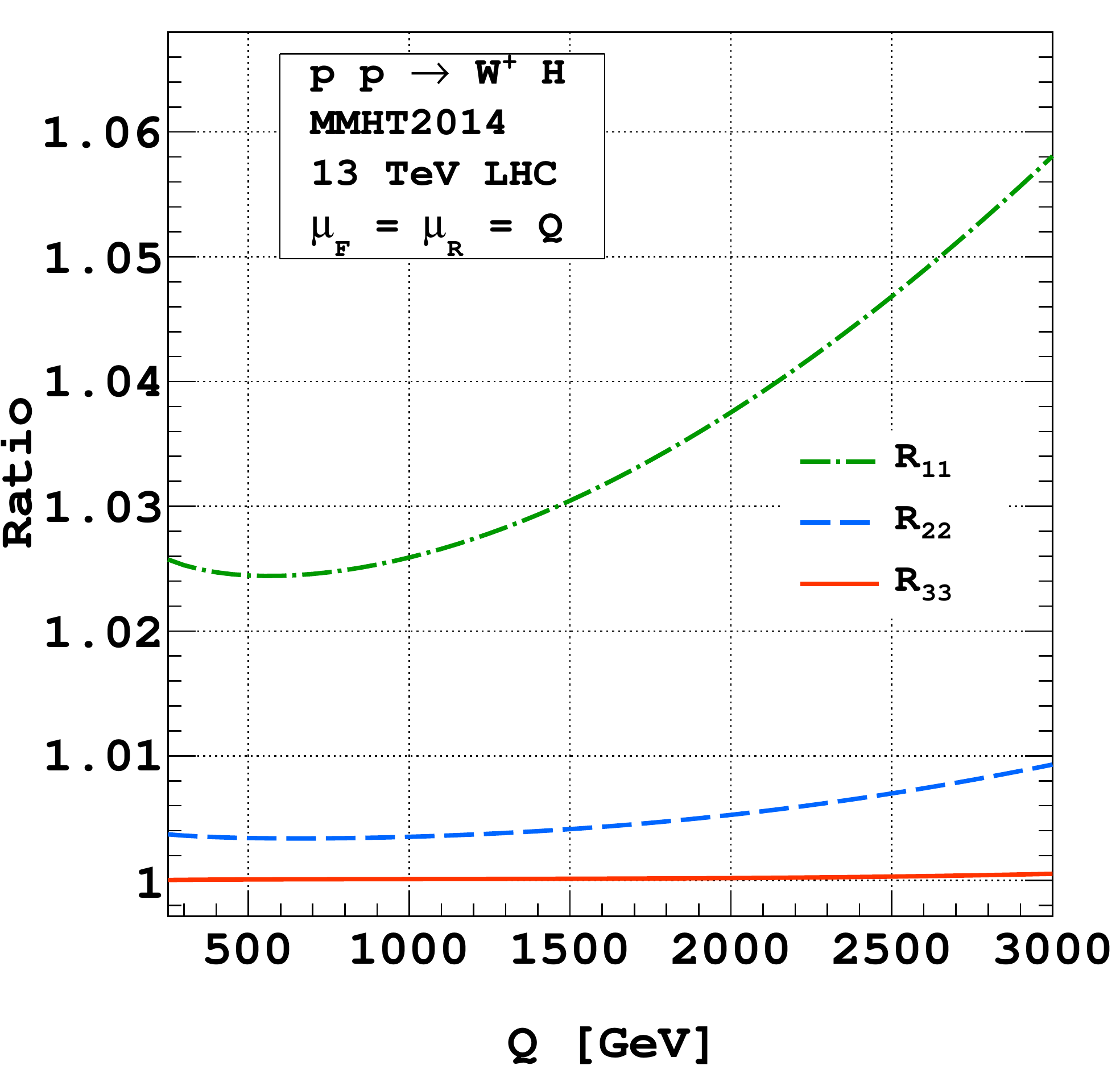}
	}
	\vspace{-2mm}
	\caption{\small{ The enhancement in the resummed cross section 
			over fixed order are shown here for different 
			Higgs associated production ZH (left panel), 
			W$^-$H (middle panel) and W$^+$H (right panel) through $R_{ij}$ is defined in 
			\eq{eq:ratio}.
			}}
	\label{fig:matched_r0fac_comp}
\end{figure}

\begin{figure}[ht!]
        \centerline{
                \includegraphics[width=5.5cm, height=5.5cm]{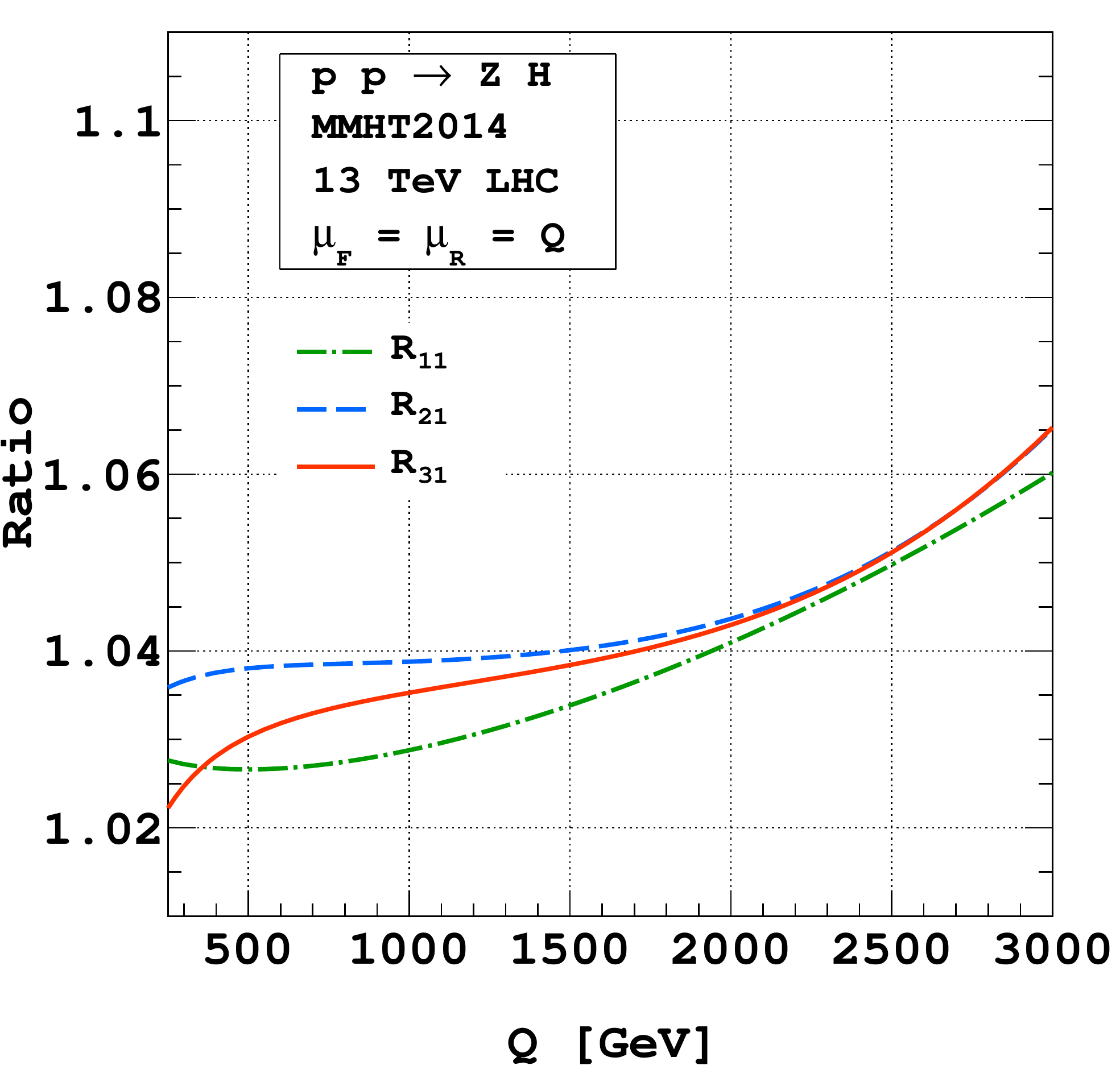}
                \includegraphics[width=5.5cm, height=5.5cm]{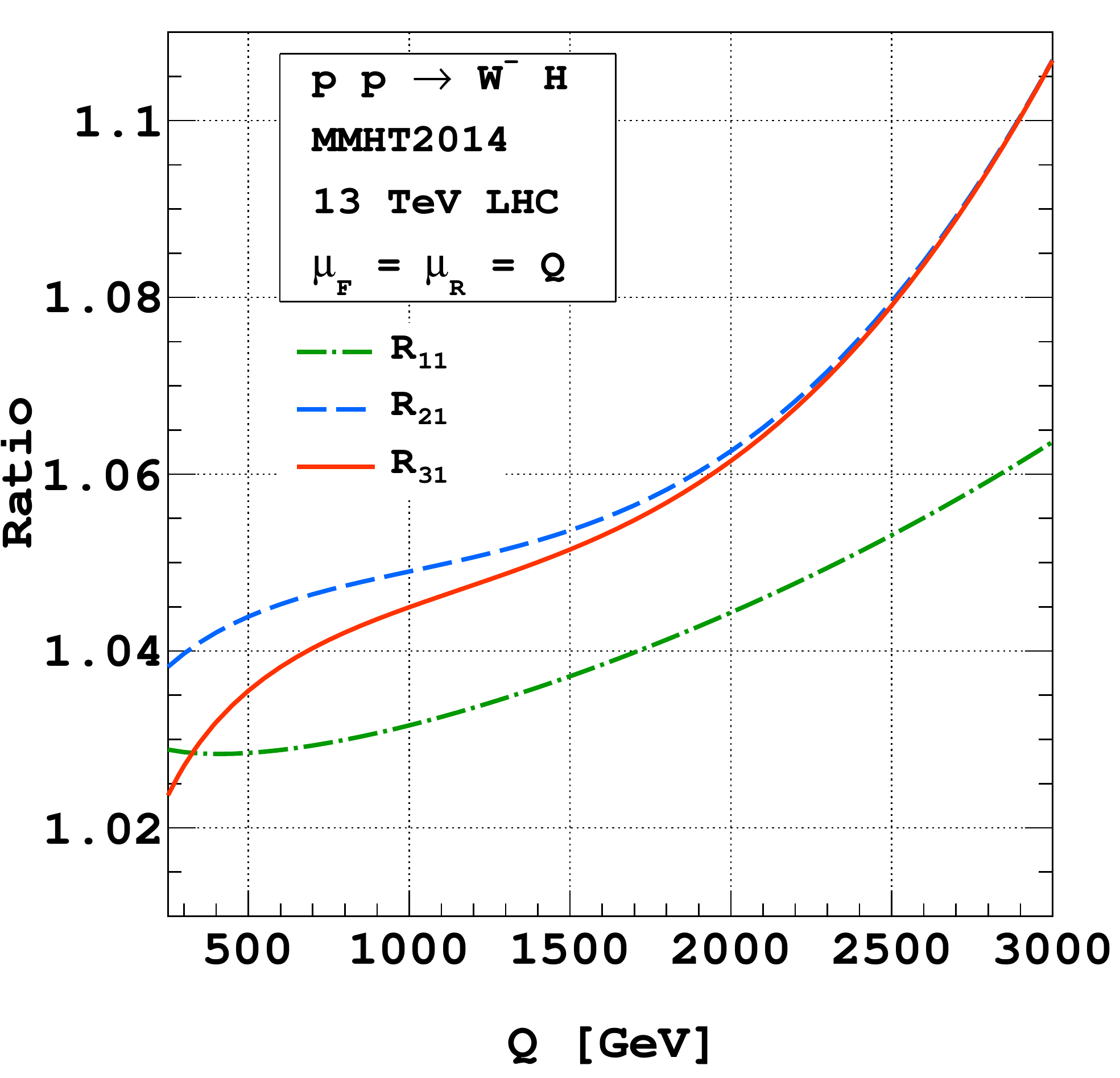}
                \includegraphics[width=5.5cm, height=5.5cm]{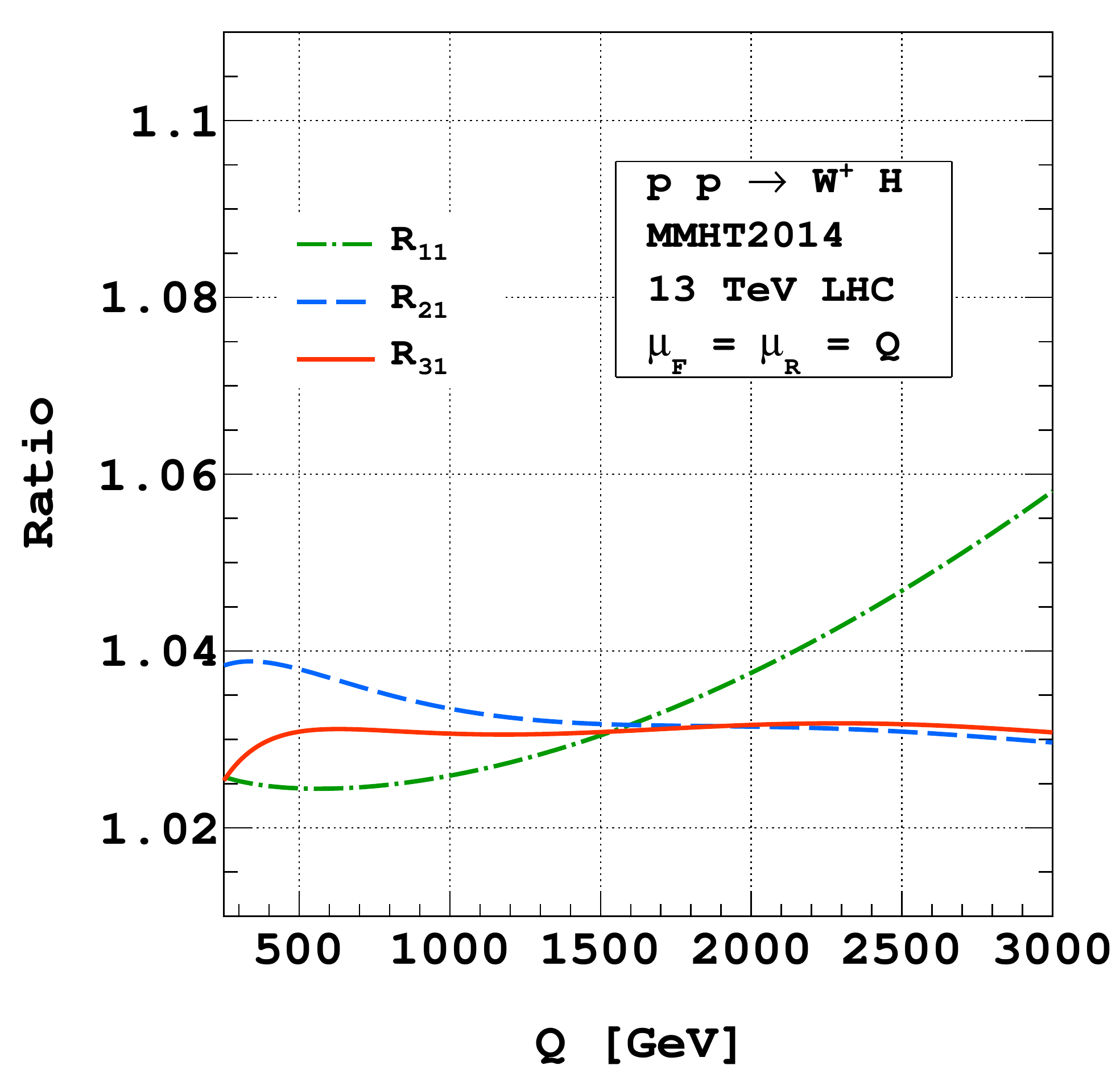}
        }
        \vspace{-2mm}
        \caption{\small{ Enhancement resummed cross section
			over NLO are shown here for different Higgs 
			associated production ZH (left panel), W$^-$H 
			(middle panel) and W$^+$H (right panel). The 
			factor $R_{ij}$ is defined in \eq{eq:ratio}.
			}}
        \label{fig:matched_r1fac_comp}
\end{figure}

We also present in \fig{fig:matched_r0fac_comp}, the 
ratios $R_{ii}$ for $i=1,2,3$ for $ZH$, $W^-H$ and $W^+H$ 
cases. 
It is also worth noting that with the automation of most 
of the NLO calculations 
\cite{Frederix:2018nkq,Alwall:2014hca}, the
NLO results are readily available for these processes and 
hence it will be particularly useful to estimate the 
$R_{i1}$ for $i=1,2,3$. We plot these $R_{i1}$ factors for 
all these $VH$ processes in \fig{fig:matched_r1fac_comp}
as a function of $Q$. We notice that in 
general approximately for $Q > 2000$ GeV, both $R_{21}$ 
and $R_{31}$ merge with each other. However, their values 
are different for different processes. In the high $Q$ 
region around $3000$ GeV, $R_{31}$ being the largest for 
$W^-H$ case and is about $1.105$ while it is the smallest 
and is about $1.03$ for $W^+H$ case, whereas the 
corresponding one for $ZH$ case is about $1.065$.
\onecolumngrid

\begin{figure}[ht!]
        \centerline{
                \includegraphics[width=5.5cm, height=5.5cm]{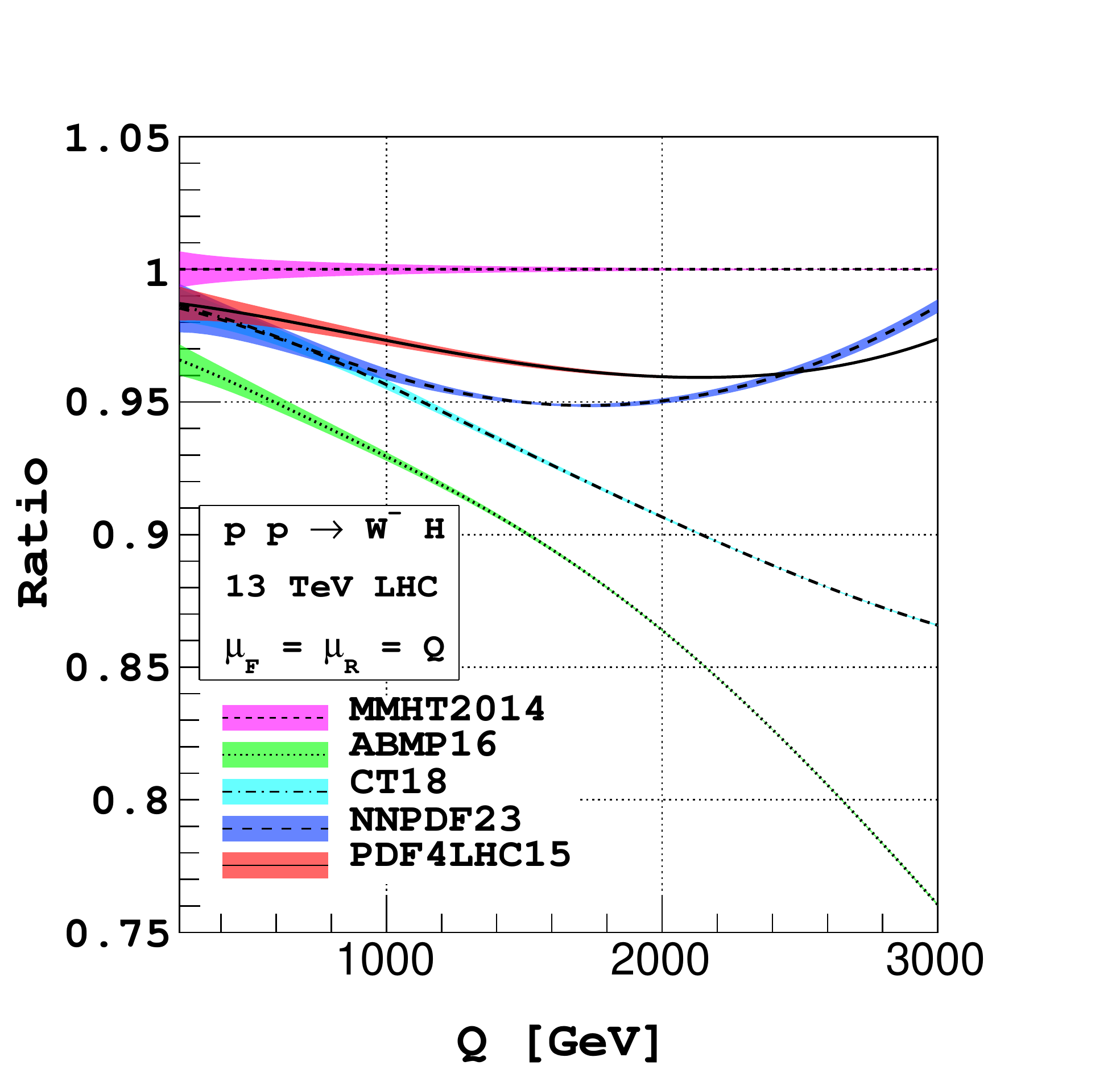}
                \includegraphics[width=5.5cm, height=5.5cm]{PDFvariation_n3ll.pdf}
                \includegraphics[width=5.5cm, height=5.5cm]{PDFvariation_n3ll.pdf}
        }
        \vspace{-2mm}
	\caption{
		\small{
			Behavior of invariant mass 
			distributions at N$^{3}$LO+N$^{3}$LL 
			for ZH (left panel), W$^-$H (middle panel) 
			and W$^+$H (right panel) with 7-point scale variation in band
			for different PDF groups 
			(central value) normalized to the default 
			choice of MMHT2014. }
			}
        \label{fig:pdf_uncertainty_comp}
\end{figure}
\begin{figure}[ht!]
	\centerline{
		\includegraphics[width=7.0cm, height=7.0cm]{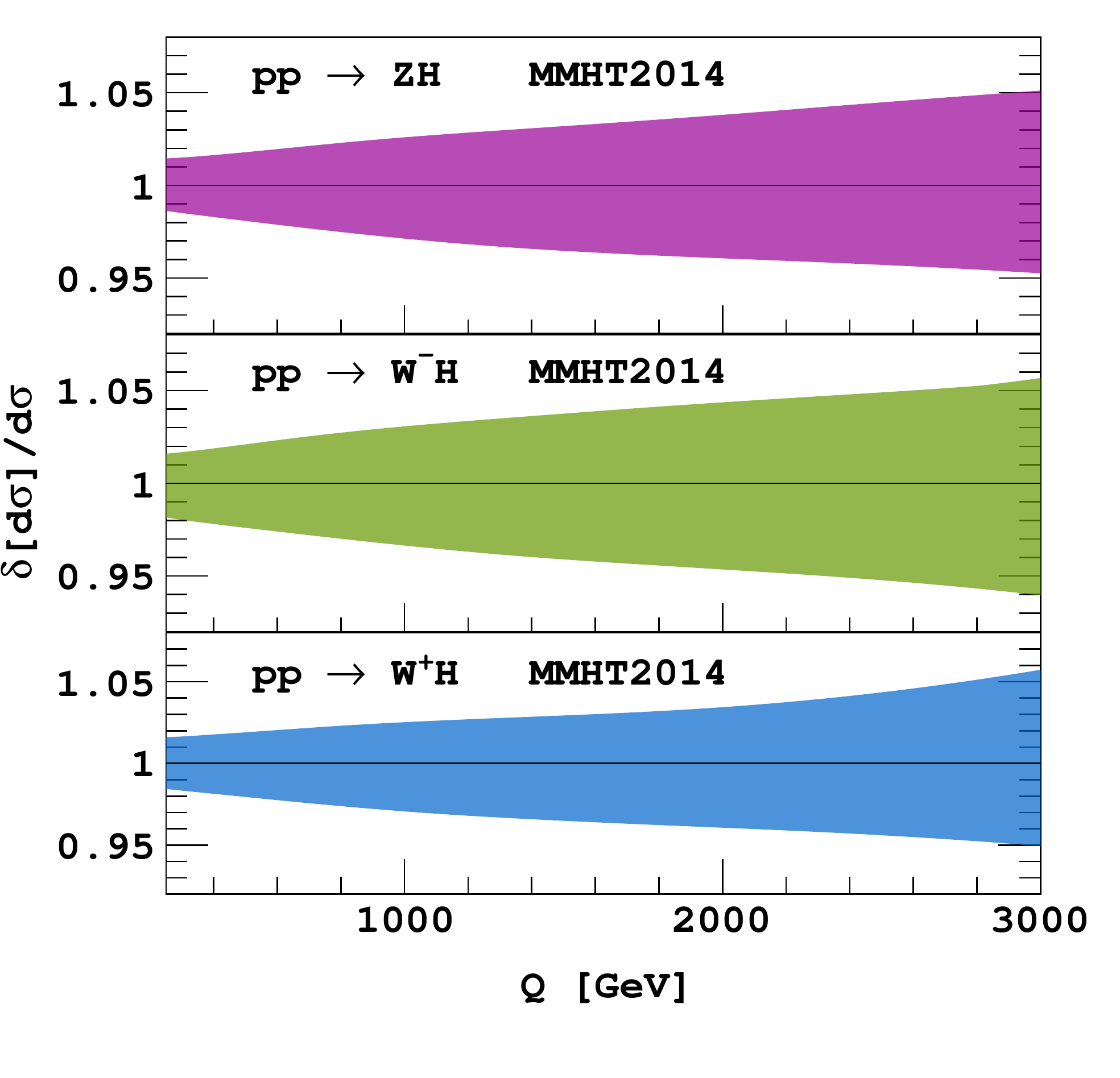}
	}
	\vspace{-2mm}
	\caption{\small{The intrinsic PDF 
	uncertainties in the 
	resummed predictions for $VH$ 
	production processes 
at N$^3$LO+N$^3$LL, obtained from 
MMHT2014nnlo68cl PDF sets.}}
	\label{fig:match_PDFuncertainty}
\end{figure}

For the $VH$ production process, we also give the total 
production cross sections obtained by integrating the 
invariant mass
distributions over the entire $Q$ region that is 
kinematically accessible. First, we give these total 
cross sections for DY type $ZH$ production processes from 
LO to N$^3$LO+N$^3$LL for different center of mass energies  
$\sqrt{S} = 7, 8, 13, 13.6 \text{ and } 100$ TeV in 
\tab{tab:tableZHnew}. 
The corresponding $7$-point scale uncertainties are also 
provided for each case.
For the total production cross sections, the bulk of the 
contribution comes essentially from the low $Q$ region and 
hence
the corresponding scale uncertainties in the resummed 
results are predominantly due to those coming from the 
FO results
that enter through matching procedure.  Hence, the scale 
uncertainties are smaller for FO case than those for the
resum case.
However, one can clearly see these scale uncertainties 
decrease for any center of mass energy
as we go from LO to N$^3$LO in FO, or as we go from LO+LL 
to N$^3$LO+N$^3$LL ~in the resummation series.
For example, for the $13.6$ TeV case, the scale uncertainties 
at LO are $4.06\%$ and get reduced to about $0.33\%$,
while those at LO+LL are $4.44\%$, and they get reduced to 
about $0.58\%$ at N$^3$LO+N$^3$LL.
In \tab{tab:tableWmH}, we present similar results for 
$W^-H$ case and in \tab{tab:tableWpH} for $W^+H$ case.
In all these results, the general observation is that 
the scale uncertainties do increase with the center of 
mass energy $\sqrt{S}$ of the incoming hadrons.

Finally, in the context of $VH$ production, we note that 
the DY type contributions do not fully give the complete 
FO predictions for $VH$ case. Starting from 
${\cal O} ({\as^2})$, particularly the $ZH$ process 
receives contributions
from the gluon fusion channel \cite{Kniehl:1990iva}, 
bottom annihilation channel as well as the heavy top-loop 
contributions \cite{Brein:2011vx}. For the gluon fusion
channel, the LO order contribution formally comes at the 
same order as NNLO of DY type $ZH$ production process and 
the results for this channel are available already. The 
higher order NLO corrections to this gluon fusion channel 
which contribute
at ${\cal O} (\as^3)$ level appear at the same order as 
the N$^3$LO of DY type corrections, and these NLO corrections
to the gluon fusion channel in the effective theory 
\cite{Altenkamp:2012sx} have been computed already. 
It is worth noting that in the recent past,
there has been a significant amount of progress toward
the computation of the NLO corrections to this gluon 
fusion process
considering the finite top quark mass effects 
\cite{Davies:2020drs,Chen:2020gae,Wang:2021rxu,Alasfar:2021ppe,Bellafronte:2022jmo,Chen:2022rua,Degrassi:2022mro}. 
The top-loop contributions are available at the $\as^2$ 
level \cite{Brein:2011vx} and are 
implemented in the {\tt vh@nnlo 2.1} code.  In the present 
context, we consider all these three contributions to 
the $ZH$ 
production process and define the total production 
cross section defined as
\begin{eqnarray}
	\sigma^{tot,ZH}_{\text{N}^3\text{LO}} = \text{\ncfodyzh} + \sigma^{gg} (\as^3) + \sigma^{\text{top}} (\as^2) + \sigma^{b\bar{b}} \, 
	\label{eq:ZHtotal_fo}
\end{eqnarray} 
\begin{eqnarray}
	\sigma^{tot,ZH}_{\text{N}^3\text{LO+N}^3\text{LL}} = \text{\nclldyzh} + \sigma^{gg} (\as^3) + \sigma^{\text{top}} (\as^2) + \sigma^{b\bar{b}} \,
	\label{eq:ZHtotal_resum}
\end{eqnarray} 
where the power of $\as$ in the parenthesis denotes the 
highest order up to which the respective contribution
has been taken into account. We present these results for 
the $ZH$ case in \tab{tab:tableZHcompare}. 
For the case $WH$ production, there will be no 
contribution from gluon fusion as well as bottom 
annihilation processes, however,
there will be contribution from top-loops from NNLO 
onwards. Hence, we define the total production 
cross sections for $WH$ case as
\begin{eqnarray}
\sigma^{tot,WH}_{\text{N}^3\text{LO}} = \text{\ncfodywh} + \sigma^{\text{top}} (\as^2) \,
	\label{eq:WHtotal_fo}
\end{eqnarray} 
\begin{eqnarray}
\sigma^{tot,WH}_{\text{N}^3\text{LO+N}^3\text{LL}} = \text{\nclldywh} + \sigma^{\text{top}} (\as^2) \cdot
	\label{eq:WHtotal_resum}
\end{eqnarray} 
These production cross sections up to N$^3$LO+N$^3$LL are 
given in \tab{tab:tableWmHcompare} for $W^-H$ and in 
\tab{tab:tableWpHcompare} for $W^+H$ cases.  
We also note that at this level, the electroweak 
corrections \cite{Ciccolini:2003jy,Denner:2011id} 
are also competitive. For the current LHC energies,
they are about $-5.28\%$ for $ZH$ and $-6.88\%$ for 
$WH$ total production cross sections. This can easily 
be implemented in the analysis, however, we did not 
consider them for simplicity and focus only on the QCD 
corrections.

\begin{table}[h!]
	\begin{center}
{\scriptsize		
\resizebox{15.0cm}{2.5cm}{
		\begin{tabular}{|c|c|c|c|c|c|}
\hline
			$\sqrt{S}$ (TeV) & $7.0$  & $8.0$  &  $13.0$  & 13.6  
			&  $100.0$  \\

\hline
\hline
LO                       &  $0.1625 \pm 0.46\%$  &  $0.2021 \pm 1.27\%$  &  $0.4262 \pm 4.04\%$  & $0.4553 \pm 4.28\%$  
			& $6.1486 \pm 13.49\%$\\
\hline
NLO                      &  $0.2144 \pm 1.55\%$  &  $ 0.2660 \pm 1.50\%$  &  $0.5524 \pm 1.37\%$  & $0.5891 \pm 1.44\%$ 
			&  $6.9628 \pm 4.39\%$\\
\hline
			NNLO                     &  $0.2234 \pm 0.39\%$  &  $0.2769 \pm 0.35\%$  &  $0.5716 \pm 0.38\%$  & $0.6093 \pm 0.39 \%$ 
			&  $6.9645 \pm 0.97\%$\\
\hline
			N$^3$LO           &  $0.2223 \pm 0.28\%$  &  $0.2752 \pm 0.30\%$  &  $0.5667 \pm 0.36\%$  & $0.6039 \pm 0.36 \%$ 
			&  $6.8460 \pm 0.54\%$\\
\hline
			LO+LL                    &  $0.1872 \pm 1.56\%$  &  $0.2320 \pm 1.68\%$  &  $0.4847 \pm 4.42\%$  & $0.5175 \pm 4.65\%$  
			&  $6.8494 \pm 13.71\%$\\
\hline
			NLO+NLL                  &  $0.2214 \pm 4.55\%$  &  $0.2744 \pm 4.53\%$  &  $0.5683 \pm 4.46\%$  & $0.6059 \pm 4.46 \%$ 
			&  $7.1248 \pm 5.39\%$\\
\hline
			NNLO+NNLL                &  $0.2245 \pm 1.49\%$  &  $0.2782 \pm 1.52\%$  &  $0.5741 \pm 1.60\%$  & $0.6118 \pm 1.61\%$ 
			&  $6.9867 \pm 1.79\%$\\
\hline
			N$^3$LO+N$^3$LL   &  $0.2223 \pm 0.53\%$  &  $0.2752 \pm 0.55\%$  &  $0.5667 \pm 0.63\%$  & $0.6039 \pm 0.63 \%$ 
			&  $6.8464 \pm 0.81\%$\\
\hline
\end{tabular}
 }
		\caption{\small{$W^{-}H$ production cross section (in pb) for different $\sqrt{S}$ with 7-point scale uncertainty.}} 
\label{tab:tableWmH}
 }
 \end{center}
\end{table}

\begin{table}[h!]
	\begin{center}
{\scriptsize		
\resizebox{13.0cm}{2.0cm}{
		\begin{tabular}{|c|c|c|c|c|}
\hline
			$\sqrt{S}$ (TeV) &
			$\sigma_{\text {N}^3\text {LO}}^{DY,W^-H}$  &  $\sigma_{\text {N}^3\text {LO+N}^3\text {LL}}^{DY,W^-H}$ & $\sigma_{\text {N}^3\text {LO}}^{tot,W^-H}$ 
			& $\sigma_{\text {N}^3\text {LO+N}^3\text {LL}}^{tot,W^-H}$  \\

\hline
\hline
7.0                       
			&  $0.2223 \pm 0.28\%$  & $0.2223 \pm 0.53\%$ & $ 0.2245 \pm 0.09\% $   & $ 0.2245 \pm 0.34\% $ \\ 
\hline
8.0                     
			&  $ 0.2752 \pm 0.30\%$ & $0.2752 \pm 0.55\%$ & $ 0.2781 \pm 0.08\% $ & $ 0.2781 \pm 0.36\%$  \\
\hline
13.0                   
			&  $0.5667 \pm 0.36\%$  & $0.5667 \pm 0.63 \%$ & $ 0.5739 \pm 0.13\%$ & $ 0.5739 \pm 0.42\%$  \\
\hline
13.6     		
			&  $0.6039 \pm 0.36\%$  & $0.6039 \pm 0.63 \%$ & $ 0.6117 \pm 0.13\%$  &  $ 0.6117 \pm 0.42\%$  \\
\hline
100.0                   
			&  $6.8460 \pm 0.54\%$  & $6.8464 \pm 0.81\%$  & $ 6.9963 \pm 0.35\%$  &  $ 6.9967 \pm 0.61\%$   \\ 
\hline
\end{tabular}
 }
		\caption{\small{$W^{-}H$ production cross section (in pb) of DY-type N$^3$LO, DY-type N$^3$LO+N$^3$LL, total N$^3$LO and total resummed N$^3$LO+N$^3$LL which are defined in \eq{eq:WHtotal_fo} and \eq{eq:WHtotal_resum} for different $\sqrt{S}$ with 7-point scale uncertainty.}} 
		\label{tab:tableWmHcompare}
 }
 \end{center}
\end{table}

\begin{table}[h!]
	\begin{center}
{\scriptsize		
\resizebox{15.0cm}{2.5cm}{
		\begin{tabular}{|c|c|c|c|c|c|}
\hline
			$\sqrt{S}$ (TeV) & $7.0$  & $8.0$  &  $13.0$  & 13.6  
			&  $100.0$  \\

\hline
\hline
LO                       &  $0.2820 \pm 0.30\%$  &  $0.3409 \pm 1.08\%$  &  $0.6583 \pm 3.77\%$  & $0.6983 \pm 4.00\%$  
			& $7.7331 \pm 13.06\%$\\
\hline
NLO                      &  $0.3805 \pm 1.57\%$  &  $ 0.4582 \pm 1.53\%$  &  $0.8676 \pm 1.40\%$  & $0.9183 \pm 1.40\%$ 
			&  $8.7881 \pm 4.38\%$\\
\hline
			NNLO                     &  $0.3944 \pm 0.42\%$  &  $0.4750 \pm 0.39\%$  &  $0.8975 \pm 0.37\%$  & $0.9497 \pm 0.38 \%$ 
			&  $8.7460 \pm 1.01\%$\\
\hline
			N$^3$LO           &  $0.3927 \pm 0.25\%$  &  $0.4725 \pm 0.27\%$  &  $0.8906 \pm 0.33\%$  & $0.9422 \pm 0.34 \%$ 
			&  $8.6026 \pm 0.54\%$\\
\hline
			LO+LL                    &  $0.3201 \pm 1.39\%$  &  $0.3861 \pm 1.49\%$  &  $0.7409 \pm 4.14\%$  & $0.7856 \pm 4.37\%$  
			&  $8.5904 \pm 13.33\%$\\
\hline
			NLO+NLL                  &  $0.3911 \pm 4.13\%$  &  $0.4707 \pm 4.12\%$  &  $0.8897 \pm 4.12\%$  & $0.9416 \pm 4.12 \%$ 
			&  $8.9811 \pm 5.26\%$\\
\hline
			NNLO+NNLL                &  $0.3960 \pm 1.34\%$  &  $0.4768 \pm 1.37\%$  &  $0.9008 \pm 1.46\%$  & $0.9532 \pm 1.47\%$ 
			&  $8.7719 \pm 1.68\%$\\
\hline
			N$^3$LO+N$^3$LL   &  $0.3927 \pm 0.45\%$  &  $0.4725 \pm 0.48\%$  &  $0.8907 \pm 0.56\%$  & $0.9423 \pm 0.57 \%$ 
			&  $8.6032 \pm 0.79\%$\\
\hline
\end{tabular}
 }
		\caption{\small{$W^{+}H$ production cross section (in pb) for different $\sqrt{S}$} with 7-point scale uncertainty.} 
\label{tab:tableWpH}
 }
 \end{center}
\end{table}
\begin{table}[h!]
	\begin{center}
{\scriptsize		
\resizebox{13.0cm}{2.00cm}{
		\begin{tabular}{|c|c|c|c|c|}
\hline
			$\sqrt{S}$ (TeV) &
			$\sigma_{\text {N}^3\text {LO}}^{DY,W^+H}$  &  $\sigma_{\text {N}^3\text {LO+N}^3\text {LL}}^{DY,W^+H}$ & $\sigma_{\text {N}^3\text {LO}}^{tot,W^+H}$ 
			& $\sigma_{\text {N}^3\text {LO+N}^3\text {LL}}^{tot,W^+H}$  \\

\hline
\hline
7.0                       
			&  $0.3927 \pm 0.25\%$  & $0.3927 \pm 0.45\%$ & $ 0.3965 \pm 0.11\% $   & $ 0.3965 \pm 0.27\% $ \\ 
\hline
8.0                     
			&  $ 0.4725 \pm 0.27\%$ & $0.4725 \pm 0.48\%$ & $ 0.4774 \pm 0.09\% $ & $ 0.4774 \pm 0.29\%$  \\
\hline
13.0                   
			&  $0.8906 \pm 0.33\%$  & $0.8907 \pm 0.56 \%$ & $ 0.9017 \pm 0.13\%$ & $ 0.9017 \pm 0.36\%$  \\
\hline
13.6     		
			&  $0.9422 \pm 0.34\%$  & $0.9423 \pm 0.57 \%$ & $ 0.9541 \pm 0.13\%$  &  $ 0.9542 \pm 0.36\%$  \\
\hline
100.0                   
			&  $8.6026 \pm 0.54\%$  & $8.6032 \pm 0.79\%$  & $ 8.7865 \pm 0.37\%$  &  $ 8.7871 \pm 0.60\%$   \\ 
\hline
\end{tabular}
 }
		\caption{\small{$W^{+}H$ production cross section (in pb) of DY-type N$^3$LO, DY-type N$^3$LO+N$^3$LL, total N$^3$LO and total resummed N$^3$LO+N$^3$LL which are defined in \eq{eq:WHtotal_fo} and \eq{eq:WHtotal_resum} for different $\sqrt{S}$} with 7-point scale uncertainty.} 
\label{tab:tableWpHcompare}
 }
 \end{center}
\end{table}

%
Finally, we estimate the uncertainties in our predictions 
for $VH$ invariant mass distributions due to the choice
of the PDFs used in our analysis. For this, we compute the 
invariant mass distributions at N$^3$LO+N$^3$LL using
different choice of PDFs : ABMP16 
\cite{Alekhin:2017kpj}, CT18 \cite{Hou:2019qau}, 
NNPDF23 \cite{Ball:2012cx} and 
PDF4LHC15 \cite{Butterworth:2015oua}. All these sets are 
considered at NNLO level
and the invariant mass distributions are obtained for 
central set (iset=0).
We obtain these results normalized with respect to 
those obtained from our default choice of MMHT2014nnlo 
PDFs (iset=0) and
for the central scale choice and present them in 
\fig{fig:pdf_uncertainty_comp} for $ZH$ (left panel), 
$W^-H$ (middle panel) and 
for $W^+H$ (right panel). The bands for each of the PDF 
sets in these figures
represent the corresponding $7$-point scale uncertainties, 
which are found to get reduced with $Q$ for all the PDF 
sets 
considered here except those for NNPDF23 sets. For the 
latter choice of PDF sets, the scale uncertainties are 
found to decrease
first with $Q$ up to about $1500$ GeV, and then they slowly 
increase with $Q$.
The uncertainty in these predictions due to the choice 
of different
PDF sets is smaller in the low $Q$-region and is at the 
most $4\%$, but these uncertainties tend to increase with 
$Q$.
In the high $Q$ region $Q > 2000$ GeV, the deviations 
from MMHT2014 results are in general larger for ABMP16
and CT18 PDF sets for $ZH$ and $W^-H$ processes. 
The deviations at $Q=2000$ GeV for $ZH$ case
are about $4.2\%$, $6.6\%$, $5.3\%$ and $4.0\%$ for ABMP16, CT18, NNPDF23
and PDF4LHC15 PDF sets respectively. The similar numbers
for $Q=3000$ GeV are about $9.0\%$, $8.3\%$, $2.9\%$ and $4.3\%$ respectively.
The deviations are largest for the 
case $W^-H$ in the high $Q$ region and are about 
$20\%$ for ABMP16 sets.

It is worth studying the intrinsic PDF uncertainties in our resummed 
predictions due to the parametrization of the PDFs themselves. For this,
we use the MMHT2014nnlo68cl PDF sets and compute the cross sections due
to all the $51$ different sets and estimate the asymmetric
uncertainties as provided by the {\tt LHAPDF} routines. We present these
uncertainties in our resummed results for $VH$ productin cross sections
at N$^3$LO+N$^3$LL accuracy normalized with respect to those obtained with the central set.
These results are shown in \fig{fig:match_PDFuncertainty} for the $ZH$, $W^-H$ and $W^+H$ processes, 
with uncertainties reaching about $5$\% in the high invariant mass region.

\subsection{$b\bar{b}H$ production}

\begin{figure}[ht!]
	\centerline{
		\includegraphics[width=10.0cm, height=10.0cm]{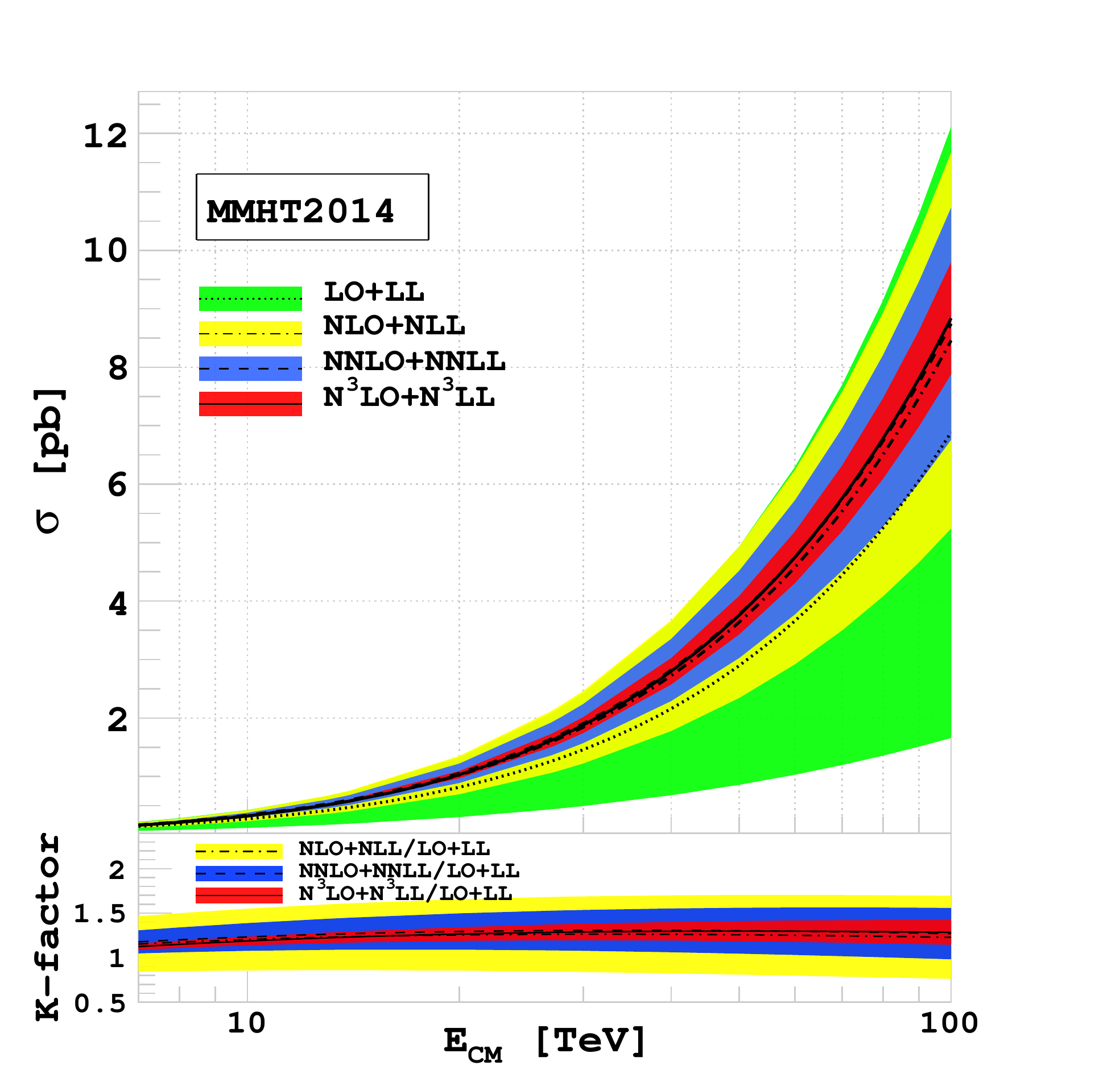}
	}
	\vspace{-2mm}
	\caption{\small{The $b\bar{b} H$ cross section 
	for different center of mass energy at 
	different order and the corresponding 
	K-factor with respect to LO+LL.}}
	\label{fig:match_cme_bbH}
\end{figure}
As a final process, we consider the Higgs boson production 
in bottom quark annihilation process at the LHC. This 
process has already been studied up to NNLO+NNLL by some 
of the authors in \cite{Ajjath:2019neu} including a 
detailed uncertainty analysis.
In \cite{Ajjath:2019neu}, the work also has been extended 
to N$^3$LO+N$^3$LL including the estimation of 
uncertainties only due to unphysical
renormalization scale. However, in this work, we complement 
these studies by considering the $b\bar{b} \to H$ production
cross sections along with a detailed $7$-point scale 
uncertainties for the choice of parameters given before.
We present in \fig{fig:match_cme_bbH} these production 
cross sections from LO+LL to N$^3$LO+N$^3$LL along with 
the respective scale 
uncertainties at hadron colliders for different center of 
mass energy $\sqrt{S}$ values from $7$ TeV to $100$ TeV. 
For $\sqrt{S} = 13.6$ TeV,  we find that these $7$-point 
scale uncertainties is $5.26\%$ at N$^3$LO and 
is $4.98\%$ at N$^3$LO+N$^3$LL. 
The scale uncertainties are found to get reduced with higher 
logarithmic accuracy in resummation while at any given order 
in perturbation theory they are in general found to increase 
with $\sqrt{S}$.

\section{Conclusions}\label{sec:conclusion}
We have studied the threshold effect on the color singlet 
processes DY, associated $VH$ productions and Higgs 
production through bottom quark annihilation. To achieve 
this, we have used the universal and process dependent 
resum coefficients that are known by some of us up to 
N$^3$LL. Thanks to the recent publicly available 
\texttt{n3loxs} code that we are able to perform the
complete matching at the third order in QCD thus 
incorporating missing regular contributions at this order. 
We performed a detailed phenomenology for neutral DY, 
charged DY production of dileptons, Higgs production in 
association with massive vector boson $(V=Z/W)$ in the 
DY-type process as well as for the Higgs production in 
bottom quark annihilation process. We presented our results 
for the dilepton and $VH$ invariant mass distributions
as well as the total production cross sections for $VH$ and 
$b\bar{b}H$ processes. For the case of $VH$ production 
process, the resummed corrections over the fixed order 
N$^3$LO results are found to be very small, thus 
demonstrating a very good convergence of the perturbation 
series. We observe that the FO results have smaller 
scale uncertainties in the low $Q$ region. However, 
the resummation gives significant reduction of the 
conventional $7$-point scale uncertainties to as small 
as $0.1\%$ in the high invariant mass region 
$Q \ge 1500$ GeV of $VH$ production processes.
For the associated Higgs production with a $Z$ boson, 
the gluon fusion channel and top-loop contributions from 
NNLO onward are also important 
\cite{Davies:2020drs,Chen:2020gae,Wang:2021rxu,Alasfar:2021ppe,
Bellafronte:2022jmo,Chen:2022rua,Degrassi:2022mro} 
at the accuracy that we considered here. We have included 
these contributions in our inclusive total production 
cross sections for $VH$ processes. To tame down the 
uncertainties further, we note that the resummation in 
gluon fusion channel is equally important. 
For the DY and DY-type $VH$ production processes we have 
considered here, the theory uncertainties after the 
resummation to N$^3$LL accuracy has been achieved are 
very much under control.  However, the PDF uncertainties are still large which can 
be reduced with the availability of N$^3$LO level PDFs and more
experimental data. Further, at this precision level the electroweak corrections 
are also important to bring the total theory uncertainties under control.

\section*{Acknowledgments}
We thank V.\ Ravindran for useful discussion. G.D. also 
thanks J.\ Baglio for the discussion related to 
\texttt{n3loxs}. C.D. would like to thank the cluster 
computing facility \textit{Param Ishan} at IIT Guwahati, 
where most of the computational work has been carried out. 
The research work of M.C.K. is supported by SERB Core 
Research Grant (CRG) under the project No. CRG/2021/005270.
The research of G.D. is supported by the 
Deutsche Forschungsgemeinschaft (DFG, German Research Foundation) 
under grant  No. 396021762 - TRR 257 
(\textit{Particle Physics Phenomenology after Higgs discovery.}).
The research work of K.S. is supported by Shanghai 
Natural Science Foundation under Grant No. 21ZR1406100.
\onecolumngrid
\appendix
\section{RESUMMATION COEFFICIENTS}
\label{appendixa}
The process-dependent $g_0$ coefficients defined in \eq{eq:g0} are given as 
(defining $L_{qr} = \ln \left(Q^2/\mur^2 \right), L_{fr} = \ln \left( \muf^2/\mur^2\right)$),
\begin{align} 
	\begin{autobreak} 
		\g01DY = 
		\Cf    \bigg\{ 
		- 16
		+ 16  \z2
		+ \bigg(-6\bigg)  \Lfr
		+ \bigg(6\bigg)  \Lqr \bigg\} ,   
	\end{autobreak} 
	\\ 
	\begin{autobreak} 
		\g02DY = 
		\Cf  \nf    \bigg\{ \frac{127}{6}
		+ \frac{8}{9}  \z3
		- \frac{64}{3}  \z2
		+ \bigg(
		- \frac{34}{3}
		+ \frac{16}{3}  \z2\bigg)  \Lqr
		+ \bigg(\frac{2}{3}
		+ \frac{16}{3}  \z2\bigg)  \Lfr
		+ \bigg(-2\bigg)  \Lfr^2
		+ \bigg(2\bigg)  \Lqr^2 \bigg\}      
		+ \Cf^2    \bigg\{ \frac{511}{4}
		- 60  \z3
		- 198  \z2
		+ \frac{552}{5}  \z2^2
		+ \bigg(
		- 93
		+ 48  \z3
		+ 72  \z2 \bigg)  \Lqr
		+ \bigg(93
		- 48  \z3
		- 72  \z2\bigg)  \Lfr
		+ \bigg(-36\bigg)  \Lqrfr
		+ \bigg(18\bigg)  \Lfr^2
		+ \bigg(18\bigg)   \Lqr^2 \bigg\}      
		+ \Ca  \Cf    \bigg\{ 
		- \frac{1535}{12}
		+ \frac{604}{9}  \z3
		+ \frac{376}{3}  \z2
		- \frac{92}{5}  \z2^2
		+ \bigg(
		- \frac{17}{3}
		+ 24  \z3
		- \frac{88}{3}  \z2\bigg)  \Lfr
		+ \bigg(\frac{193}{3}
		- 24  \z3
		- \frac{88}{3}  \z2\bigg)  \Lqr
		+ \bigg(-11\bigg)  \Lqr^2
		+ \bigg(11\bigg)  \Lfr^2 \bigg\} ,   
	\end{autobreak} 
	\\ 
	\begin{autobreak} 
		\g03DY = 
		\Cf  \nf^2    \bigg\{ 
		- \frac{7081}{243}
		+ \frac{16}{81}  \z3
		+ \frac{1072}{27}  \z2
		+ \frac{448}{135}  \z2^2
		+ \bigg(
		- \frac{68}{9}
		+ \frac{32}{9}  \z2\bigg)  \Lqr^2
		+ \bigg(
		- \frac{8}{9}\bigg)  \Lfr^3
		+ \bigg(\frac{4}{9}
		+ \frac{32}{9}  \z2\bigg)  \Lfr^2
		+ \bigg( \frac{8}{9}\bigg)  \Lqr^3
		+ \bigg(\frac{34}{9}
		+ \frac{32}{9}  \z3
		- \frac{160}{27}  \z2\bigg)  \Lfr
		+ \bigg(\frac{220}{9}
		- \frac{64}{27}  \z3
		- \frac{608}{27}  \z2\bigg)  \Lqr \bigg\}      
		+ \Cf^2  \nf    \bigg\{ 
		- \frac{421}{3}
		- \frac{608}{9}  \z5
		+ \frac{9448}{27}  \z3
		+ \frac{9064}{27}  \z2
		- \frac{256}{3}  \z2   \z3
		- \frac{36208}{135}  \z2^2
		+ \bigg(
		- 92
		+ 32  \z3
		+ 48  \z2\bigg)  \Lqr^2
		+ \bigg(
		- \frac{275}{3}
		+ \frac{256}{3}  \z3
		+ 40  \z2
		+ \frac{272}{5}  \z2^2\bigg)  \Lfr
		+ \bigg(20
		- 32  \z3
		- 48  \z2\bigg)  \Lfr^2
		+ \bigg(230
		- \frac{496}{3}  \z3
		- 272  \z2
		+ \frac{464}{5}  \z2^2\bigg)  \Lqr
		+ \bigg(-12\bigg)  \Lqrfrt
		+ \bigg(-12\bigg)  \Lqrtfr
		+ \bigg( 12\bigg)  \Lfr^3
		+ \bigg(12\bigg)  \Lqr^3
		+ \bigg(72\bigg)  \Lqrfr \bigg\}      
		+ \Cf^3    \bigg\{ 
		- \frac{5599}{6}
		+ 1328  \z5
		- 460  \z3
		+ 32  \z3^2
		+ \frac{2936}{3}  \z2
		- 400  \z2   \z3
		- \frac{5972}{5}  \z2^2
		+ \frac{169504}{315}  \z2^3
		+ \bigg(
		- \frac{1495}{2}
		+ 480  \z5
		+ 992  \z3
		+ 720  \z2
		- 704  \z2  \z3
		- \frac{1968}{5}  \z2^2\bigg)  \Lfr
		+ \bigg(
		- 270
		+ 288  \z3
		+ 144  \z2\bigg)   \Lfr^2
		+ \bigg(
		- 270
		+ 288  \z3
		+ 144  \z2\bigg)  \Lqr^2
		+ \bigg(540
		- 576  \z3
		- 288  \z2\bigg)   \Lqrfr
		+ \bigg(\frac{1495}{2}
		- 480  \z5
		- 992  \z3
		- 720  \z2
		+ 704  \z2  \z3
		+ \frac{1968}{5}  \z2^2\bigg)   \Lqr
		+ \bigg(-108\bigg)  \Lqrtfr
		+ \bigg(-36\bigg)  \Lfr^3
		+ \bigg(36\bigg)  \Lqr^3
		+ \bigg(108\bigg)  \Lqrfrt \bigg\}      
		+ \Ca  \Cf  \nf    \bigg\{ \frac{110651}{243}
		- 8  \z5
		- \frac{24512}{81}  \z3
		- \frac{44540}{81}  \z2
		+ \frac{880}{9}  \z2   \z3
		+ \frac{1156}{135}  \z2^2
		+ \bigg(
		- \frac{3052}{9}
		+ \frac{3440}{27}  \z3
		+ \frac{7504}{27}  \z2
		- \frac{344}{15}   \z2^2\bigg)  \Lqr
		+ \bigg(
		- 40
		- \frac{400}{9}  \z3
		+ \frac{2672}{27}  \z2
		- \frac{8}{5}  \z2^2\bigg)  \Lfr
		+ \bigg(
		- \frac{146}{9}
		+ 16  \z3
		- \frac{352}{9}  \z2\bigg)  \Lfr^2
		+ \bigg(
		- \frac{88}{9}\bigg)  \Lqr^3
		+ \bigg(\frac{88}{9}\bigg)  \Lfr^3
		+ \bigg(\frac{850}{9}
		- 16  \z3
		- \frac{352}{9}  \z2\bigg)  \Lqr^2 \bigg\}      
		+ \Ca  \Cf^2    \bigg\{ \frac{74321}{36}
		- \frac{5512}{9}  \z5
		- \frac{51508}{27}  \z3
		+ \frac{592}{3}  \z3^2
		- \frac{66544}{27}  \z2
		+ \frac{3680}{3}  \z2  \z3
		+ \frac{258304}{135}  \z2^2
		- \frac{123632}{315}  \z2^3
		+ \bigg(
		- \frac{3439}{2}
		+ 240  \z5
		+ \frac{5368}{3}  \z3
		+ 1552  \z2
		- 352  \z2  \z3
		- \frac{2912}{5}  \z2^2\bigg)  \Lqr
		+ \bigg(
		- 420
		+ 288  \z3\bigg)  \Lqrfr
		+ \bigg(
		- 131
		+ 32  \z3
		+ 264  \z2\bigg)  \Lfr^2
		+ \bigg(551
		- 320  \z3
		- 264  \z2\bigg)  \Lqr^2
		+ \bigg(\frac{2348}{3}
		- 240  \z5
		- \frac{4048}{3}  \z3
		- 100  \z2
		+ 352  \z2  \z3
		- \frac{1136}{5}  \z2^2\bigg)  \Lfr
		+ \bigg(-66\bigg)  \Lfr^3
		+ \bigg(-66\bigg)  \Lqr^3
		+ \bigg(66\bigg)  \Lqrfrt
		+ \bigg(66\bigg)   \Lqrtfr \bigg\}      
		+ \Ca^2  \Cf    \bigg\{ 
		- \frac{1505881}{972}
		- 204  \z5
		+ \frac{139345}{81}  \z3
		- \frac{400}{3}  \z3^2
		+ \frac{130295}{81}  \z2
		- \frac{7228}{9}  \z2  \z3
		- \frac{23357}{135}  \z2^2
		+ \frac{7088}{63}  \z2^3
		+ \bigg(
		- \frac{2429}{9}
		+ 88  \z3
		+ \frac{968}{9}  \z2\bigg)  \Lqr^2
		+ \bigg(
		- \frac{242}{9}\bigg)  \Lfr^3
		+ \bigg(\frac{242}{9}\bigg)  \Lqr^3
		+ \bigg(\frac{493}{9}
		- 88  \z3
		+ \frac{968}{9}  \z2\bigg)  \Lfr^2
		+ \bigg(\frac{1657}{18}
		- 80  \z5
		+ \frac{3104}{9}  \z3
		- \frac{8992}{27}  \z2
		+ 4  \z2^2\bigg)  \Lfr
		+ \bigg(\frac{3082}{3}
		+ 80  \z5
		- \frac{22600}{27}  \z3
		- \frac{20720}{27}   \z2
		+ \frac{1964}{15}  \z2^2\bigg)  \Lqr \bigg\}      
		+ N_{4}  \nfv  \Cf    \bigg\{ 8
		- \frac{160}{3}  \z5
		+ \frac{28}{3}  \z3
		+ 20  \z2
		- \frac{4}{5}  \z2^2 \bigg\} \,. 
	\end{autobreak} 
\end{align}
Here $N_4 = (n_c^2-4)/n_c$ and $n_{fv}$ is proportional to the charge weighted sum of quark 
flavors \cite{Gehrmann:2010ue}.

The process-independent universal resum exponent defined in \eq{eq:gn} which 
are used for DY-type processes are given as,
\begin{align} 
	\begin{autobreak} 
		\gNB1 =
		\bigg[ \AAo  ~  \bigg\{ 2
		- 2 ~ \LogmW1
		+ 2 ~ \LogmW1 ~ \iW \bigg\}
		\bigg],   
	\end{autobreak} 
	\\ 
	\begin{autobreak} 
		\gNB2 =
		\bigg[ \DDo  ~  \bigg\{ \frac{1}{2} ~ \LogmW1 \bigg\}      
		+ \AAt  ~  \bigg\{ 
		- \LogmW1
		- \w \bigg\}      
		+ \AAo  ~  \bigg\{ \bigg(\LogmW1
		+ \frac{1}{2} ~ \LogmW1^2
		+ \w\bigg) ~ \bigg(\frac{\beta_{1}}{\beta_0^{2}}\bigg)
		+ \bigg(\w\bigg) ~ \Lfr
		+ \bigg(\LogmW1\bigg) ~ \Lqr \bigg\}
		\bigg],   
	\end{autobreak} 
	\\ 
	\begin{autobreak} 
		\gNB3 =
		\bigg[ \btzAIII  ~  \bigg\{ 
		- \WbimW
		+ \w \bigg\}      
		+ \btzAII  ~  \bigg\{ \bigg(2 ~ \WbimW\bigg) ~ \Lqr
		+ \bigg(3 ~ \WbimW
		+ 2 ~ \LogomWtIMW
		- \w\bigg) ~ \bigg(\frac{\beta_{1}}{\beta_0^{2}}\bigg)
		+ \bigg(
		- 2 ~ \w\bigg) ~  \Lfr \bigg\}      
		+ \btzAI  ~  \bigg\{ 
		- 4 ~ \z2 ~ \WbimW
		+ \bigg(
		- \LogtmWtIMW
		- \WbimW
		- 2 ~ \LogomWtIMW
		+ 2 ~  \LogmW1
		+ \w\bigg) ~ \bigg(\frac{\beta_{1}}{\beta_0^{2}}\bigg)^2
		+ \bigg(
		- \WbimW\bigg) ~ \Lqr^2
		+ \bigg(
		- \WbimW
		- 2 ~ \LogmW1
		- \w\bigg) ~ \btoo
		+ \bigg(\bigg(
		- 2 ~ \WbimW
		- 2 ~ \LogomWtIMW\bigg) ~ \bigg(\frac{\beta_{1}}{\beta_0^{2}}\bigg)\bigg) ~ \Lqr
		+ \bigg(\w\bigg) ~ \Lfr^2 \bigg\}      
		+ \btzDII  ~  \bigg\{ \WbimW \bigg\}      
		+ \btzDI  ~  \bigg\{ \bigg(
		- \WbimW\bigg) ~ \Lqr
		+ \bigg(
		- \WbimW
		- \LogomWtIMW\bigg) ~ \bigg(\frac{\beta_{1}}{\beta_0^{2}}\bigg) \bigg\}
		\bigg],   
	\end{autobreak} 
	\\ 
	\begin{autobreak} 
		\gNB4 =
		\bigg[ \btztAIV  ~  \bigg\{ \frac{1}{6} ~ \WttmWtimWt
		- \frac{1}{3} ~ \w \bigg\}      
		+ \btztAIII  ~  \bigg\{ \bigg(
		- \frac{1}{2} ~ \WttmWtimWt\bigg) ~ \Lqr
		+ \bigg(
		- \frac{5}{12} ~ \WttmWtimWt
		- \frac{1}{2} ~  \LogomWtIMWt
		+ \frac{1}{3} ~ \w\bigg) ~ \bigg(\frac{\beta_{1}}{\beta_0^{2}}\bigg)
		+ \bigg(\w\bigg) ~ \Lfr \bigg\}      
		+ \btztAII  ~  \bigg\{ 2 ~ \z2 ~ \WttmWtimWt
		+ \bigg(\frac{1}{2} ~ \LogtmWtIMWt
		- \frac{1}{12} ~ \WtbimWt
		+ \frac{5}{6} ~  \WbimW
		+ \frac{1}{2} ~ \LogomWtIMWt
		- \frac{1}{3} ~ \w\bigg) ~ \bigg(\frac{\beta_{1}}{\beta_0^{2}}\bigg)^2
		+ \bigg(\frac{1}{2} ~ \WttmWtimWt\bigg) ~ \Lqr^2
		+ \bigg(\frac{1}{3}  ~ \WtbimWt
		- \frac{1}{3} ~ \WbimW
		+ \frac{1}{3} ~ \w\bigg) ~ \btoo
		+ \bigg(
		- \w\bigg) ~ \Lfr^2
		+ \bigg(\bigg(\frac{1}{2} ~ \WttmWtimWt
		+ \LogomWtIMWt\bigg) ~ \bigg(\frac{\beta_{1}}{\beta_0^{2}}\bigg)\bigg) ~ \Lqr \bigg\}      
		+ \btztAI  ~  \bigg\{ \frac{8}{3} ~ \z3 ~ \WttmWtimWt
		+ \bigg(
		- \frac{1}{6} ~ \LogttmWtIMWt
		+ \frac{1}{3} ~ \WtbimWt
		- \frac{1}{3} ~ \WbimW
		+ \frac{1}{2} ~ \LogomWtIMWt
		- \LogomWtIMW
		+ \frac{1}{2} ~ \LogmW1
		+ \frac{1}{3} ~ \w\bigg) ~ \bigg(\frac{\beta_{1}}{\beta_0^{2}}\bigg)^3
		+ \bigg(
		- \frac{1}{6} ~ \WttmWtimWt\bigg) ~ \Lqr^3
		+ \bigg(\frac{1}{12} ~ \WttmWtimWt
		+ \frac{1}{2} ~ \LogmW1
		+ \frac{1}{3} ~ \w\bigg) ~ \bthr
		+ \bigg(
		- \frac{5}{12} ~ \WtbimWt
		+ \frac{1}{6} ~ \WbimW
		- \frac{1}{2} ~ \LogomWtIMWt
		+ \LogomWtIMW
		- \LogmW1
		- \frac{2}{3} ~ \w\bigg) ~ \bobt
		+ \bigg(\frac{1}{3} ~ \w\bigg) ~ \Lfr^3
		+ \bigg(
		- 2 ~ \z2 ~ \WttmWtimWt
		+ \bigg(
		- \frac{1}{2} ~  \LogtmWtIMWt
		+ \frac{1}{2} ~ \WtbimWt\bigg) ~ \bigg(\frac{\beta_{1}}{\beta_0^{2}}\bigg)^2
		+ \bigg(
		- \frac{1}{2} ~ \WtbimWt\bigg) ~ \btoo\bigg) ~ \Lqr
		+ \bigg(
		- 2 ~ \z2  ~ \LogomWtIMWt\bigg) ~ \bigg(\frac{\beta_{1}}{\beta_0^{2}}\bigg)
		+ \bigg(\bigg(
		- \frac{1}{2} ~ \LogomWtIMWt\bigg) ~ \bigg(\frac{\beta_{1}}{\beta_0^{2}}\bigg)\bigg) ~ \Lqr^2
		+ \bigg(\bigg(
		- \frac{1}{2} ~ \w\bigg) ~ \bigg(\frac{\beta_{1}}{\beta_0^{2}}\bigg)\bigg) ~  \Lfr^2 \bigg\}      
		+ \btztDIII  ~  \bigg\{ 
		- \frac{1}{4} ~ \WttmWtimWt \bigg\}      
		+ \btztDII  ~  \bigg\{ \bigg(\frac{1}{4} ~ \WttmWtimWt
		+ \frac{1}{2} ~ \LogomWtIMWt\bigg) ~ \bigg(\frac{\beta_{1}}{\beta_0^{2}}\bigg)
		+ \bigg(\frac{1}{2} ~ \WttmWtimWt\bigg) ~  \Lqr \bigg\}      
		+ \btztDI  ~  \bigg\{ 
		- \z2 ~ \WttmWtimWt
		+ \bigg(
		- \frac{1}{4} ~ \LogtmWtIMWt
		+ \frac{1}{4} ~ \WtbimWt\bigg) ~ \bigg(\frac{\beta_{1}}{\beta_0^{2}}\bigg)^2
		+ \bigg(
		- \frac{1}{4} ~ \WttmWtimWt\bigg) ~ \Lqr^2
		+ \bigg(
		- \frac{1}{4} ~ \WtbimWt\bigg) ~ \btoo
		+ \bigg(\bigg(
		- \frac{1}{2} ~  \LogomWtIMWt\bigg) ~ \bigg(\frac{\beta_{1}}{\beta_0^{2}}\bigg)\bigg) ~ \Lqr \bigg\}
		\bigg].
	\end{autobreak} 
\end{align}
Here $A_i$ are the universal cusp anomalous dimensions, $D_i$ are the threshold 
noncusp anomalous dimensions, and $\omega= 2 a_S \beta_0 \ln \overbar{N} $. 
Note that all the perturbative quantities are expanded in powers of $a_S$.
The cusp anomalous dimensions $A_i$ are given as (with the recently known four-loops results \cite{Henn:2019swt, Huber:2019fxe, vonManteuffel:2020vjv}) ,
\begin{align} 
	\begin{autobreak} 
		\A1 = C_{F}
		\bigg\{ 4
		\bigg\},   
	\end{autobreak} 
	\\ 
	\begin{autobreak} 
		\A2 = C_{F}
		\bigg\{ \nf    \bigg( 
		- \frac{40}{9} \bigg)      
		+ \Ca    \bigg( \frac{268}{9}
		- 8  \z2 \bigg)
		\bigg\},   
	\end{autobreak} 
	\\ 
	\begin{autobreak} 
		\A3 = C_{F}
		\bigg\{ \nf^2    \bigg( 
		- \frac{16}{27} \bigg)      
		+ \Cf  \nf    \bigg( 
		- \frac{110}{3}
		+ 32  \z3 \bigg)      
		+ \Ca  \nf    \bigg( 
		- \frac{836}{27}
		- \frac{112}{3}  \z3
		+ \frac{160}{9}  \z2 \bigg)      
		+ \Ca^2    \bigg( \frac{490}{3}
		+ \frac{88}{3}  \z3
		- \frac{1072}{9}  \z2
		+ \frac{176}{5}  \z2^2 \bigg)
		\bigg\}, 
	\end{autobreak} 
	\\
	\begin{autobreak} 
		\A4 = C_{F}
		\bigg\{ 
		\nf^3    \bigg( 
		- \frac{32}{81}
		+ \frac{64}{27}  \z3 \bigg)      
		+ \Cf  \nf^2    \bigg( \frac{2392}{81}
		- \frac{640}{9}  \z3
		+ \frac{64}{5}  \z2^2 \bigg)      
		+ \Cf^2  \nf    \bigg( \frac{572}{9}
		- 320  \z5
		+ \frac{592}{3}  \z3 \bigg)      
		+ \Ca  \nf^2    \bigg( \frac{923}{81}
		+ \frac{2240}{27}  \z3
		- \frac{608}{81}  \z2
		- \frac{224}{15}  \z2^2 \bigg)      
		+ \Ca  \Cf  \nf    \bigg( 
		- \frac{34066}{81}
		+ 160  \z5
		+ \frac{3712}{9}  \z3
		+ \frac{440}{3}  \z2
		- 128  \z2  \z3
		- \frac{352}{5}  \z2^2 \bigg)      
		+ \Ca^2  \nf    \bigg( 
		- \frac{24137}{81}
		+ \frac{2096}{9}  \z5
		- \frac{23104}{27}  \z3
		+ \frac{20320}{81}  \z2
		+ \frac{448}{3}  \z2  \z3
		- \frac{352}{15}  \z2^2 \bigg)      
		+ \Ca^3    \bigg( \frac{84278}{81}
		- \frac{3608}{9}  \z5
		+ \frac{20944}{27}  \z3
		- 16  \z3^2
		- \frac{88400}{81}  \z2
		- \frac{352}{3}  \z2  \z3
		+ \frac{3608}{5}  \z2^2
		- \frac{20032}{105}  \z2^3 \bigg)
		\bigg\}
		+ \dRAoNR     \bigg( \frac{3520}{3}  \z5
		+ \frac{128}{3}  \z3
		- 384  \z3^2
		- 128  \z2
		- \frac{7936}{35}   \z2^3 \bigg)      
		+ \nf  \dRFoNR    \bigg( 
		- \frac{1280}{3}  \z5
		- \frac{256}{3}  \z3
		+ 256  \z2 \bigg)      \,. 
	\end{autobreak} 
\end{align}

The quartic casimirs are given as
\begin{align}
\frac{d_A^{abcd}d_A^{abcd}}{N_A} &= \frac{n_c^2 (n_c^2 + 36)}{24},
\frac{d_A^{abcd}d_F^{abcd}}{N_A} = \frac{n_c (n_c^2 + 6)}{48}, \nn\\
\frac{d_F^{abcd}d_A^{abcd}}{N_F} &=  \frac{(n_c^2-1)(n_c^2+6)}{48},
\frac{d_F^{abcd}d_F^{abcd}}{N_F} = \frac{(n_c^2-1)(n_c^4 - 6 n_c^2 + 18)}{96n_c^3},
\end{align}
with $N_A = n_c^2 -1$ and $N_F = n_c$ where $n_c = 3$ for QCD.
The coefficients $D_i$ are given as,
\begin{align} 
	\begin{autobreak} 
		D_1 = C_{F}
		\bigg\{0
		\bigg\},   
	\end{autobreak} 
	\\ 
	\begin{autobreak} 
		D_2 = C_{F}
		\bigg\{ \nf    \bigg( \frac{224}{27}
		- \frac{32}{3}  \z2 \bigg)      
		+ \Ca    \bigg( 
		- \frac{1616}{27}
		+ 56  \z3
		+ \frac{176}{3}  \z2 \bigg)
		\bigg\},   
	\end{autobreak} 
	\\ 
	\begin{autobreak} 
		D_3 = C_{F}
		\bigg\{ \nf^2    \bigg( 
		- \frac{3712}{729}
		+ \frac{320}{27}  \z3
		+ \frac{640}{27}  \z2 \bigg)      
		+ \Cf  \nf    \bigg( \frac{3422}{27}
		- \frac{608}{9}  \z3
		- 32  \z2
		- \frac{64}{5}  \z2^2 \bigg)      
		+ \Ca  \nf    \bigg( \frac{125252}{729}
		- \frac{2480}{9}  \z3
		- \frac{29392}{81}  \z2
		+ \frac{736}{15}  \z2^2 \bigg)      
		+ \Ca^2    \bigg( 
		- \frac{594058}{729}
		- 384  \z5
		+ \frac{40144}{27}  \z3
		+ \frac{98224}{81}  \z2
		- \frac{352}{3}   \z2  \z3
		- \frac{2992}{15}  \z2^2 \bigg)
		\bigg\}\,.
	\end{autobreak} 
\end{align}

\bibliographystyle{apsrev4-2}
\bibliography{n3ll}
\end{document}